\documentclass[fleqn,usenatbib]{mnras}

\usepackage{newtxtext,newtxmath}
\usepackage{xspace}

\usepackage[T1]{fontenc}

\DeclareRobustCommand{\VAN}[3]{#2}
\let\VANthebibliography\thebibliography
\def\thebibliography{\DeclareRobustCommand{\VAN}[3]{##3}\VANthebibliography}

\usepackage{graphicx}	
\usepackage{booktabs}
\usepackage{xcolor}

\newcommand{\nickel}{\ensuremath{^{56}\mathrm{Ni}}}

\newcommand{\artis}{\textsc{artis}\xspace}
\newcommand{\artisnlte}{\textsc{artis-nlte}\xspace}
\newcommand{\artiscl}{\textsc{artis-classic}\xspace}

\newcommand{\msun}{\ensuremath{\mathrm{M}_\odot}}

\newcommand{\subsubsubsection}[1]{%
\vspace{0.6em}
\noindent\textit{#1}\par
\vspace{0.5em}
}

\title[2D early-phase NLTE radiative transfer]{Multi-dimensional NLTE radiative transfer for a double detonation Type Ia explosion model in the photospheric phase}

\author[F. P. Callan et al.]{F. P. Callan,$^{1}$\thanks{E-mail: f.callan@qub.ac.uk}
L. J. Shingles,$^1$ S. A. Sim,$^{1}$ C. E. Collins,$^2$ J. M. Pollin,$^3$ R. Pakmor,$^4$ \newauthor S. Gronow,$^5$ F. K. R\"opke,$^{5,6,7}$ A. Holas$^{5}$ and I. R. Seitenzahl$^{8,5}$
\\
$^{1}$School of Mathematics and Physics, Queen's University Belfast, University Road, Belfast BT7 1NN, UK \\
$^{2}$School of Physics, Trinity College Dublin, The University of Dublin, Dublin 2, Ireland\\
$^{3}$Department of Physics, Oregon State University, 301 Weniger Hall, Corvallis, OR 97331-6507, USA\\
$^4$Max-Planck-Institut f\"{u}r Astrophysik, Karl-Schwarzschild-Str. 1, D-85748, Garching, Germany\\
$^5$Heidelberger Institut f\"ur Theoretische Studien, Schloss-Wolfsbrunnenweg 35, D-69118, Heidelberg, Germany\\
$^6$Zentrum f{\"u}r Astronomie der Universit{\"a}t Heidelberg, Institut f{\"u}r Theoretische Astrophysik, Philosophenweg 12, D-69120 Heidelberg, Germany\\
$^7$Zentrum f{\"u}r Astronomie der Universit{\"a}t Heidelberg, Astronomisches Rechen-Institut, M{\"o}nchhofstr, 12--14, 69120 Heidelberg, Germany\\
$^8$Research School of Astronomy and Astrophysics, Australian National University, Canberra ACT 2611, Australia\\
}

\date{Accepted 22/10/2026}

\pubyear{2026}

\begin{document}
\label{firstpage}
\pagerange{\pageref{firstpage}--\pageref{lastpage}}
\maketitle

\begin{abstract}
Previous radiative transfer calculations have shown that both multi-dimensionality and NLTE (non local thermodynamic equilibrium) effects impact the synthetic observables predicted for Type Ia supernova explosion models. Here we carry out a 2D NLTE radiative transfer simulation in the photospheric phase for a double detonation model with a 1\,\msun\ carbon--oxygen core and 0.02\,\msun\ helium shell. The predicted observables demonstrate that departures from both spherical symmetry and local thermodynamic equilibrium are significant on scales relevant to comparisons with observations. The NLTE treatment leads to a bluer spectral energy distribution, more slowly declining optical-band light curves, changes to the near-infrared light curve shapes, and differences in the evolution of key spectral features. Although substantial viewing-angle variation is predicted for the model, the scale of the variation is not significantly impacted by NLTE effects up to peak. However, after peak, our 2D NLTE simulation retains a strong viewing-angle dependence in its spectra whereas our simulation of the same ejecta model using an approximate NLTE treatment predicts significantly reduced spectroscopic viewing-angle variation. The NLTE treatment leads to improved agreement with normal Type Ia supernovae, primarily due to the increased ionisation state of the simulation, although the absence of a secondary near-infrared maximum suggests that the ejecta remain over-ionised relative to observations after peak. Despite the improved agreement, many viewing angles still show excessive line blanketing, suggesting that the 0.02\,\msun\ helium-shell mass of the current model is too large to reproduce normal Type Ia supernovae over the majority of lines of sight.
\end{abstract}




\section{Introduction}
\label{sec:intro}
Type Ia supernovae (SNe~Ia) are the thermonuclear explosions of white dwarfs (WDs) in close binary systems. Despite decades of theoretical and observational efforts, their explosion mechanisms and progenitor systems remain a mystery, with many competing explosion models proposed (see e.g. \citealt{liu2023a, ruiter2025a} for recent reviews). 

The \textit{double detonation} (see e.g.\,\citealt{nomoto1980a, nomoto1982a, taam1980a, livne1990a, woosley1994a, hoeflich1996a, nugent1997a}) is a widely studied sub-Chandrasekhar mass (sub-$M_{\mathrm{Ch}}$) explosion mechanism that can potentially explain SNe~Ia across a range of luminosities. In this scenario, a He detonation ignites in a surface He layer (He-shell) on a carbon–oxygen (CO) WD, triggering a secondary off-centre detonation in the core that completely unbinds the star. Recent double detonation simulations invoking low He-shell masses (< 0.1\,\msun) have shown broad agreement with the observed properties of normal SNe~Ia \citep{townsley2019a, shen2021b}. However, the significant abundances of heavy elements synthesised in the He-shell detonation in the outer ejecta layers--particularly Ti, Cr, and iron-group elements (IGEs; here defined as Fe, Co, and Ni)--can cause strong line blanketing at blue wavelengths, leading to overly red light curves inconsistent with normal SNe Ia at early times (e.g. \citealt{kromer2010a, woosley2011a, sim2012a, polin2019a, gronow2020a, gronow2021a, shen2021b, collins2022a, pollin2024a, holas2025a}). 

Early 1D NLTE (non local thermodynamic equilibrium) radiative transfer simulations of double detonation models adopted large He-shell masses (${\sim}$0.2\,\msun) resulting in photospheric phase light curves and spectra that are incompatible with normal SNe Ia, due to over production of IGEs in the outer ejecta \citep{hoeflich1996a, nugent1997a}. More recent 1D radiative transfer simulations of double detonation models with lower He-shell masses (<\,0.1\,\msun) have demonstrated that using an NLTE treatment leads to reduced absorption at blue wavelengths, resulting in bluer colours and improved agreement with normal SNe~Ia \citep{collins2025a, boos2025a}. However, hydrodynamic explosion simulations of double detonation explosion models exhibit substantial ejecta asymmetries (e.g.\,\citealt{fink2010a, gronow2021a, boos2021a}) and radiative transfer calculations of these models predict significant variation with viewing angle (e.g.\,\citealt{kromer2010a, shen2021a, collins2022a}). Additionally, 1D radiative transfer simulations of SNe Ia models predict substantial departures from local thermodynamic equilibrium (LTE), already during the photospheric phase (e.g.\,\citealt{dessart2014b, shen2021a, collins2025a, boos2025a}). 

In this paper, we present a 2D NLTE photospheric-phase radiative transfer simulation for a double detonation model from \cite{gronow2021a} with a 0.02\,\msun\ He-shell. We aim to investigate the impact of simultaneously accounting for multi-dimensionality and detailed NLTE physics in radiative transfer simulations when evaluating explosion models against observed SNe~Ia in the photospheric phase. We therefore compare our 2D NLTE simulation to (i) a radiative transfer calculation of the same 2D ejecta model using an approximate NLTE treatment, (ii) a 1D NLTE simulation constructed from the equatorial direction of the same double detonation model, (iii) normal SNe~Ia, primarily the well-observed SN~2011fe \citep{nugent2011a}. As discussed above, radiative transfer simulations of double detonation models predict significant viewing angle variations and are impacted by NLTE effects, making the scenario well suited for this study. 

We describe the ejecta structures of the models and radiative transfer simulation set-ups in Section~\ref{sec:numerical_methods}. In Section~\ref{sec:results} we present comparisons between the light curves and spectra predicted by our radiative transfer simulations and discuss how variations in the plasma conditions in the simulations drive differences in the synthetic observables. We also present comparisons with the light curves and spectra of SN~2011fe as well as the width-luminosity relation of the normal SNe~Ia population in B- and V-band. Finally, in Section~\ref{sec:conclusions} we summarise the main results and present our conclusions.

\section{Numerical Methods}  
\label{sec:numerical_methods}

\begin{figure*}
	\includegraphics[width=1.0\linewidth,trim={0.4cm 1.6cm 0.4cm 0.4cm},clip]
    {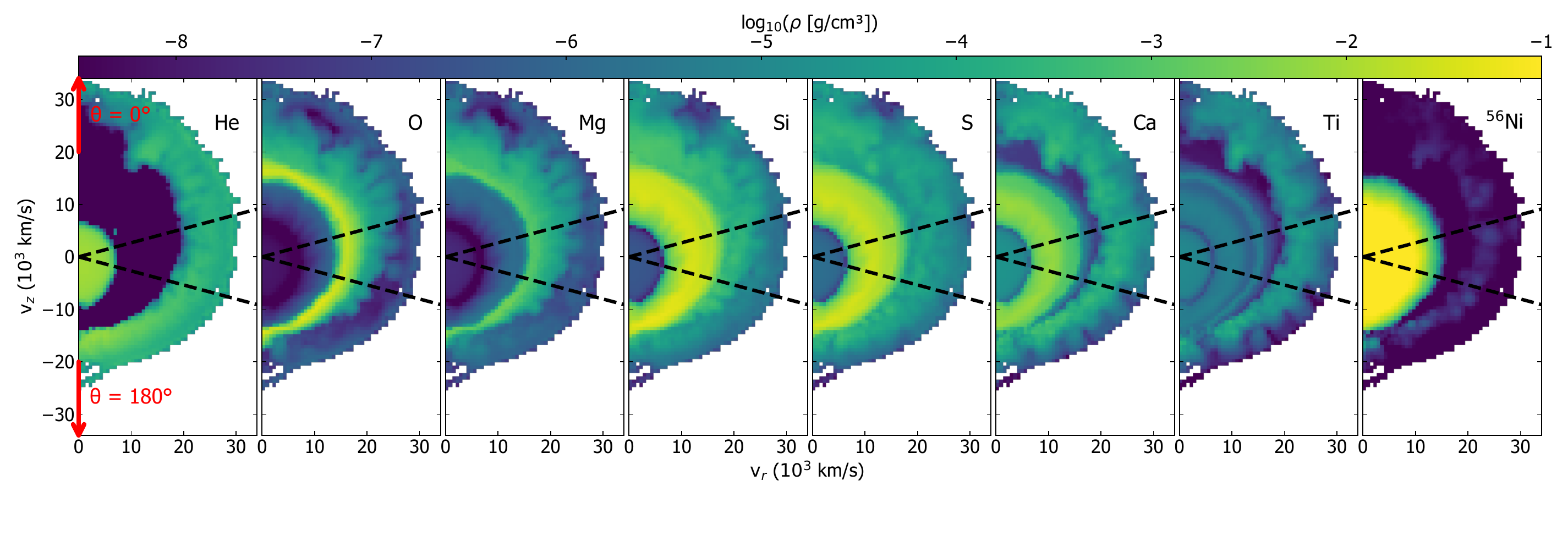}  
    \includegraphics[width=1.0\linewidth,trim={0.0cm 0.0cm 0.2cm 0.0cm},clip]
    {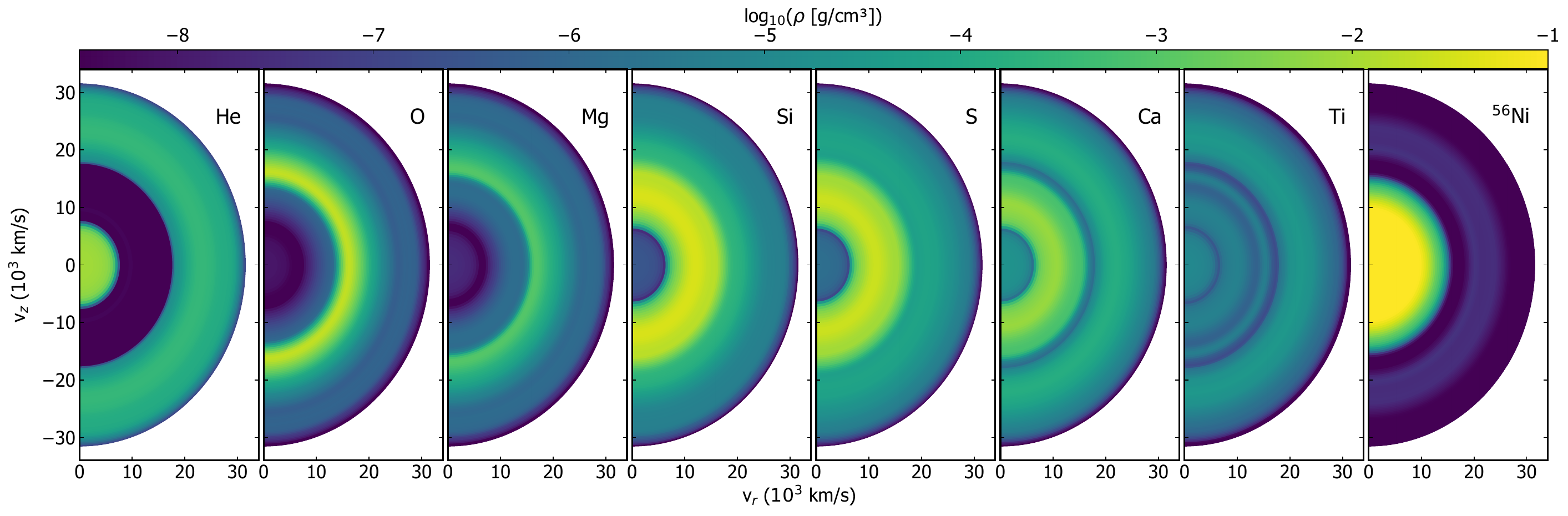}  
    \caption{\textit{Top}: Mass density of key species at 100\,s after explosion for the 2D ejecta model. The boundaries of the solid angle cone that the 1D ejecta model is generated from is displayed with dashed lines for reference. \textit{Bottom}: Mass density of key species for the 1D ejecta model shown on the same 2D coordinate system for ease of comparison. The $\theta=0^\circ$ and $\theta=180^\circ$ viewing angles are displayed for reference.}
    \label{fig:2D_ejecta_composition}
\end{figure*}

\begin{figure}
\centering
	\includegraphics[width=0.9\linewidth,trim={0.0cm 0.0cm 0.0cm 0.0cm},clip]{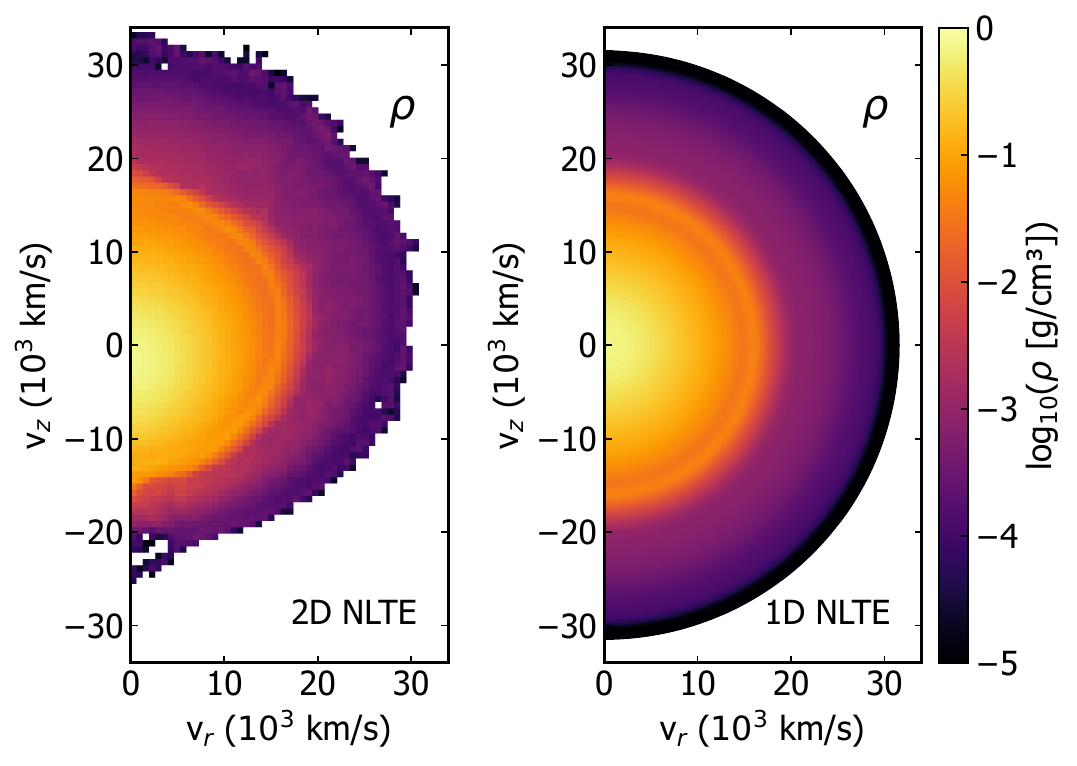} 
    \caption{2D and 1D ejecta model densities at 100\,s after explosion. As in Figure~\ref{fig:2D_ejecta_composition} the 1D density profile is shown on the same coordinate system as the 2D model.}
    \label{fig:density_1D_2D}
\end{figure}

\subsection{NLTE and non-thermal radiative transfer}
\label{subsec:RT_method}
The simulations presented here are carried out with the multi-dimensional time-dependent Monte Carlo radiative transfer code \textsc{artis} \citep{sim2007b, kromer2009a, shingles2020a}. Following the methods of \cite{lucy2002a, lucy2003a, lucy2005a}, \textsc{artis} divides the radiation field into indivisible energy packet Monte Carlo quanta (packets). 

We present a 1D and a 2D simulation that enable the NLTE and non-thermal capabilities of \textsc{artis} \citep{shingles2020a}, hereafter referred to as \artisnlte. The 2D \artisnlte simulation represents the first multi-dimensional \artisnlte simulation of the photospheric phase. \artisnlte includes a full solution to the NLTE equations of statistical equilibrium for atomic level populations and treatment for collisions with non-thermal leptons. To follow the energy distribution of high-energy leptons from nuclear decays and Compton scattering of $\gamma$-rays, \artis~solves the Spencer--Fano equation (as framed by \citealt{kozma1992a}). Auger electrons are allowed to contribute to heating, ionisation and excitation. Excitation of bound electrons by non-thermal collisions is also included. \artis\ uses the full Monte Carlo packet trajectories to obtain a rate estimator for each photoionisation transition from an NLTE level, while a parameterised radiation field is used to estimate the rates of bound-bound radiative excitation and any remaining photoionisation transitions. Treating all levels in NLTE for every ion has a significant computational and memory cost. Therefore, for ions with $Z\,{<}\,20$, we treat the first 100 levels in NLTE and for ions with $Z \geq 20$, we treat the first 200 levels in NLTE. All other levels are placed into an additional level called a `superlevel' \citep{anderson1989a}, which has an absolute population determined by the NLTE solver while the relative populations of states within the superlevel are set by a Boltzmann distribution at the electron temperature. To reduce the memory requirements associated with the simulation we follow \cite{Pollin2026a} and only store detailed photoionisation Monte Carlo estimators for levels treated in NLTE.  

NLTE radiative transfer simulations at photospheric phases are computationally challenging, particularly when simulating multi-dimensional models. At early phases, the ejecta conditions evolve rapidly and material across a wide range of velocities is relevant to the spectral formation and plasma conditions. This means a significant number of ionisation stages can be relevant to the plasma state (here we consider ion stages \textsc{ii} to \textsc{vi} for Fe, Co and Ni). The \cite{shingles2020a} implementation of the NLTE solver required that all ionisation states of each element present in the simulation be included in the solution for that element. When carrying out multi-dimensional early-phase \artisnlte simulations this resulted in numerical problems related to ion stages with extremely small populations. We therefore implemented an adaptive NLTE solution that removes ion stages with negligible populations when they cause the solution to fail, and then recalculates the level populations for the element. These developments are described in detail in Appendix \ref{appendix:NLTE_solution_developments}.

We compare our 2D \artisnlte simulation to a radiative transfer simulation of the same 2D model that adopts the approximate NLTE treatment of \artis\ described by \cite{kromer2009a} (hereafter \artiscl). \artiscl allows for departures from LTE ionisation balance by tracking one ground-state photoionisation estimator per ion that is used to calculate a departure coefficient that is applied to excited state photoionisation. \artiscl parametrises the radiation field with a dilute blackbody at the radiation temperature ($T\mathrm{_R}$). Level populations are determined using the Boltzmann formula evaluated at the temperature corresponding to the local energy density of the radiation field ($T\mathrm{_J}$). Although it approximates an NLTE ionisation balance, it does not allow for departures from Boltzmann distributed level populations, does not track excited state photoionisation directly or include non-thermal ionisation. 

We use the same atomic data for all \artis\ simulations presented here. The atomic data set is based on the compilation of \textsc{cmfgen} \citep{hillier1990a, hillier1998a} and is similar to that described by \cite{shingles2020a}, but with some additional species and ions, including updated atomic data for \ion{Ca}{iv}, \ion{Ti}{ii} and \ion{Ti}{iii} from the most recent compilation of \textsc{cmfgen} (see \citealt{blondin2023a}). In these simulations, we include He \textsc{i}–\textsc{iii}, C \textsc{i}–\textsc{iv}, O \textsc{i}–\textsc{iv}, Ne \textsc{i}–\textsc{iv}, Mg \textsc{ii}–\textsc{v}, Al \textsc{ii}–\textsc{v}, Si \textsc{ii}–\textsc{v}, S \textsc{ii}–v, Ar \textsc{ii}–\textsc{v}, Ca \textsc{ii}–\textsc{v}, Ti \textsc{ii}–\textsc{iv}, Cr \textsc{ii}–\textsc{vi}, Fe \textsc{ii}–\textsc{vi}, Co \textsc{ii}–\textsc{vi}, and Ni \textsc{ii}–\textsc{vi}. Preliminary simulations demonstrated that the neutral state had sub-dominant populations and negligible impact on the synthetic observables predicted for all but the lightest species. Therefore, to reduce the size of our atomic data set and thus the computational resources required for each simulation, we do not include the neutral species in our atomic data set for elements above Ne. 

To improve the signal-to-noise ratio of the line-of-sight dependent synthetic observables, both the 2D \artisnlte and 2D \artiscl simulations use the virtual packet scheme \citep{bulla2015a} in \artis. For the 1D \artisnlte simulation we used $3 \times 10^{7}$ Monte Carlo packets while for both 2D simulations, which require more packets for the plasma state to converge, we used $5 \times 10^{7}$ packets. All simulations described here evolve from 1.5 to 45\,d post explosion with logarithmically spaced time steps ($\Delta \log (t/\mathrm{days}) = 0.012$). In all cases, the first three time steps (1.6\,d) are treated in LTE before the NLTE/approximate NLTE method is enabled. For all simulations, a grey approximation is adopted in optically thick cells.

\subsection{Double detonation ejecta models}
\label{subsec:ejecta_model}
The ejecta structures we simulate here are drawn from the 3D double detonation explosion model M10\_02, described by \cite{gronow2021a}, which has a 1\,\msun\ CO core and 0.02\,\msun\ He-shell. The models in the \cite{gronow2021a} sequence are assumed to ignite due to stable mass transfer from a He-rich companion. M10\_02 is the model with the lowest He-shell mass in the sequence. The ignition location of the He detonation chosen for the models in the \cite{gronow2021a} sequence is the $+z$ pole ($\theta=0^\circ$ in Figure~\ref{fig:2D_ejecta_composition}). This produces symmetry about the $z$-axis. \cite{collins2022a} carried out \artiscl simulations for the models in this sequence and demonstrated that, while the models show a strong angle dependence on the angle from the $+z$ pole ($\theta$), there is no significant angle variation with azimuthal angle ($\phi$)\footnote{We use the convention: $x = r \sin \theta \cos \phi, y = r \sin \theta \sin \phi, z = r \cos \theta$}. The main viewing angle variation of the 3D explosion simulation can therefore be captured by a 2D ejecta model. 

For this work, we generated a 2D cylindrical model from the $\phi=90^\circ$ half-plane ($x=0,y\geq0$) slice of the 3D M10\_02 ejecta. The model has dimensions $(N_r, N_z) = (50, 100)$ and velocity extent $v_r\in[0, 33500]~\mathrm{km\,s^{-1}}, \quad
v_z \in [-33500, 33500]~\mathrm{km\,s^{-1}}$. The composition and density structure of this 2D ejecta model (and the 1D model we simulate for comparison) are shown in Figures~\ref{fig:2D_ejecta_composition} and \ref{fig:density_1D_2D} respectively. To confirm that our findings are not sensitive to the way the 2D model was generated we carried out \artiscl simulations for two other 2D ejecta models, one based on a $\phi=270^\circ$ ($x=0,y<=0$) slice in the 3D model and one generated by azimuthally averaging the 3D model about the $z$-axis in a way that preserves the mass of each species in each 2D zone, and therefore the total ejecta mass. All the 2D ejecta models had total and \nickel\ masses within 1\% of the 3D ejecta model and the differences between the synthetic observables predicted by these simulations were negligible, making the choice of the specific 2D ejecta model unimportant.

We also carried out a 3D \artiscl simulation for the M10\_02 model with the same atomic data set as all other simulations in this paper. Comparing the observables predicted by this simulation to our 2D \artiscl simulations confirmed that the 3D model, including its viewing angle variation, is well represented by a 2D model. A 3D \artisnlte\ simulation is therefore not necessary to represent this model, especially given the increased number of Monte Carlo packets--and hence computational resources--it would require relative to our already computationally expensive 2D \artisnlte simulation (see Appendix~\ref{appendix:computational_resources}). 3D simulations are, however, essential to accurately represent the ejecta structures of explosion models with no clear symmetry axis. Exploring such models with 3D early-phase \artisnlte simulations will be the focus of future work.

To investigate how successfully a 1D model can reproduce the properties of a line-of-sight in our 2D \artisnlte radiative transfer simulation, we generated a 1D ejecta model based on the equatorial direction of the multi-dimensional model. To construct this spherically symmetric model we radially averaged cells in the multi-dimensional model that are contained within a solid angle cone with full opening angle of $30^\circ$ centred on the equatorial plane (see Figure~\ref{fig:2D_ejecta_composition}). The 1D model has total and \nickel\ masses of 0.99 and 0.51\,\msun\ compared to 1.03 and 0.55\,\msun\ for the 2D ejecta model. 

\section{Results}
\label{sec:results}
In this section, we present the synthetic observables from our 2D \artisnlte simulation and compare to the 2D \artiscl and 1D \artisnlte simulations described in Section~\ref{sec:numerical_methods}. We also discuss how the ejecta conditions in the different simulations impact the synthetic observables predicted. To provide context for the simulation comparisons we compare to the well-observed normal SN~Ia, SN~2011fe and the B- and V-band width-luminosity distribution of the normal SNe~Ia population.

\subsection{2D \artisnlte versus 2D \artiscl comparisons}
\label{subsec:2D_comparisons}

\begin{figure*}
	\includegraphics[width=0.85\linewidth,trim={0.0cm 0.0cm 0.0cm 0.0cm},clip]{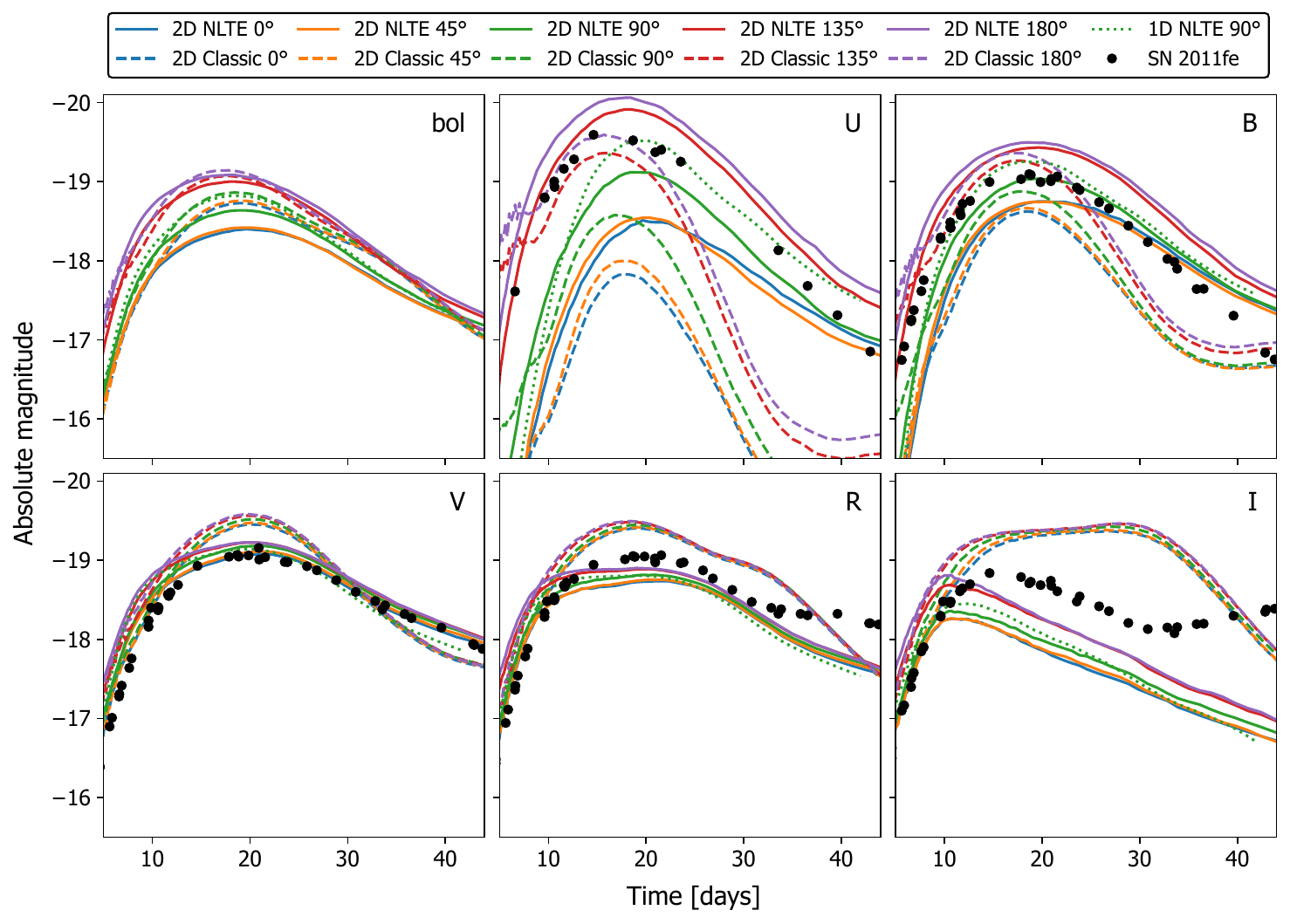} 
    \caption{Simulated light curves for five different viewing angles from our 2D \artisnlte and \artiscl  simulations as well as our 1D \artisnlte simulation. The normal SN~Ia 2011fe is plotted for reference \citep{richmond2012a, tsvetkov2013a}.}
    \label{fig:band_lightcurves}
\end{figure*}

\begin{figure*}
	\includegraphics[width=1.0\linewidth,trim={0.0cm 0.0cm 0.0cm 0.0cm},clip]{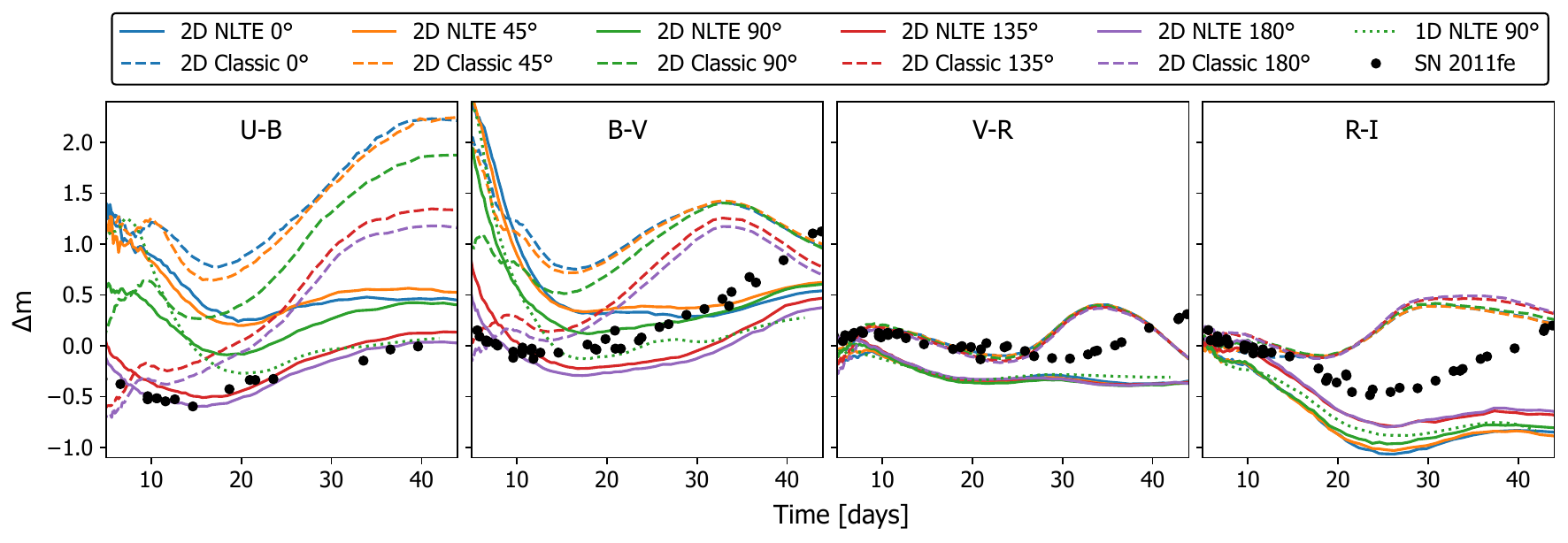}    
    \caption{Same as Figure~\ref{fig:band_lightcurves} for colour evolution.}
    \label{fig:colour_evolution}
\end{figure*}

\begin{figure*}
    \centering
    
    \includegraphics[height=0.27\textwidth, trim={0.8cm 0.3cm 2.5cm 1.5cm}, clip]{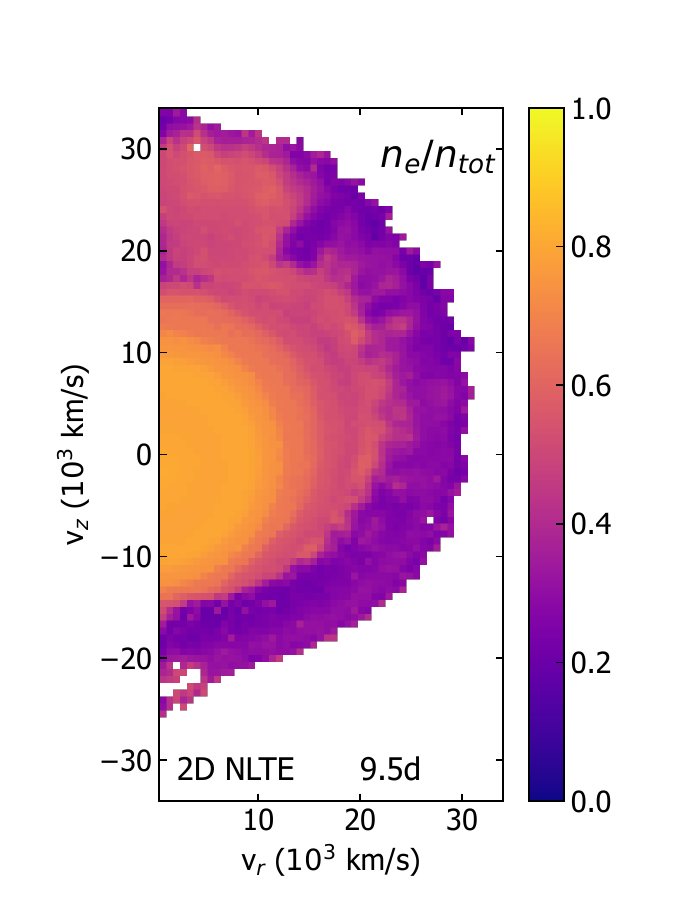}
    \includegraphics[height=0.27\textwidth, trim={2.65cm 0.3cm 2.5cm 1.5cm}, clip]{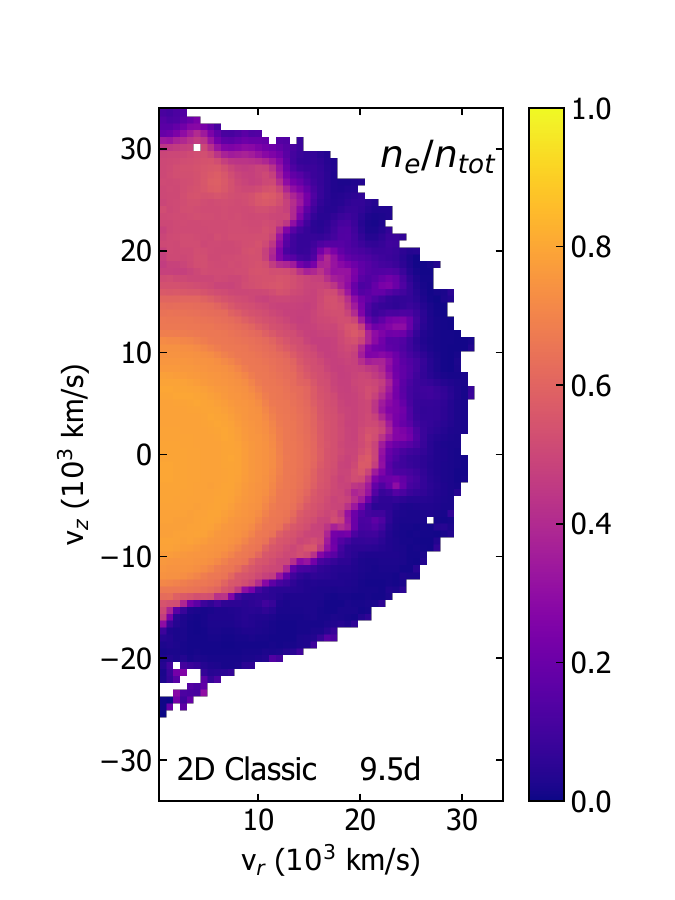}
    \includegraphics[height=0.27\textwidth, trim={2.65cm 0.3cm 0.2cm 1.5cm}, clip]{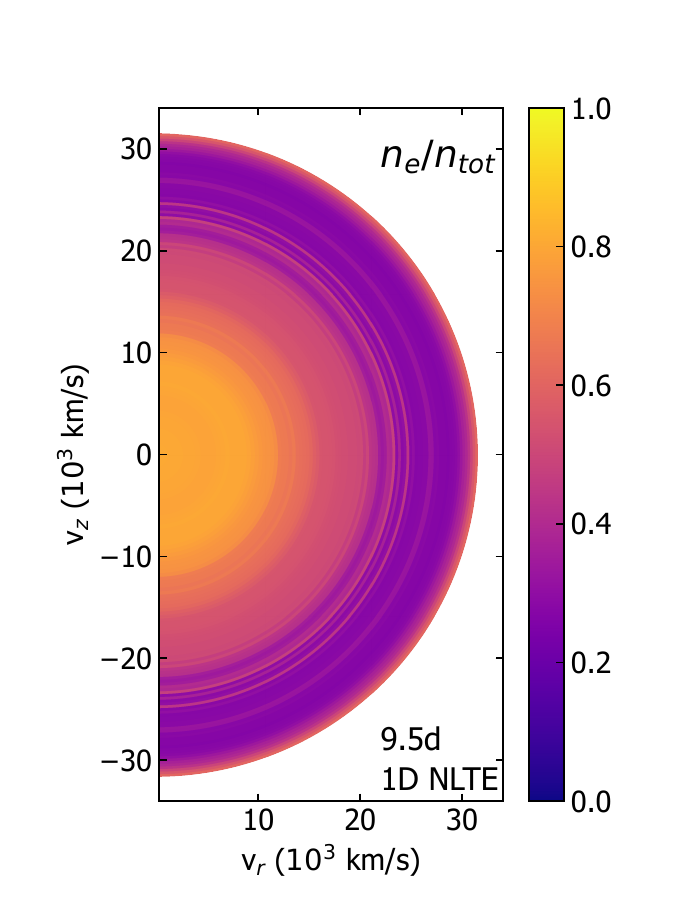}
    \includegraphics[height=0.27\textwidth, trim={1.6cm 0.3cm 3.0cm 1.5cm}, clip]{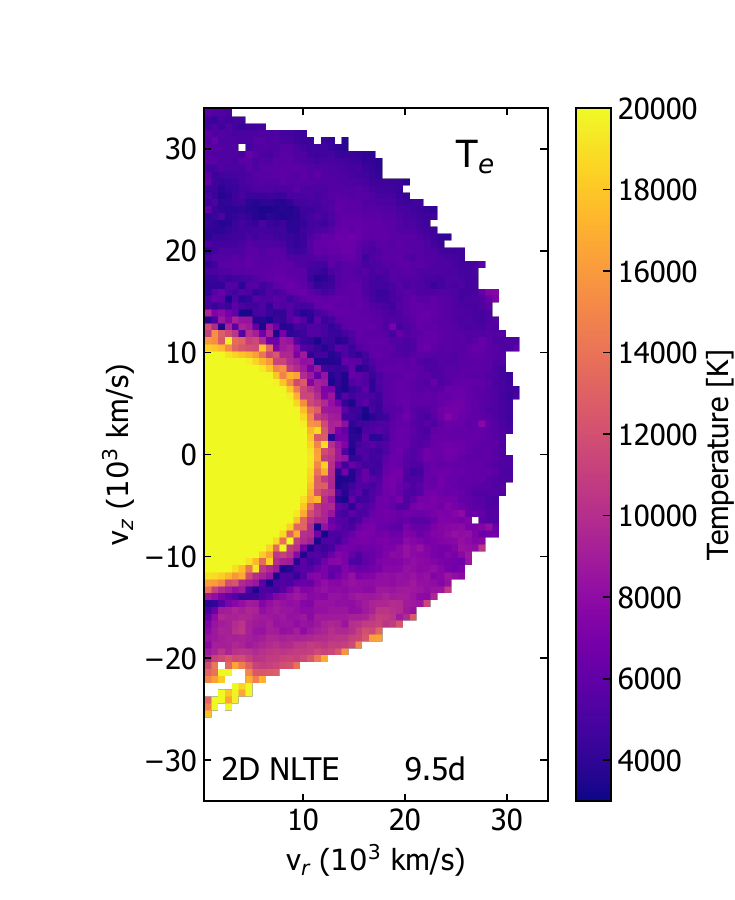}
    \includegraphics[height=0.27\textwidth, trim={3.4cm 0.3cm 3.0cm 1.5cm}, clip]{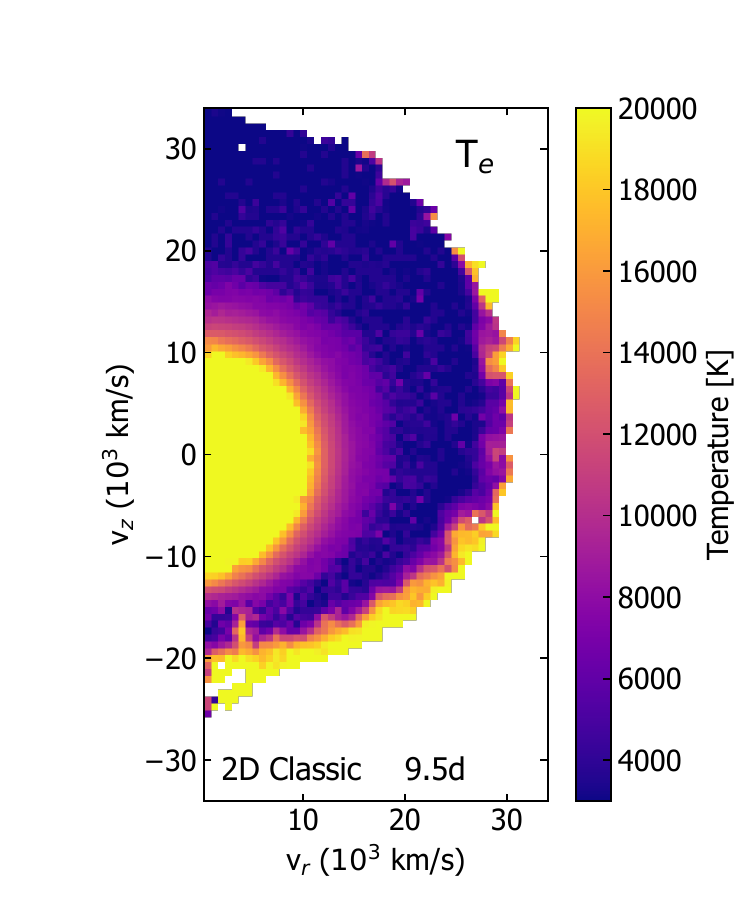}
    \includegraphics[height=0.27\textwidth, trim={3.4cm 0.3cm 0.0cm 1.5cm}, clip]{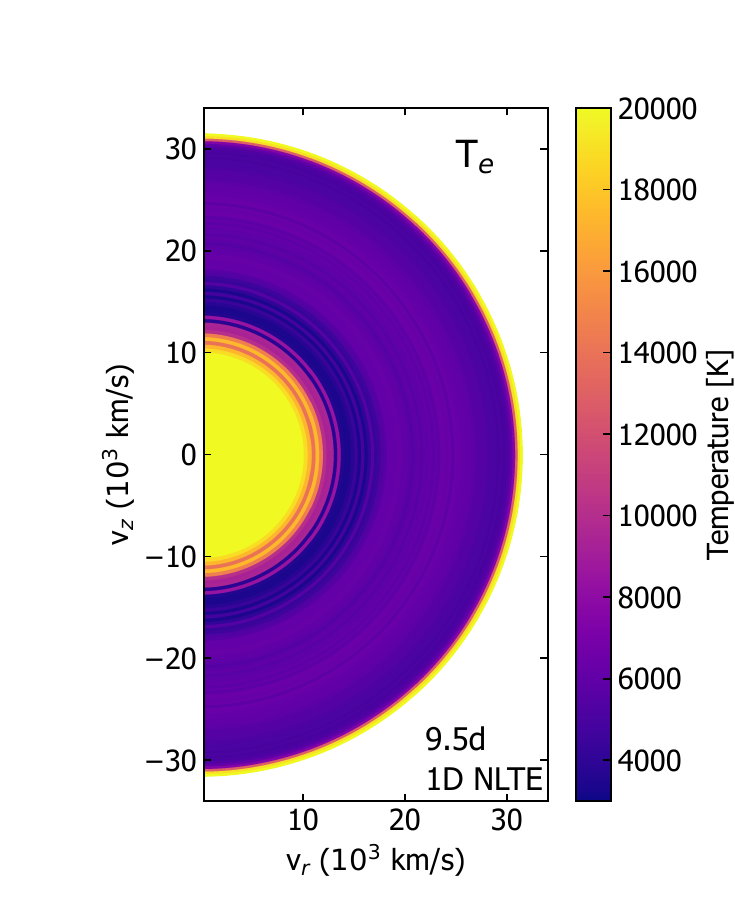}
    
    \includegraphics[height=0.27\textwidth, trim={0.8cm 0.3cm 2.5cm 1.5cm}, clip]{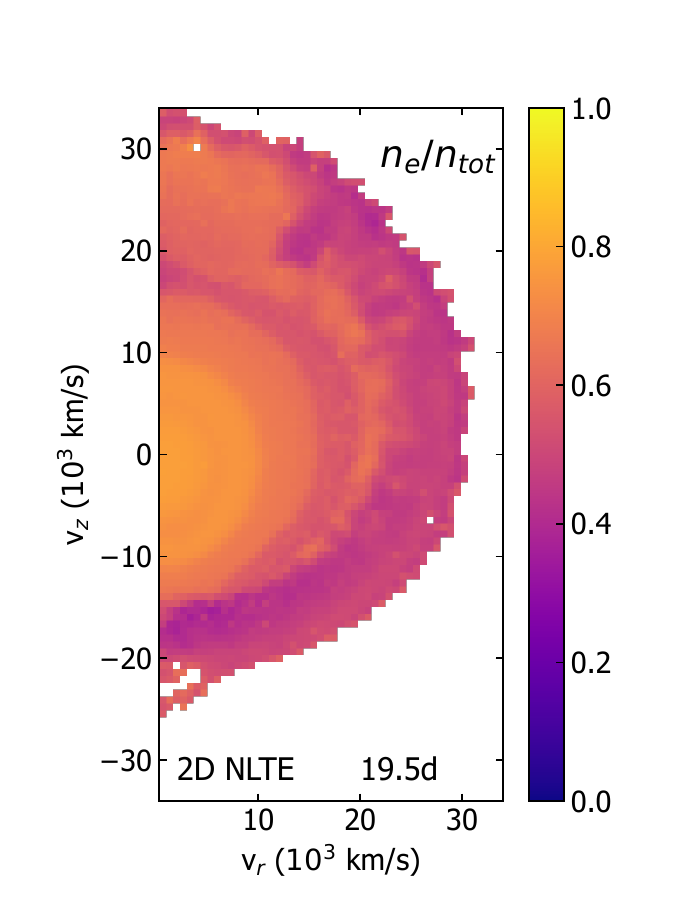}
    \includegraphics[height=0.27\textwidth, trim={2.65cm 0.3cm 2.5cm 1.5cm}, clip]{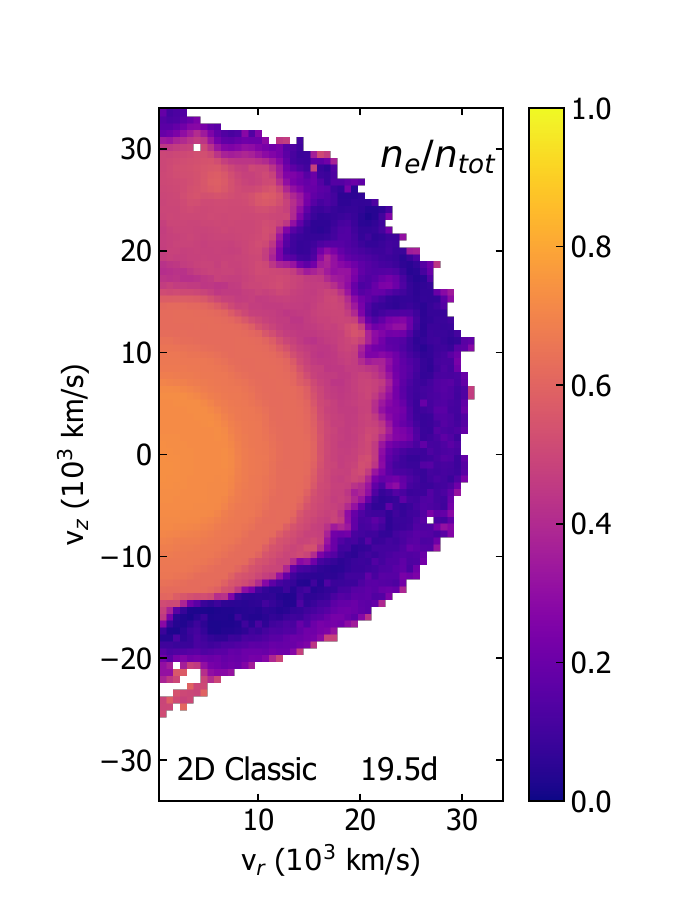}
    \includegraphics[height=0.27\textwidth, trim={2.65cm 0.3cm 0.2cm 1.5cm}, clip]{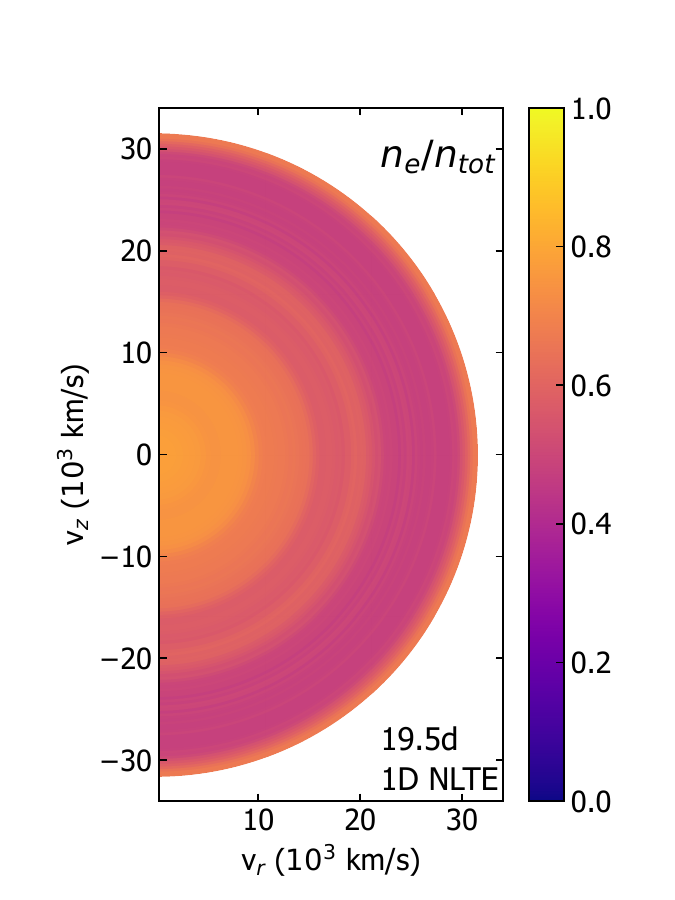}
    \includegraphics[height=0.27\textwidth, trim={1.6cm 0.3cm 3.0cm 1.5cm}, clip]{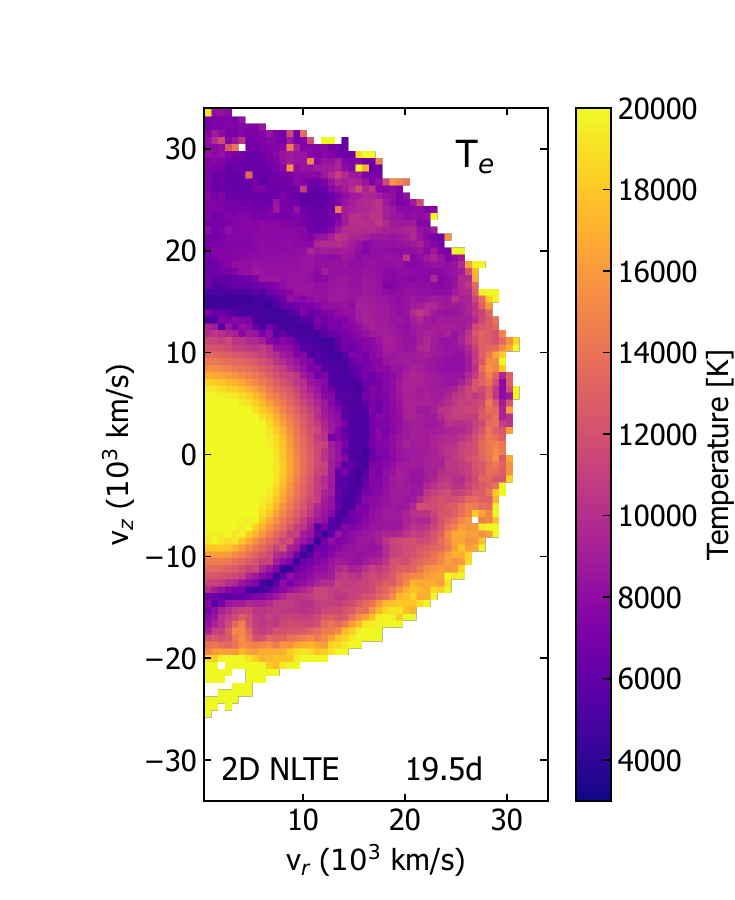}
    \includegraphics[height=0.27\textwidth, trim={3.4cm 0.3cm 3.0cm 1.5cm}, clip]{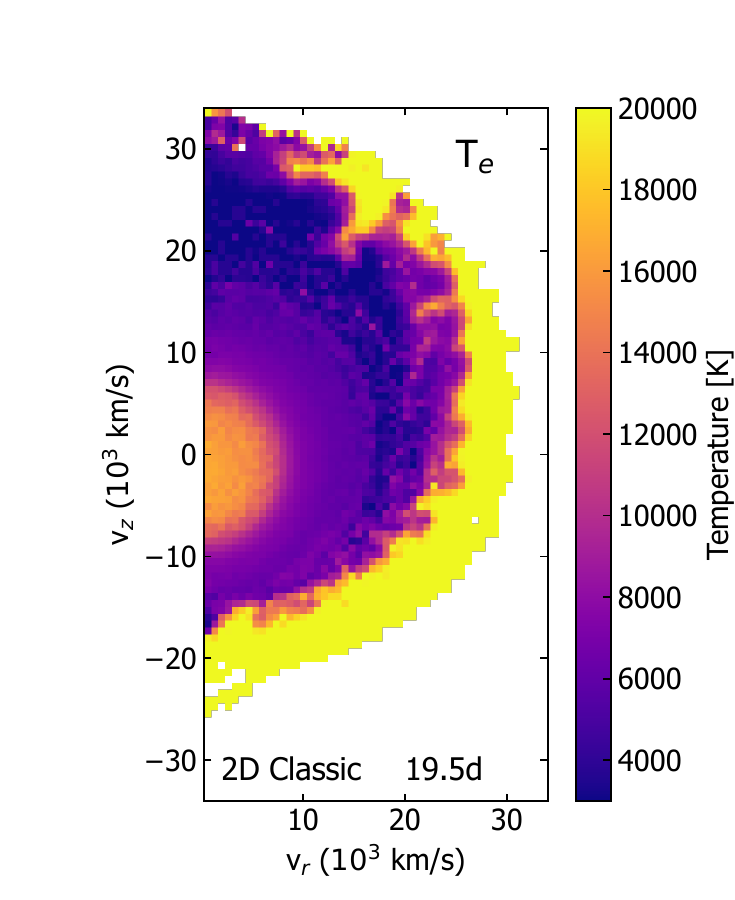}
    \includegraphics[height=0.27\textwidth, trim={3.4cm 0.3cm 0.0cm 1.5cm}, clip]{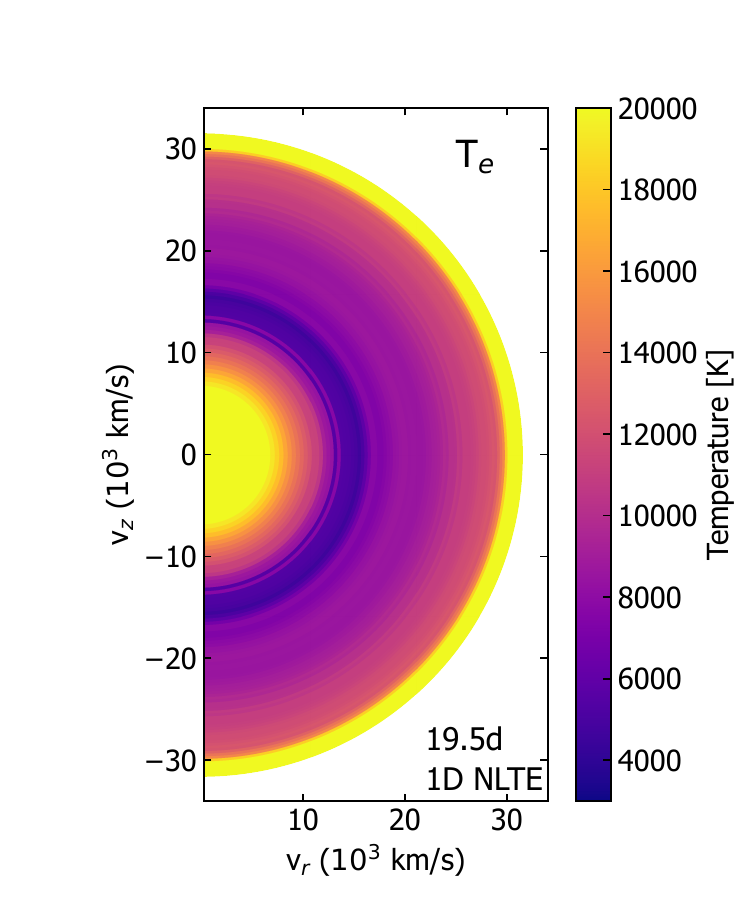}

    \includegraphics[height=0.27\textwidth, trim={0.8cm 0.3cm 2.5cm 1.5cm}, clip]{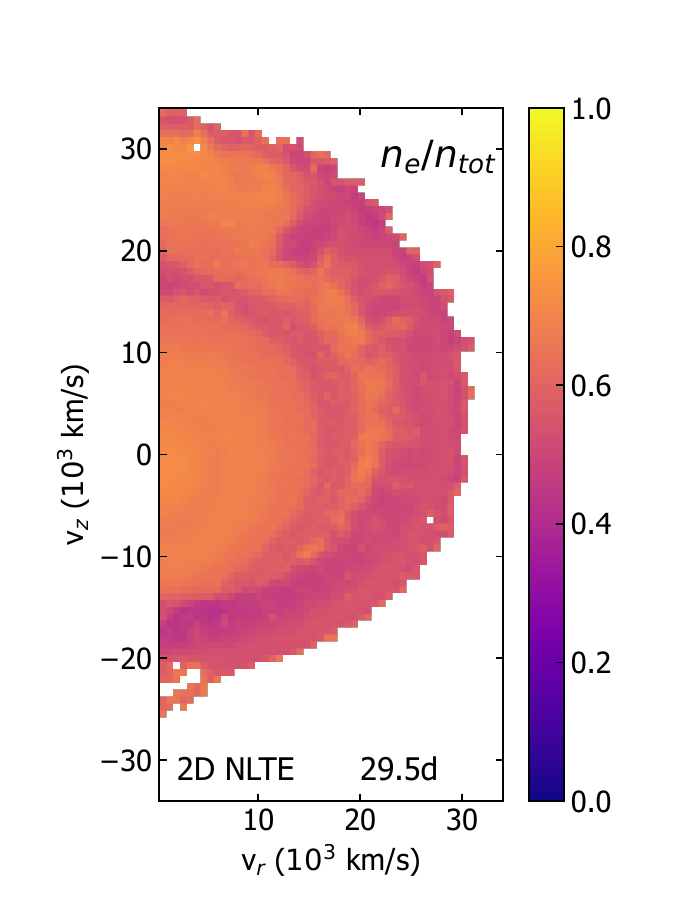}
    \includegraphics[height=0.27\textwidth, trim={2.65cm 0.3cm 2.5cm 1.5cm}, clip]{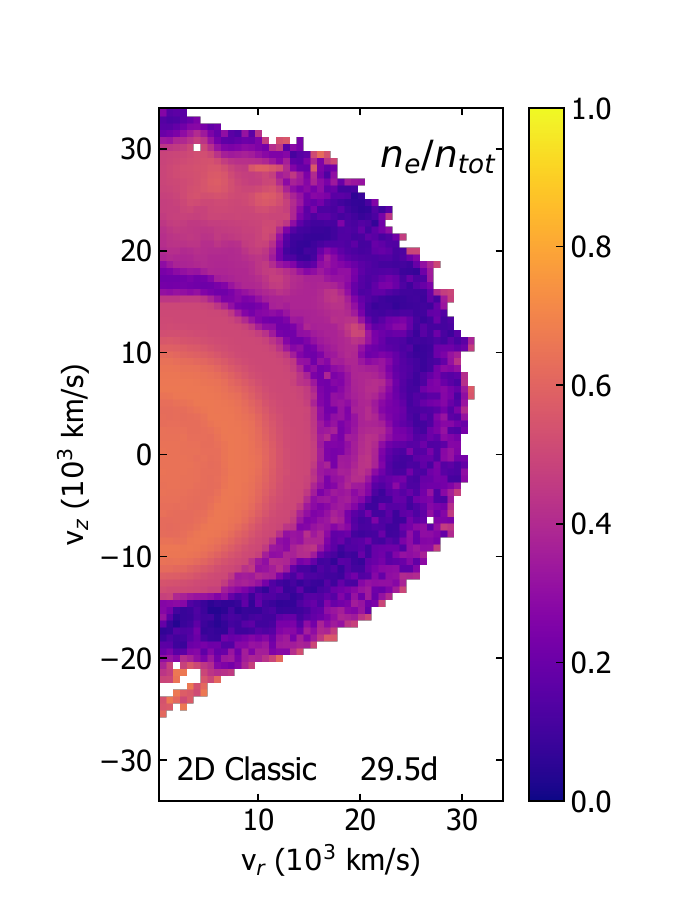}
    \includegraphics[height=0.27\textwidth, trim={2.65cm 0.3cm 0.2cm 1.5cm}, clip]{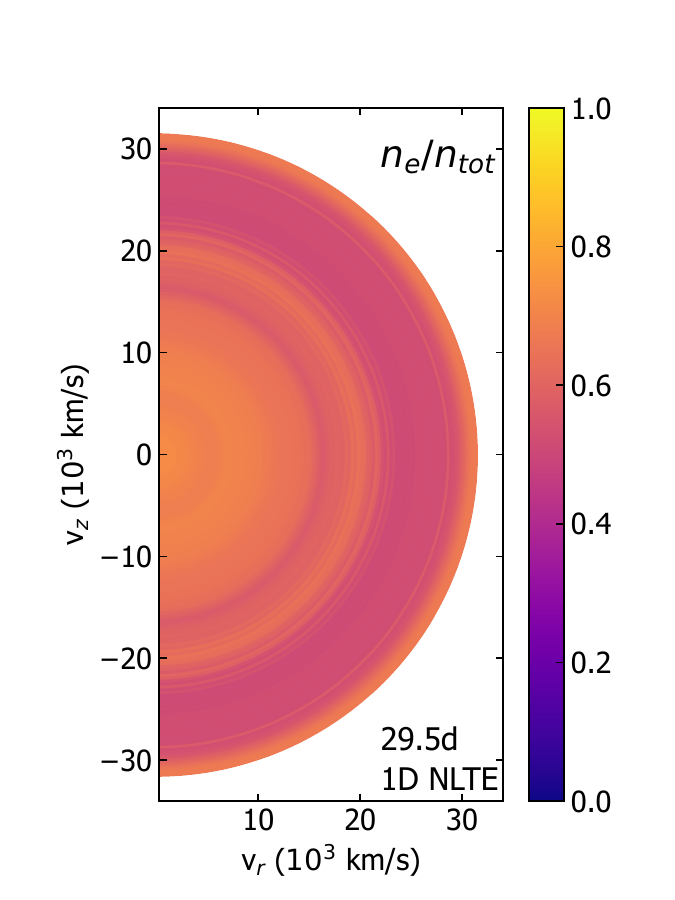}
    \includegraphics[height=0.27\textwidth, trim={1.6cm 0.3cm 3.0cm 1.5cm}, clip]{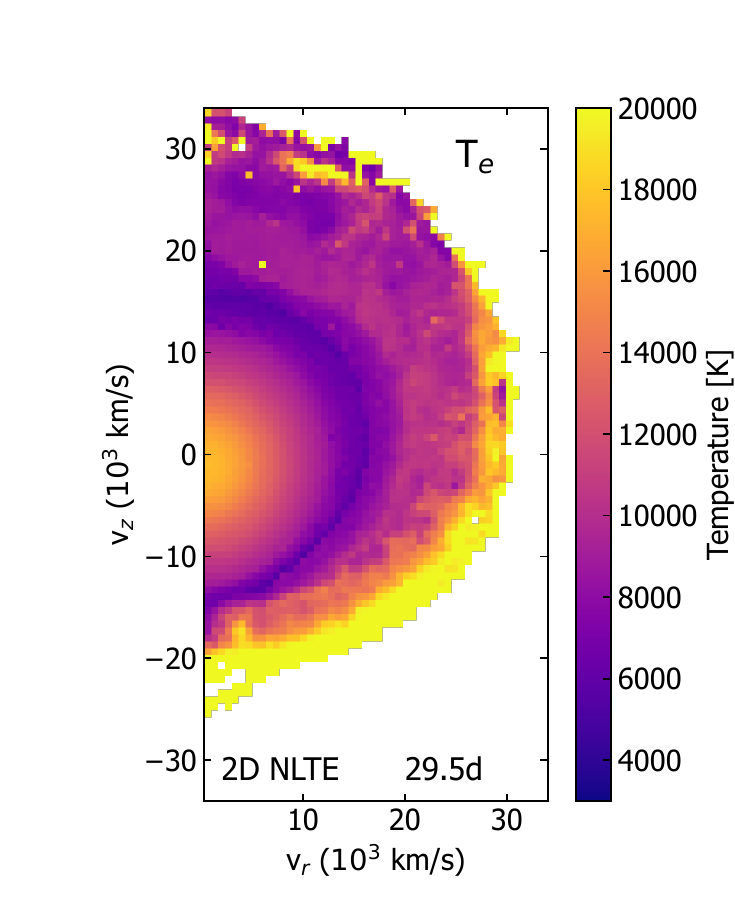}
    \includegraphics[height=0.27\textwidth, trim={3.4cm 0.3cm 3.0cm 1.5cm}, clip]{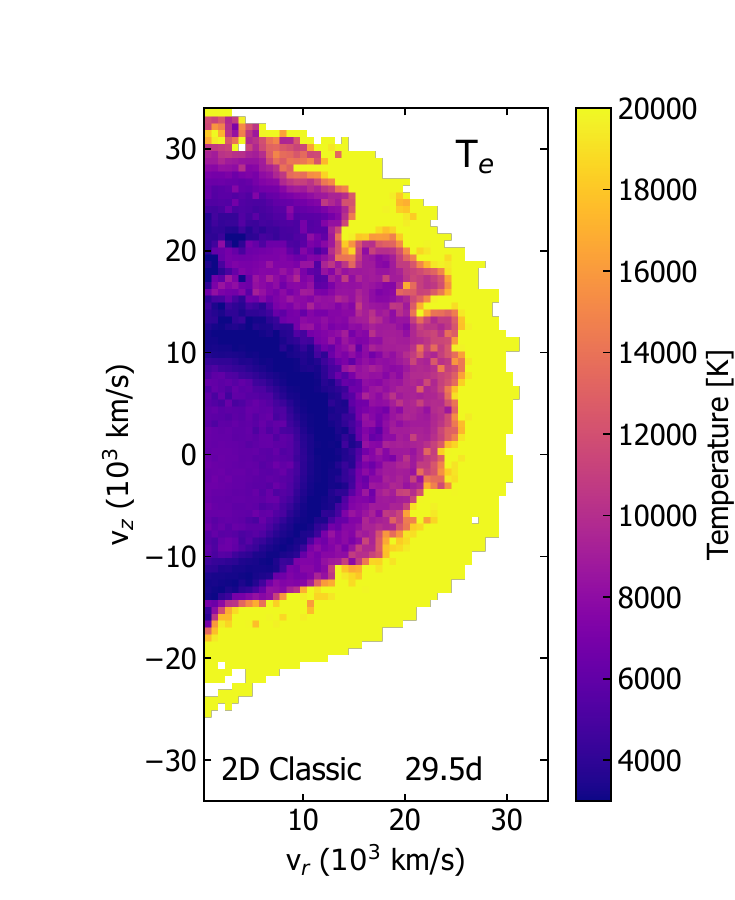}
    \includegraphics[height=0.27\textwidth, trim={3.4cm 0.3cm 0.0cm 1.5cm}, clip]{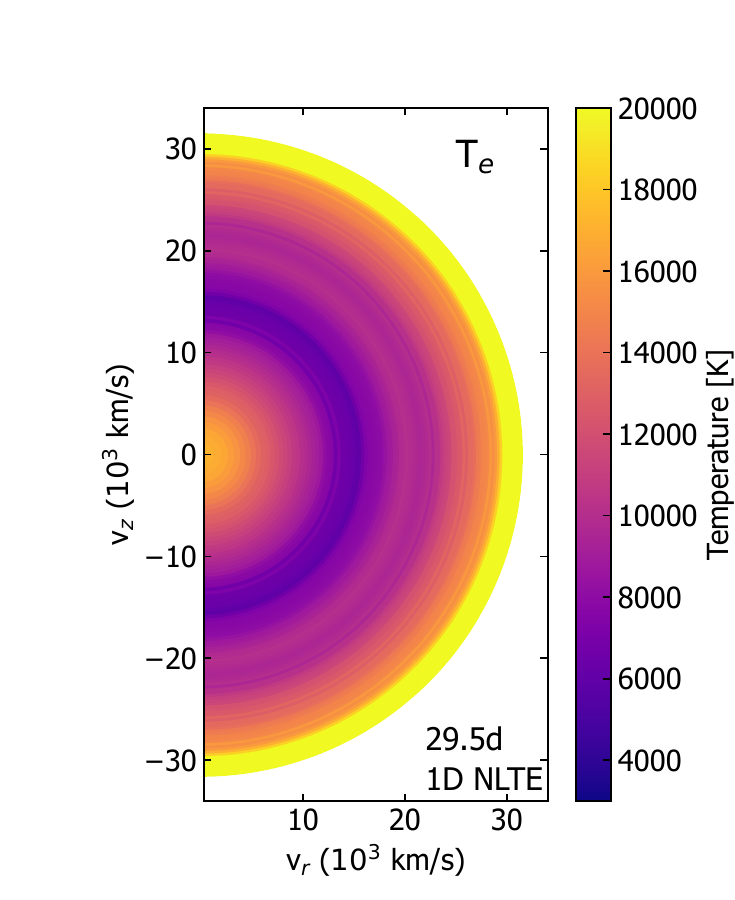}
    
    \caption{Free electron fraction, defined as the ratio of the free electron density to the total particle number density ($n_{\rm e}/n_{\rm tot}$; left three columns), where $n_{\rm tot}$ includes free electrons, ions, and neutral atoms, and electron temperature (right three columns), at 9.5, 19.5, and 29.5\,d after explosion (top to bottom) for the 2D \artisnlte, 2D \artiscl, and 1D \artisnlte simulations.}
    \label{fig:Te_nne_2D_maps}
\end{figure*}

\begin{figure*}
    \centering
    
    \includegraphics[height=0.3\textwidth, trim={0.9cm 0.3cm 2.6cm 1.5cm}, clip]{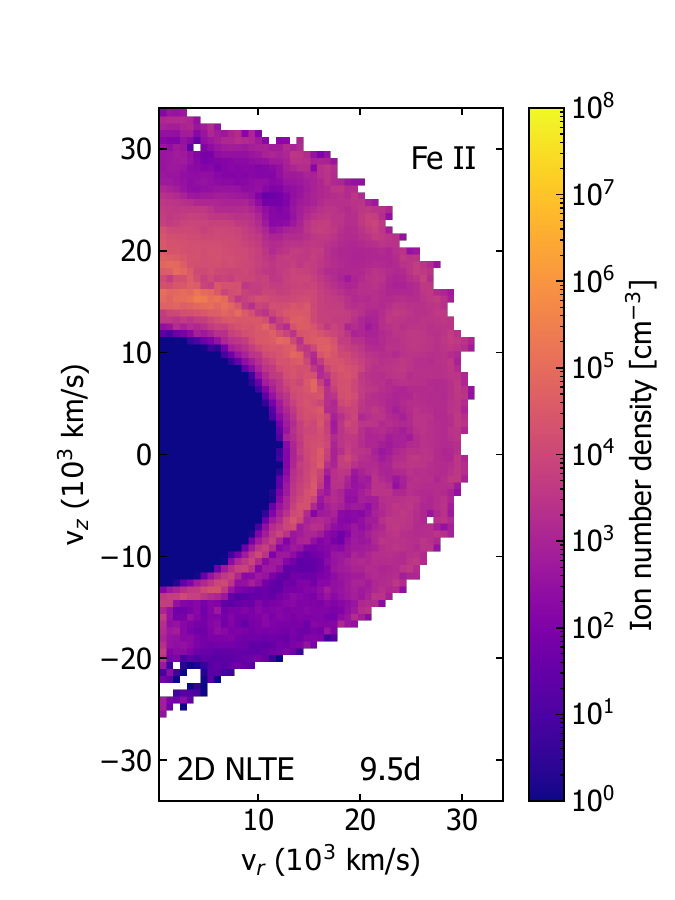}
    \includegraphics[height=0.3\textwidth, trim={2.6cm 0.3cm 2.6cm 1.5cm}, clip]{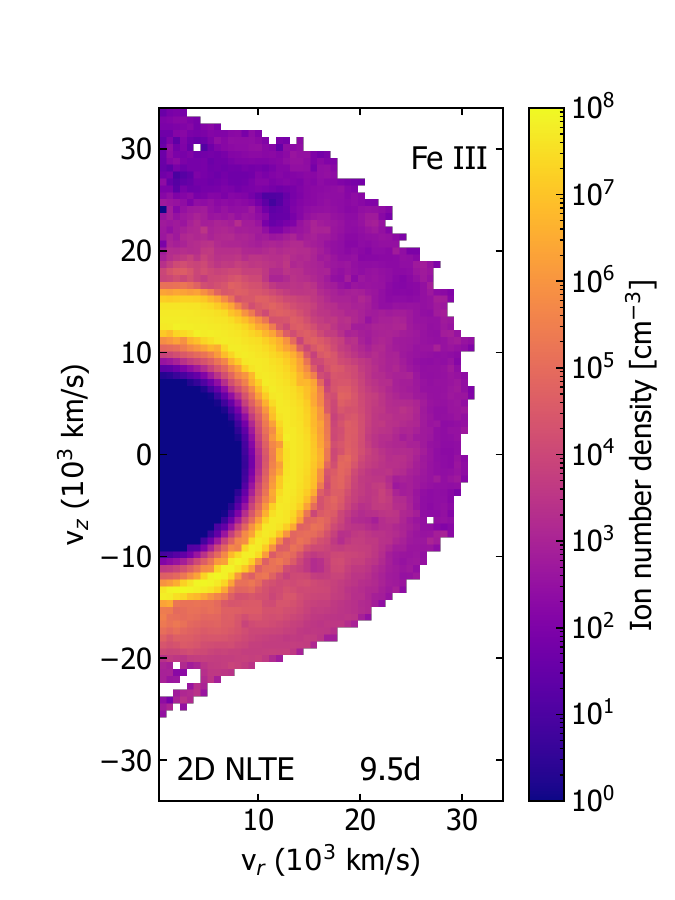}
    \includegraphics[height=0.3\textwidth, trim={2.6cm 0.3cm 2.6cm 1.5cm}, clip]{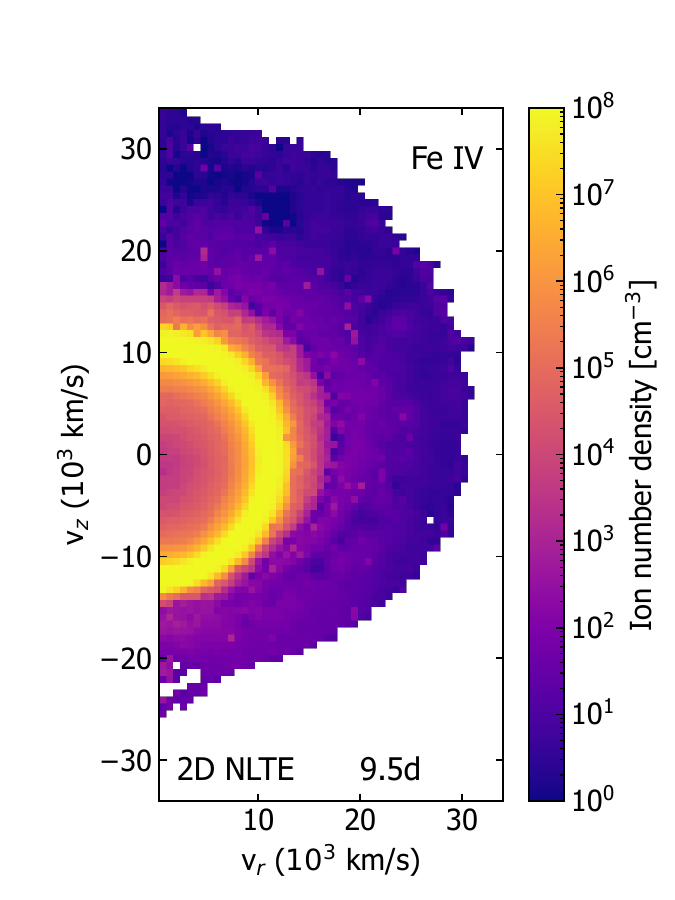}
    \includegraphics[height=0.3\textwidth, trim={2.6cm 0.3cm 2.6cm 1.5cm}, clip]{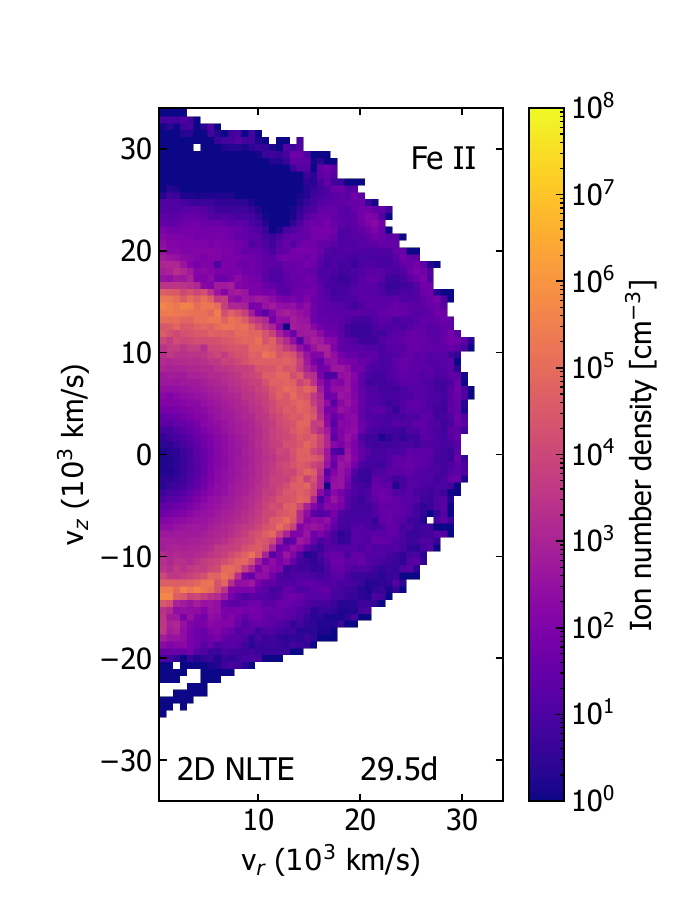}
    \includegraphics[height=0.3\textwidth, trim={2.6cm 0.3cm 2.6cm 1.5cm}, clip]{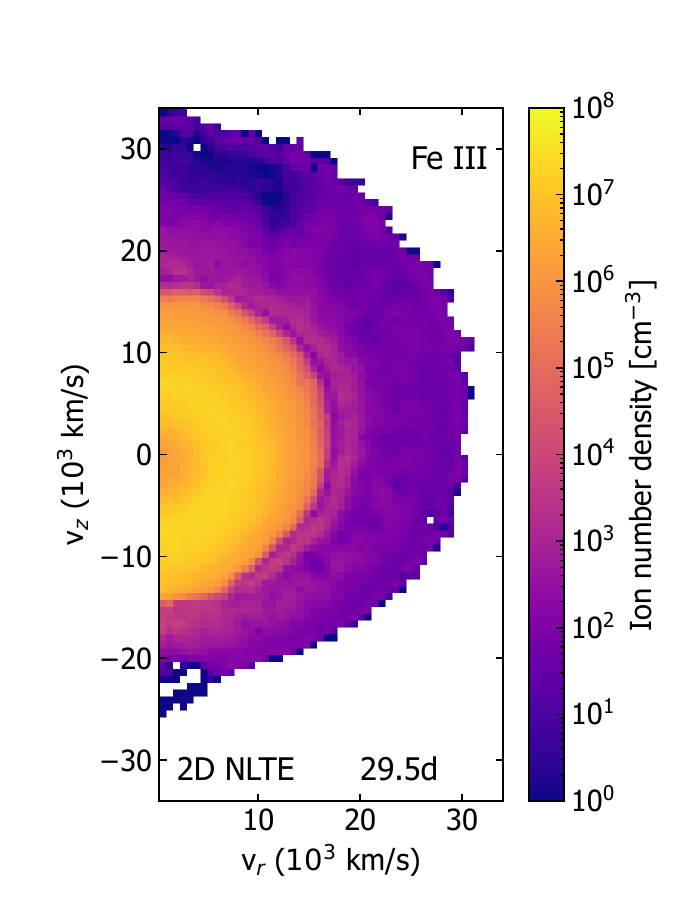}
    \includegraphics[height=0.3\textwidth, trim={2.6cm 0.3cm 0.2cm 1.5cm}, clip]{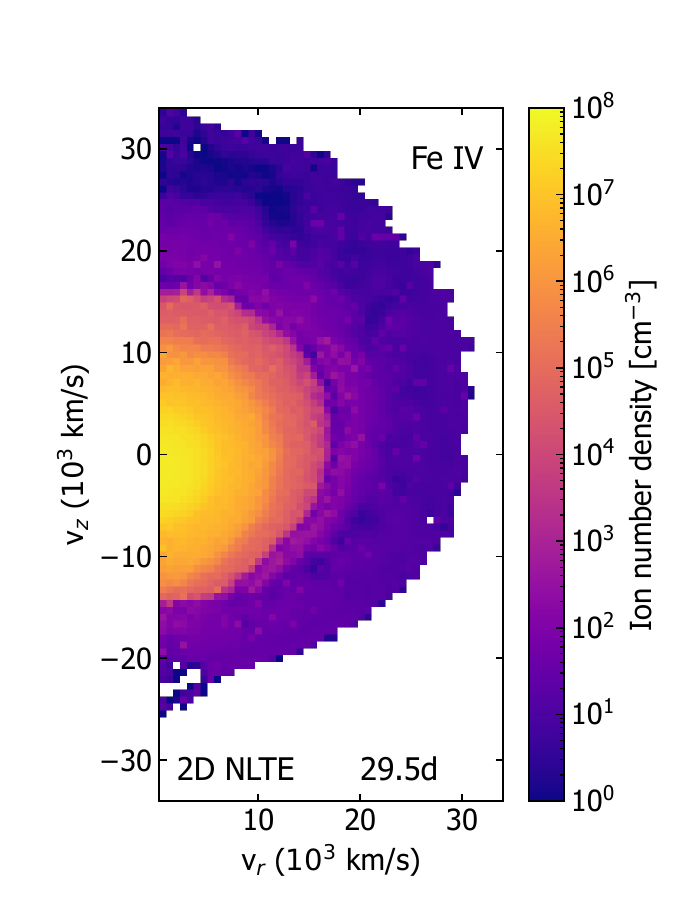}
    
    \vspace{0.5em}
    
    \includegraphics[height=0.3\textwidth, trim={0.9cm 0.3cm 2.6cm 1.5cm}, clip]{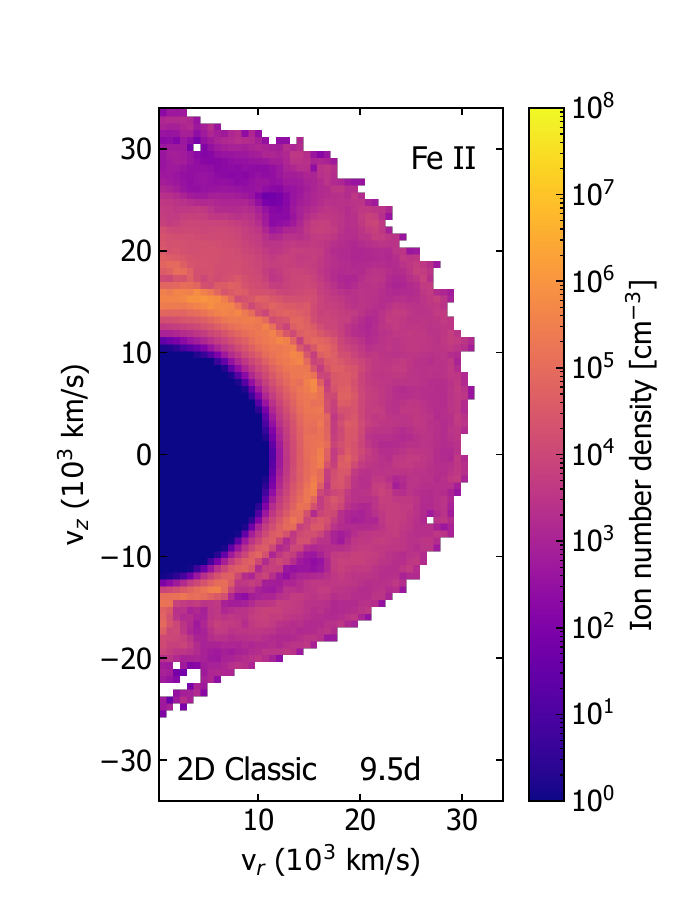}
    \includegraphics[height=0.3\textwidth, trim={2.6cm 0.3cm 2.6cm 1.5cm}, clip]{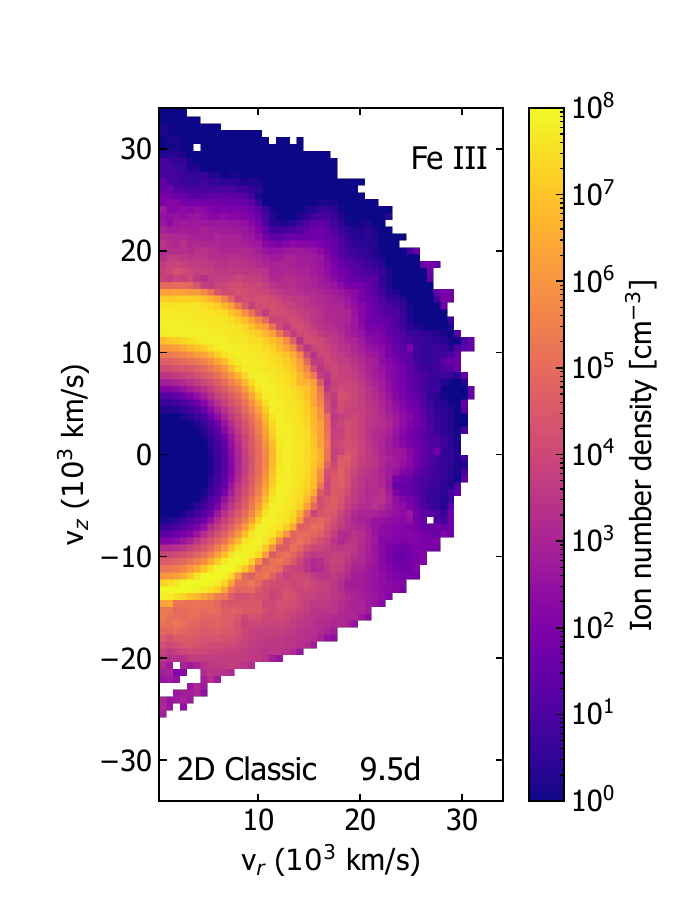}
    \includegraphics[height=0.3\textwidth, trim={2.6cm 0.3cm 2.6cm 1.5cm}, clip]{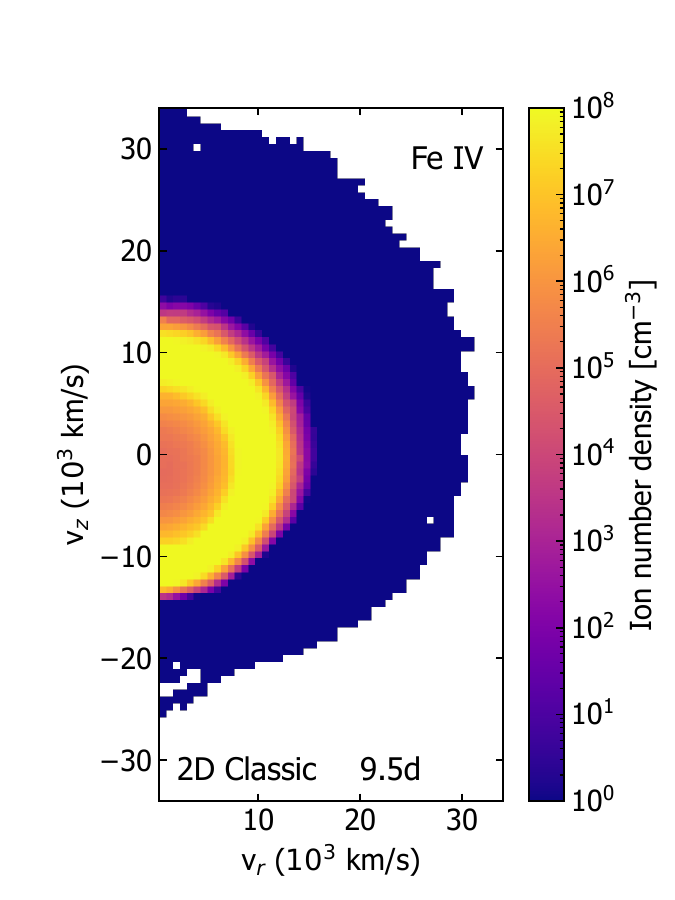}
    \includegraphics[height=0.3\textwidth, trim={2.6cm 0.3cm 2.6cm 1.5cm}, clip]{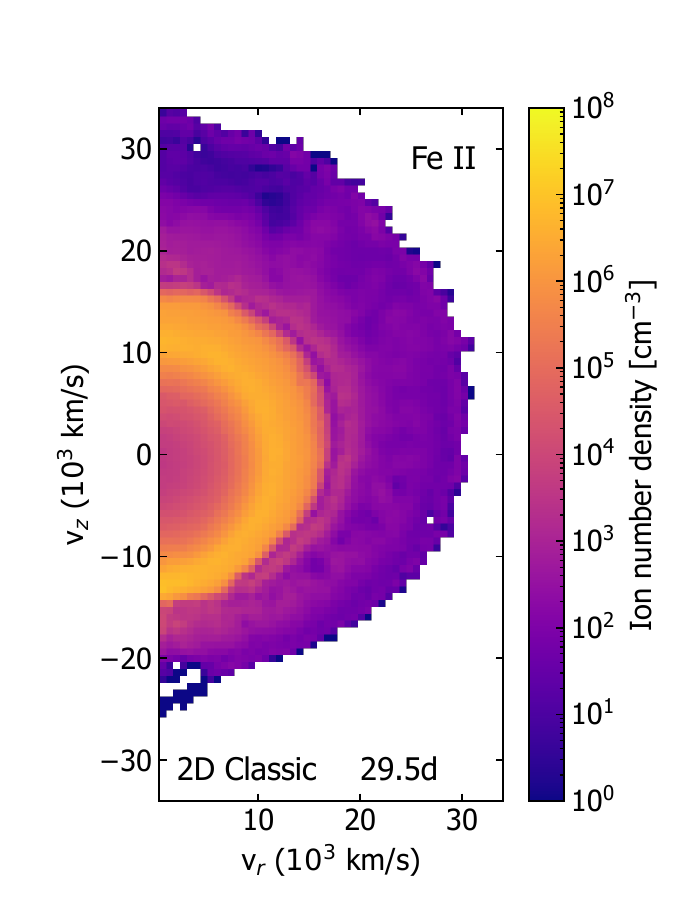}
    \includegraphics[height=0.3\textwidth, trim={2.6cm 0.3cm 2.6cm 1.5cm}, clip]{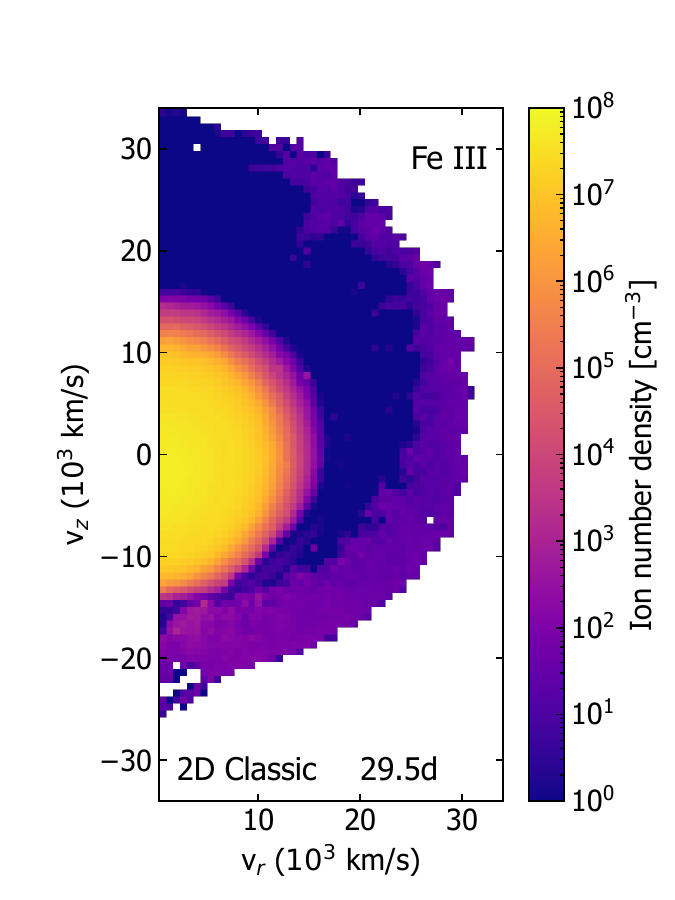}
    \includegraphics[height=0.3\textwidth, trim={2.6cm 0.3cm 0.2cm 1.5cm}, clip]{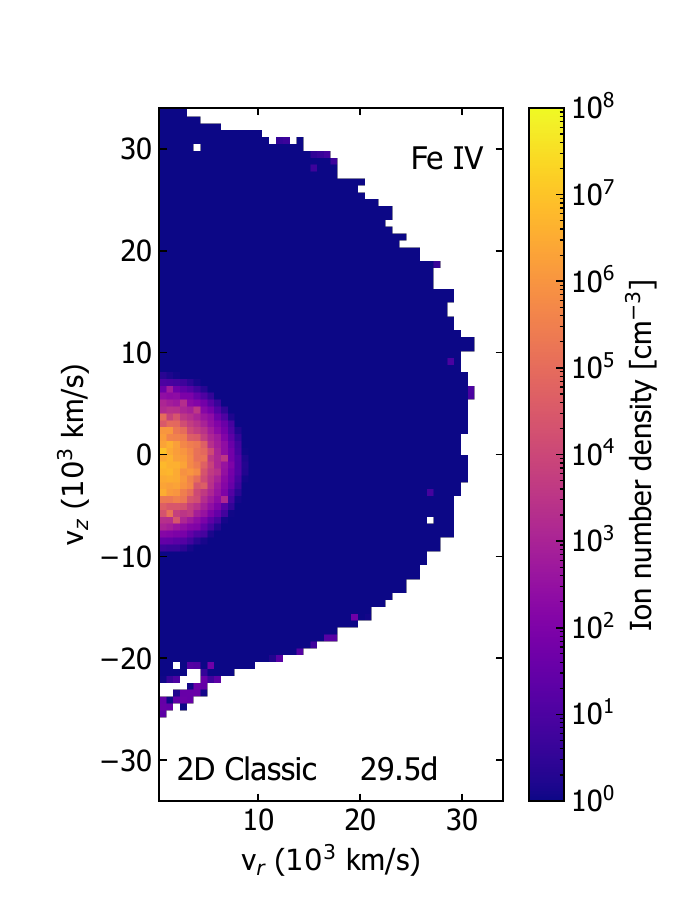}
    
    \caption{\textit{Top}: Number densities of \ion{Fe}{ii}, \ion{Fe}{iii} and \ion{Fe}{iv} at 9.5\,d and 29.5\,d after explosion for the 2D \artisnlte simulation. \textit{Bottom}: Same for the 2D \artiscl simulation.}
    \label{fig:Fe_ionisation_2D_maps}
\end{figure*}

\subsubsection{Light curve comparisons}
\label{subsec:2D_lightcurve_comparisons}
Figures~\ref{fig:band_lightcurves} and \ref{fig:colour_evolution} compare the bolometric and band light curves and the colour evolution, respectively, for five viewing angles from the 2D \artisnlte and \artiscl simulations, together with the 1D \artisnlte simulation and SN~2011fe for reference. As can be seen from Figure~\ref{fig:band_lightcurves} the 2D \artisnlte and \artiscl simulations predict substantial differences in their light curves across all bands. These differences are present in the U- and B-bands as early as 5\,d after explosion, and become apparent in the V-, R- and I-bands from around 10\,d onward (see Figure~\ref{fig:band_lightcurves}). Comparing equivalent viewing angles, the \artisnlte simulation predicts significantly bluer light curves, with brighter and more slowly evolving U-, B- and V-band light curves, fainter R- and I-band light curves, and a much faster evolution of the I-band light curve (see Figure~\ref{fig:band_lightcurves}). Consequently, the 2D \artiscl simulation evolves to redder colours earlier and more rapidly (see Figure~\ref{fig:colour_evolution}). The most striking difference between the simulations is the strong secondary I-band maximum predicted by the 2D \artiscl simulation for all viewing angles, which is entirely absent in the 2D \artisnlte simulation. This drives much of the redder evolution of the \artiscl simulation.

We find that the ionisation state is generally higher in the 2D \artisnlte simulation, as illustrated by the free electron fraction shown at different epochs for our three simulations in Figure~\ref{fig:Te_nne_2D_maps} (corresponding electron temperatures are also shown). This increase in ionisation is expected because non-thermal ionisation is included in the \artisnlte simulation while the \artiscl simulation only considers photoionisation. Non-thermal ionisation can increase the ionisation state both through direct non-thermal ionisations and through the subsequent recycling of photons (represented here by Monte Carlo packets), which enhances photoionisation (\citealt{axelrod1980a}; this behaviour is present in the nebular phase \artisnlte simulations of \citealt{shingles2022a}). Both non-thermal ionisation and photoionisation are important processes in the \artisnlte simulation. 

The light curve differences are primarily due to these differences in ionisation state between the simulations and, in particular, the more substantial recombination to singly ionised species predicted by the \artiscl simulation from 10\,d after explosion onward. In SNe~Ia, the secondary maximum in the near-infrared (NIR) bands is attributed to doubly ionised IGEs recombining to singly ionised in the Fe-rich inner ejecta regions \citep{kasen2006a,kromer2009a,dessart2014b, jack2015a}. Figure~\ref{fig:Fe_ionisation_2D_maps} shows Fe ion populations for our 2D  simulations around the time of the only I-band peak in the \artisnlte simulation (9.5\,d) and around the time of the second I-band peak (29.5\,d) in the \artiscl simulation. At 9.5\,d \ion{Fe}{ii} has a relatively similar population in both simulations. However, the \artiscl simulation predicts a very substantial increase in \ion{Fe}{ii} population at lower velocities (below $\sim$20\,000 km\,s$^{-1}$) at 29.5\,d that is not predicted by the \artisnlte simulation. The higher ionisation state of the \artisnlte simulation is further highlighted by the \ion{Fe}{iv} populations at 29.5\,d, with the \artisnlte simulation still showing a very substantial population, particularly below 20\,000 km\,s$^{-1}$, while the \artiscl simulation only has \ion{Fe}{iv} present in the very inner regions of the ejecta. 

Both simulations predict strong absorption from singly ionised species, particularly \ion{Fe}{ii}, \ion{Co}{ii}, \ion{Ti}{ii} and \ion{Cr}{ii}, blueward of $\sim$4500\,\AA, with this flux subsequently redistributed to redder wavelengths via fluorescence. From $\sim$10\,d onwards, the greater population of singly ionised species in the higher velocity ejecta (above 20\,000 km\,s$^{-1}$) of the \artiscl simulation leads to substantially increased absorption and fluorescence, further contributing to its redder colours.

The differences in ionisation are also highlighted by the role of \ion{Co}{iii} in the simulations: \ion{Co}{iii} contributes significant absorption and fluorescence until peak in both simulations, but remains important after peak only in the \artisnlte simulation, while the \artiscl simulation becomes almost entirely dominated by the contributions of singly ionised species.

\subsubsection{Spectra}
\label{subsec:2D_spectra_comparisons}
\begin{figure*}
    \centering
    
    \includegraphics[width=0.99\textwidth, trim={0cm 0.0cm 0cm 0cm}, clip]{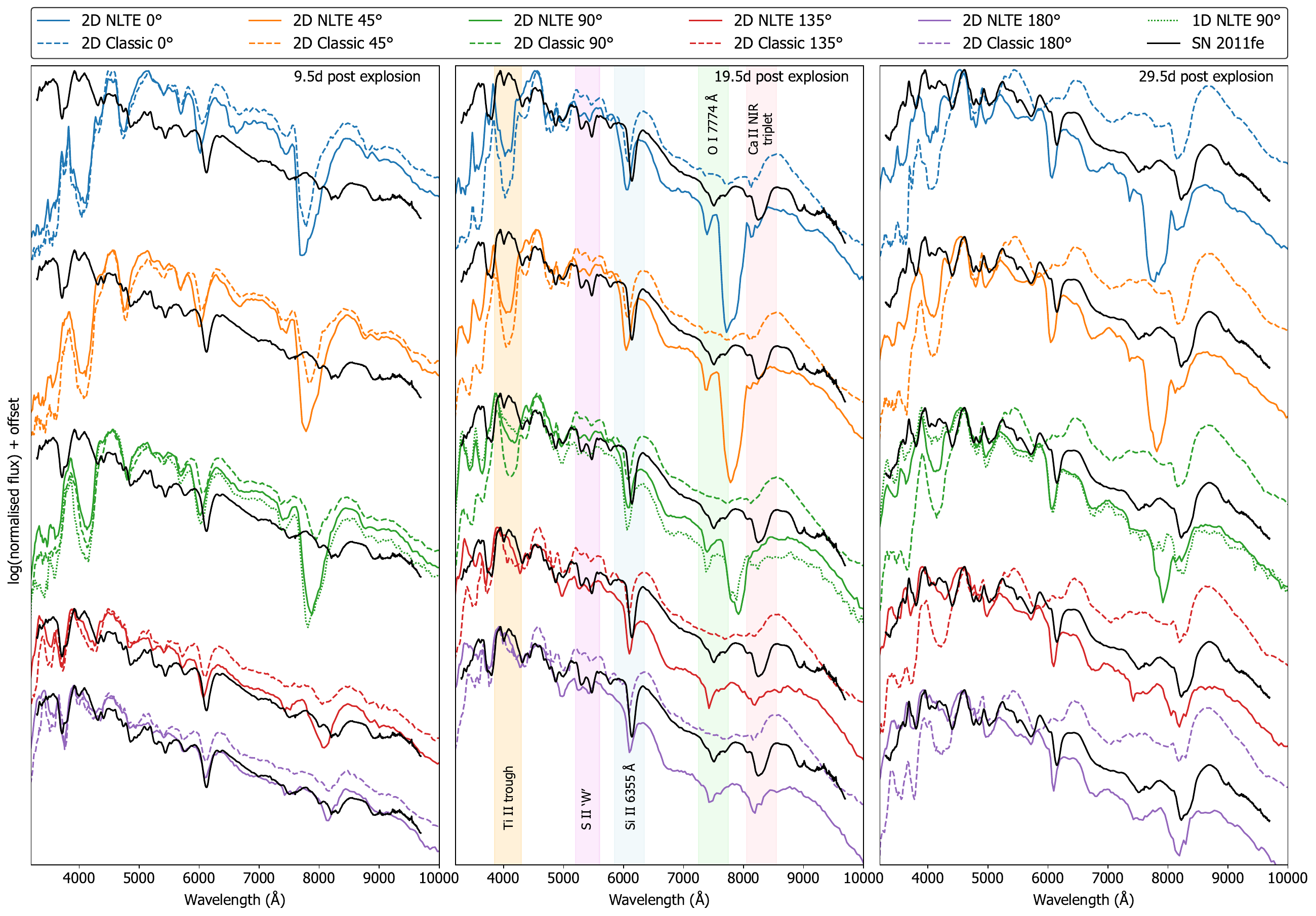}

    \caption{Simulated spectra for five different viewing angles from our 2D \artisnlte and \artiscl  simulations as well as our 1D \artisnlte simulation at 9.5, 19.5 (around peak) and 29.5\,d after explosion (left, middle and right panels). The normal SN~Ia 2011fe is plotted for reference \citep{pereira2013a}. In the spectra around peak we have highlighted key features we discuss in the text. Note, for visual clarity we highlight only the wavelength region of the observed \ion{Ca}{ii} NIR triplet. The high-velocity component predicted for $0^\circ$–$90^\circ$ viewing angles therefore appears just blueward of the highlighted region.}
    \label{fig:spectra}
\end{figure*}

\begin{figure}
    \centering
    
    \includegraphics[height=0.27\textwidth, trim={0.9cm 0.3cm 2.6cm 1.5cm}, clip]{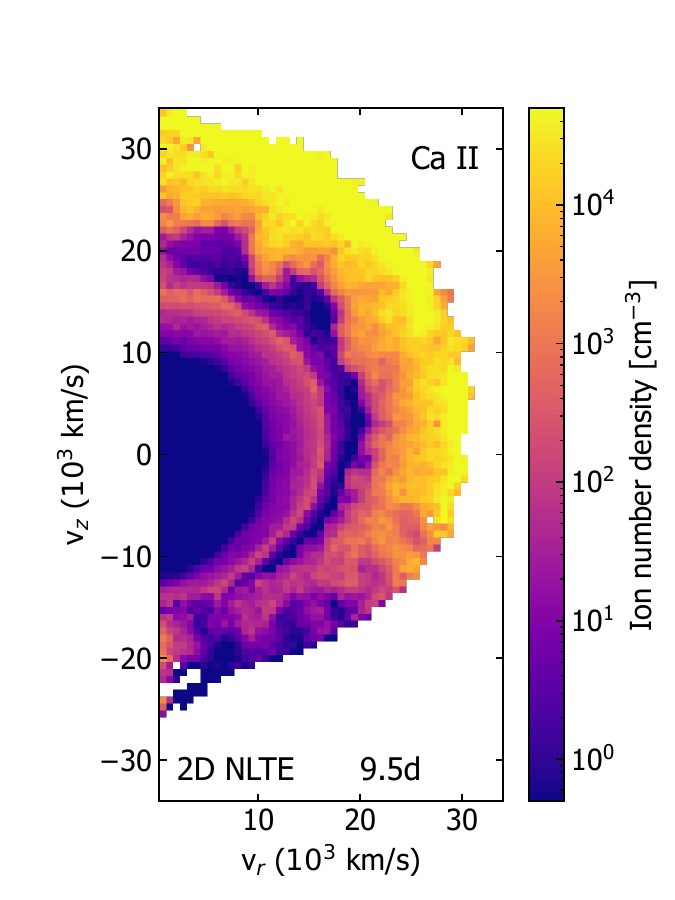}
    \includegraphics[height=0.27\textwidth, trim={2.6cm 0.3cm 2.6cm 1.5cm}, clip]{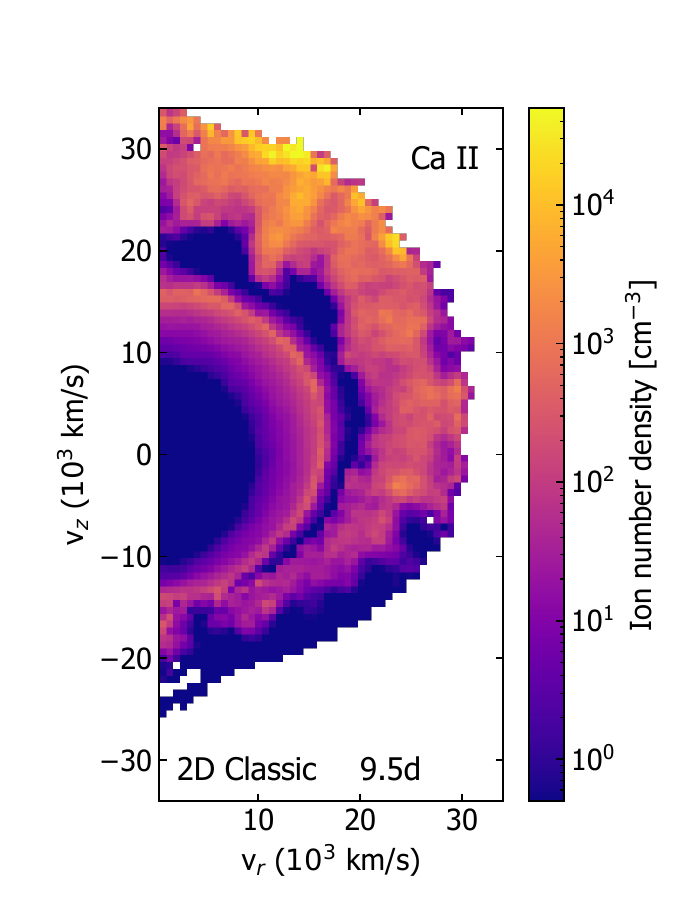}
    \includegraphics[height=0.27\textwidth, trim={2.6cm 0.3cm 0.2cm 1.5cm}, clip]{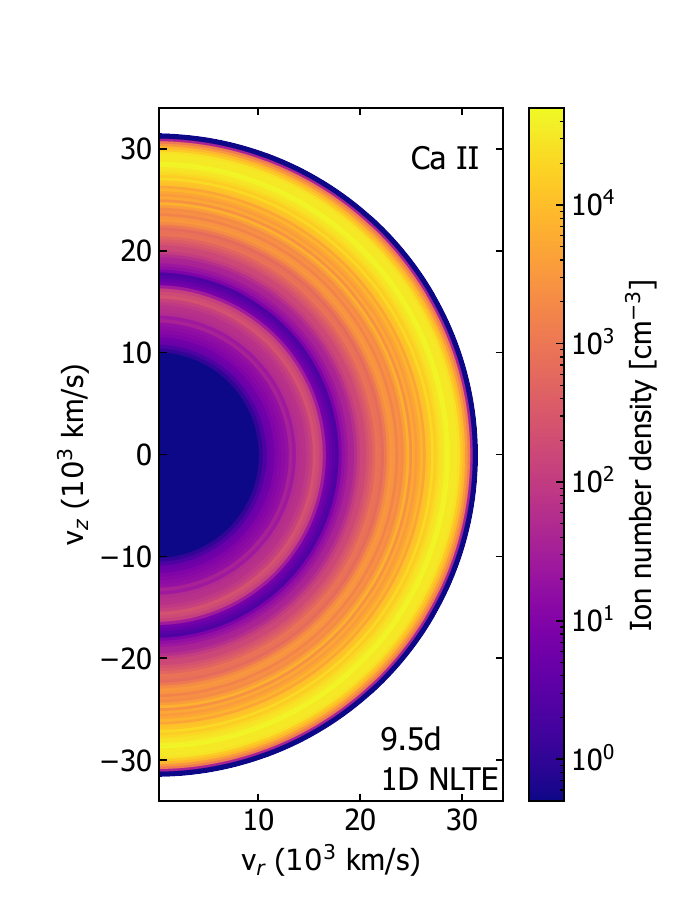}
    
    \includegraphics[height=0.27\textwidth, trim={0.9cm 0.3cm 2.6cm 1.5cm}, clip]{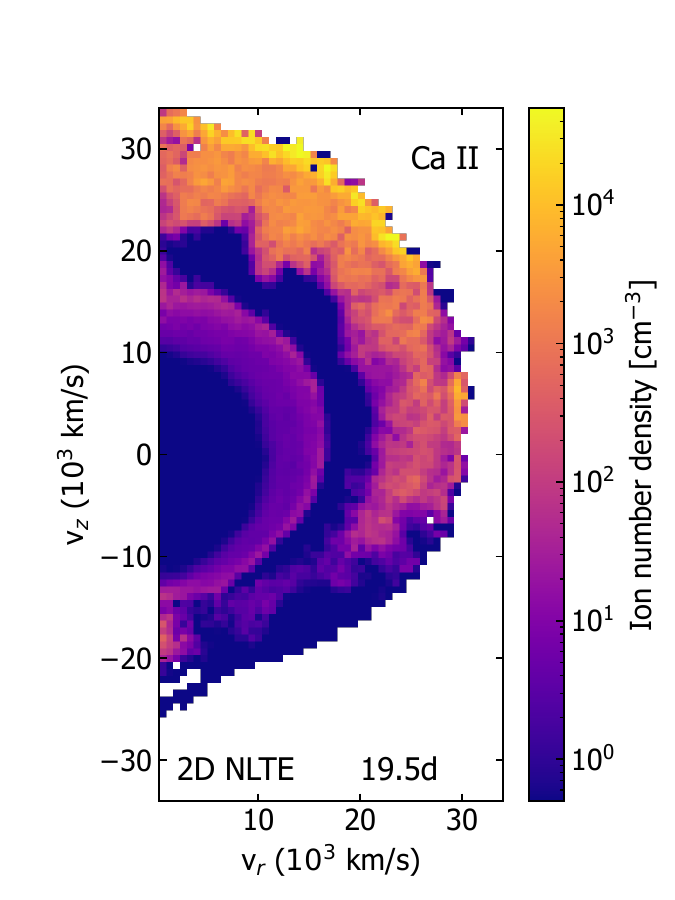}
    \includegraphics[height=0.27\textwidth, trim={2.6cm 0.3cm 2.6cm 1.5cm}, clip]{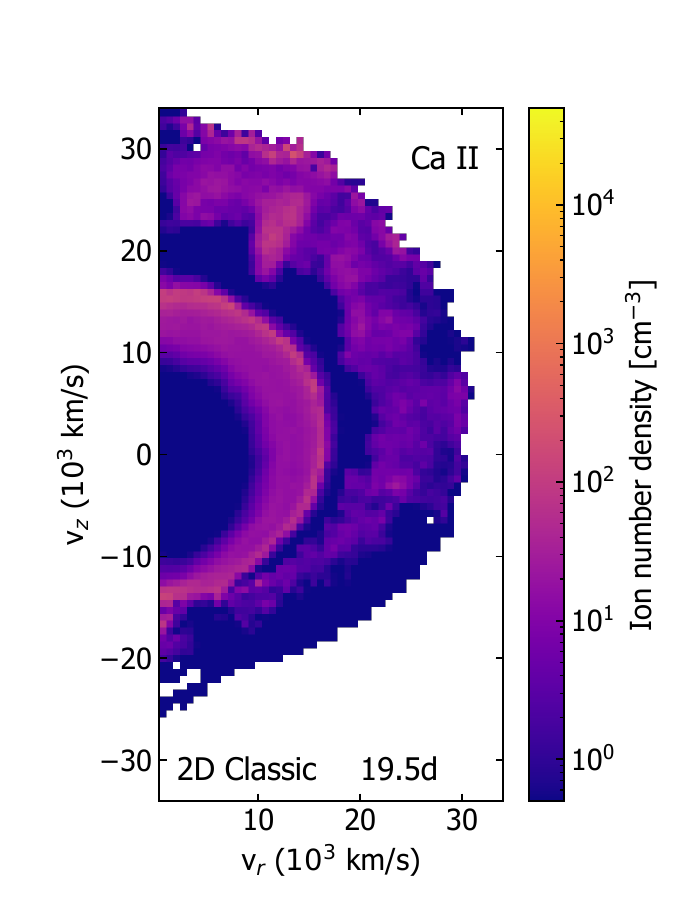}
    \includegraphics[height=0.27\textwidth, trim={2.6cm 0.3cm 0.2cm 1.5cm}, clip]{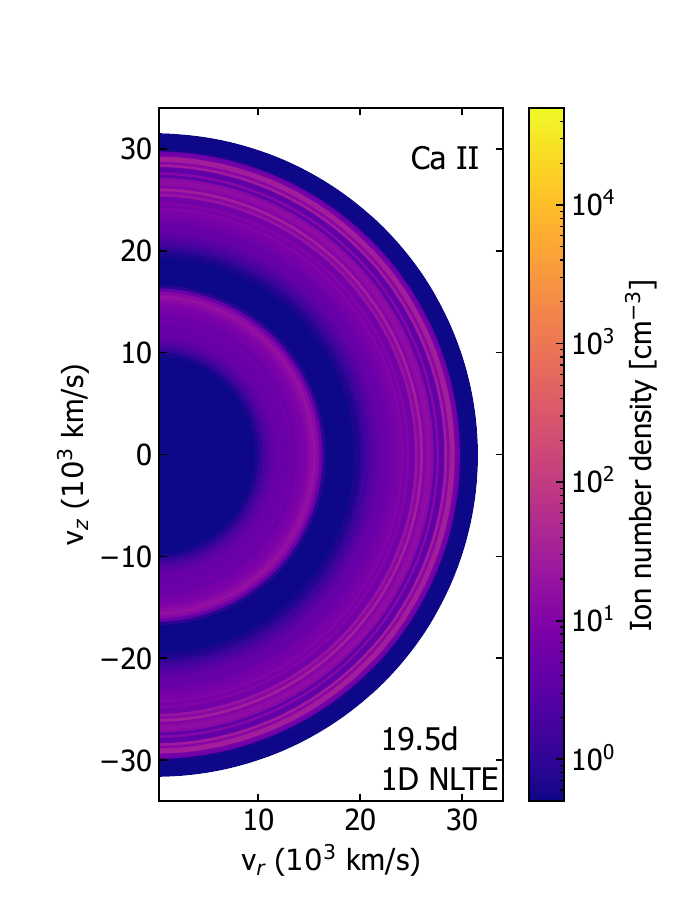}

    \includegraphics[height=0.27\textwidth, trim={0.9cm 0.3cm 2.6cm 1.5cm}, clip]{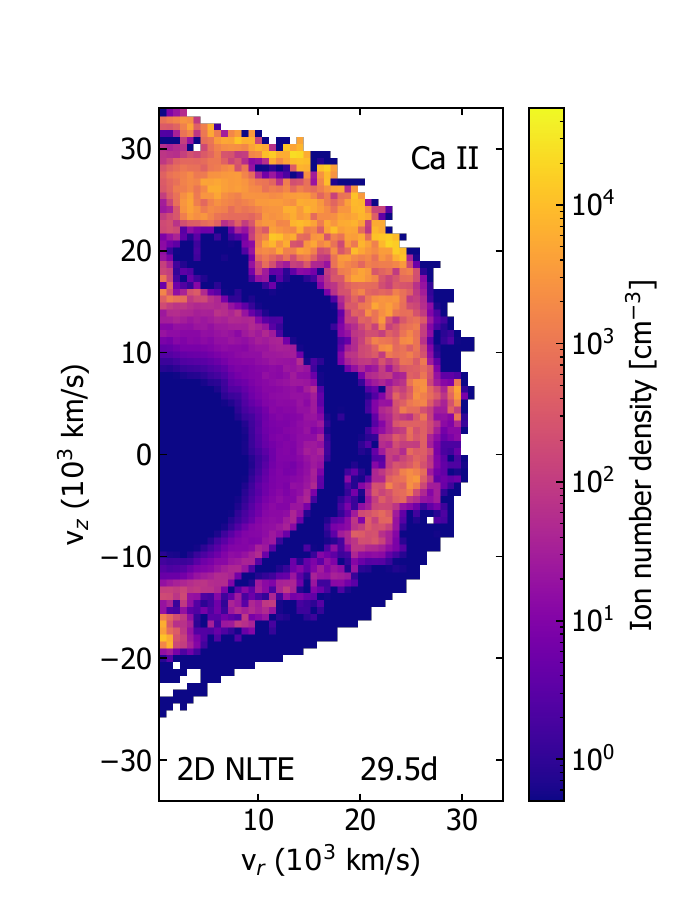}
    \includegraphics[height=0.27\textwidth, trim={2.6cm 0.3cm 2.6cm 1.5cm}, clip]{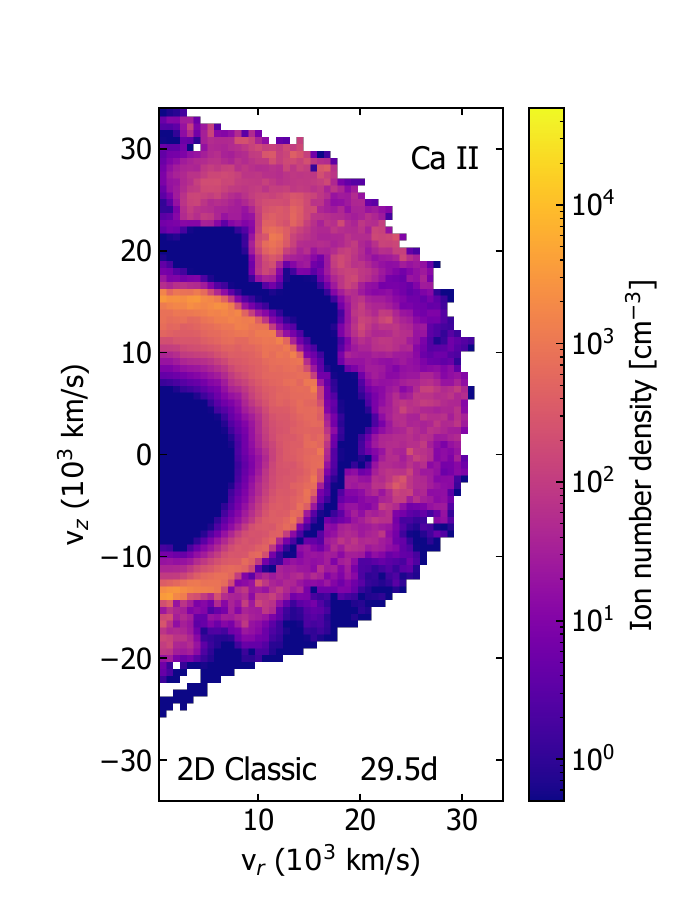}
    \includegraphics[height=0.27\textwidth, trim={2.6cm 0.3cm 0.2cm 1.5cm}, clip]{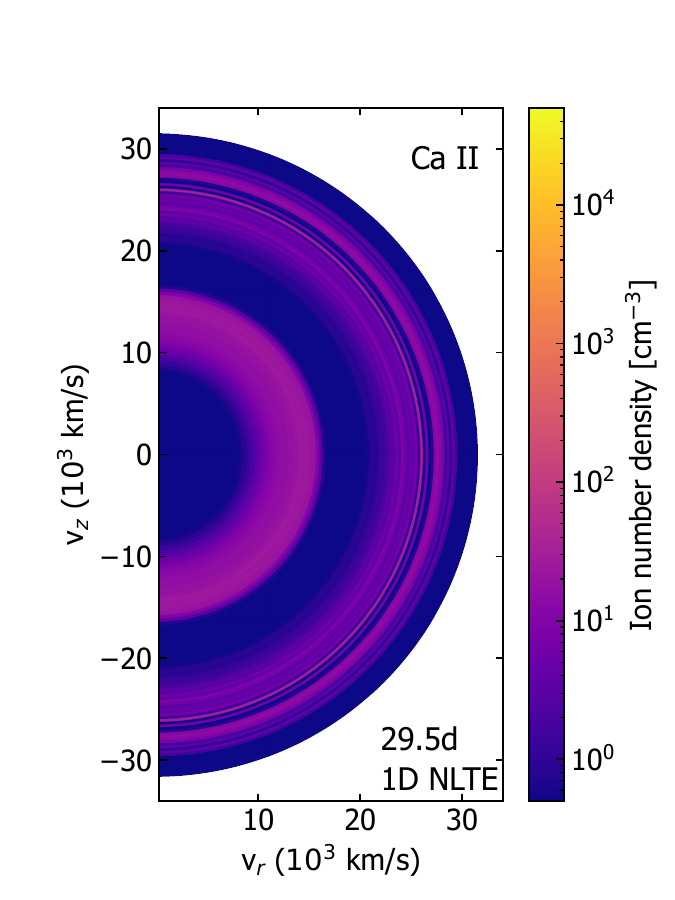}
    
    \caption{Number density of \ion{Ca}{ii} at 9.5, 19.5 and 29.5\,d after explosion for our 2D \artisnlte, 2D \artiscl and 1D \artisnlte simulations.}
    \label{fig:CaII_ionisation_2D_maps}
\end{figure}

\begin{figure}
	\includegraphics[width=0.9\linewidth,trim={0.0cm 0.0cm 0.0cm 0.0cm},clip]{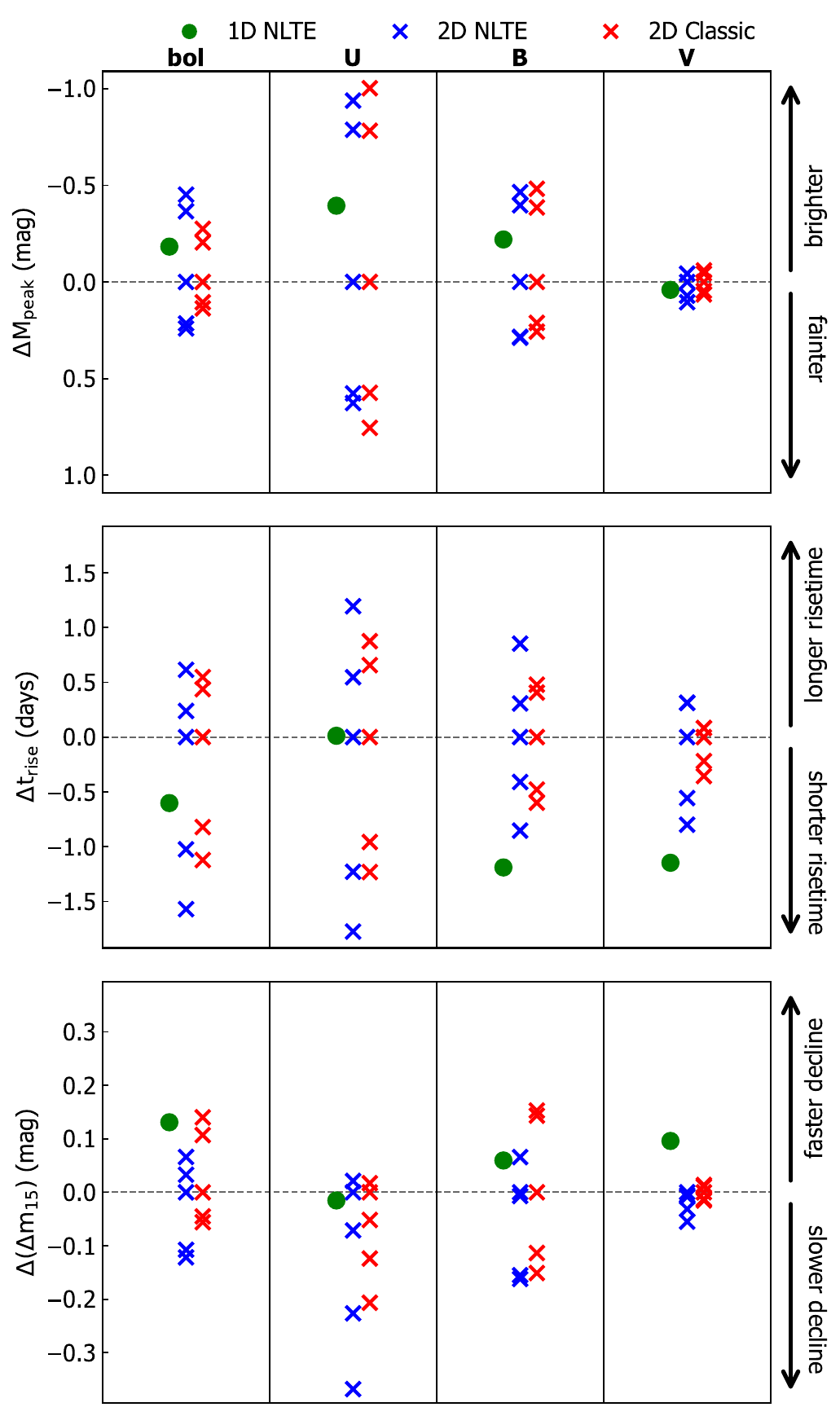}    
    \caption{Viewing angle variation of peak light curve brightness, rise time to peak and $\Delta \mathrm{{m}_{{15}}}$ in bolometric, U-, B- and V-band for the 2D \artisnlte and \artiscl simulations. The same five viewing angles as Figures~\ref{fig:band_lightcurves} and \ref{fig:spectra} are plotted with the 1D \artisnlte simulation also included for reference. The equatorial viewing angle (90$^\circ$) is plotted at zero (dashed horizontal line). The value of this viewing angle is subtracted from the other viewing angles (and the 1D model) with the differences from the equatorial ($\Delta$) then plotted for each quantity. The R- and I-band light curves show a more complex (flat or double peaked) evolution and minimal viewing angle variations in both simulations so are not included here.}
    \label{fig:bolUBV_viewing_angle_scatter}
\end{figure}

Figure~\ref{fig:spectra} shows the optical spectra before, around and after maximum light for five viewing angles from the 2D \artisnlte and \artiscl simulations, together with the 1D \artisnlte simulation and SN~2011fe for reference. Comparing equivalent viewing angles in Figure~\ref{fig:spectra}, the 2D \artisnlte and \artiscl simulations show clear differences for all epochs. The spectra are most similar before peak and increasingly diverge as we move to later times. 
The simulations also differ in their spectroscopic viewing angle dependence at later epochs (see Section~\ref{subsec:viewing_angle_variation} for details). 

As discussed in Section~\ref{subsec:2D_lightcurve_comparisons}, one of the most striking differences between the 2D \artisnlte and \artiscl simulations is the redder spectral energy distribution (SED) predicted by the \artiscl simulation, particularly from peak onward, primarily owing to the presence of a strong secondary NIR maximum in the \artiscl simulation that is absent in the \artisnlte simulation. However, the treatment of the plasma conditions also impacts the evolution of individual spectral features in the simulations. Therefore, in the following, we focus on differences in the evolution of key spectral features, specifically the \ion{Ca}{ii} NIR triplet, \ion{Ti}{ii} absorption trough, and \ion{O}{i} 7774\,\AA\ and \ion{Si}{ii} 6355\,\AA\ features.

\subsubsubsection{\ion{Ca}{ii} NIR triplet}
\noindent The Ca distribution in the 2D ejecta model shows two distinct regions: high-velocity Ca from the He-shell detonation and lower velocity Ca from the core detonation (see Figure~\ref{fig:2D_ejecta_composition}). In both simulations, for viewing angles from the north pole to the equator ($0^\circ$--$90^\circ$), these regions are sufficiently separated to produce distinct high- and low-velocity components in the \ion{Ca}{ii} NIR triplet, while for southern viewing angles ($135^\circ$, $180^\circ$) they tend to blend into a single feature. The feature is strongest and appears at higher velocities towards the north pole and weakens and shifts to lower velocities moving to the south pole, reflecting the Ca distribution in the ejecta. Outside these general trends, the evolution of the NIR triplet differs markedly when comparing equivalent viewing angles between the \artisnlte and \artiscl simulations. In the following, we discuss the strengths and velocity shifts of the high- and low-velocity components of the NIR triplet\footnote{The 8542 and 8662\,\AA\ lines have oscillator strengths ${\sim}$9--10 times greater than the 8498\,\AA\ line and therefore dominate the \ion{Ca}{ii} NIR triplet. In the simulated spectra both the high- and low-velocity components of the NIR triplet often show some degree of `double-dip' absorption associated with the contributions of the 8542\,\AA\ and 8662\,\AA\ lines.  In these cases we measure the velocity shift relative to the 8542\,\AA\ line for the shorter wavelength dip and the 8662\,\AA\ line for the longer wavelength dip and take the average as the velocity shift of the component. When only a single clear absorption minimum is present, the velocity is measured relative to the line producing the deepest absorption where identifiable, or otherwise relative to the midpoint of the two lines. None of our conclusions are sensitive to the adopted method for calculating the velocity shift.}. 

For the northern and equatorial viewing angles ($0^\circ$--$90^\circ$), the evolution of the \ion{Ca}{ii} NIR triplet shows  substantial differences between the simulations (see Figure~\ref{fig:spectra}). Before peak (e.g. 9.5\,d in Figure~\ref{fig:spectra}), both simulations show strong high-velocity absorption features, with velocities of $\sim$25\,000--28\,000 km\,s$^{-1}$ in the \artisnlte simulation and $\sim$23\,000--27\,000 km\,s$^{-1}$ in the \artiscl simulation for these viewing angles. The absorption features are stronger in the \artisnlte simulation for equivalent viewing angles. Although low-velocity absorption is present in both simulations before peak, it is obscured by emission from the high-velocity component of the NIR triplet so no clear feature appears. Around peak (19.5\,d) the behaviour diverges more significantly for the $0^\circ$--$90^\circ$ viewing angles. The \artisnlte simulation still predicts strong high-velocity NIR triplet absorption at the same velocities, with a weaker low-velocity contribution also becoming visible (strongest absorption at ${\sim}$15\,000 km\,$\mathrm{s}^{-1}$ for all cases). In contrast, the \artiscl simulation shows no high-velocity absorption but does predict a clear low-velocity feature for these viewing angles, also at $\sim$15\,000 km\,$\mathrm{s}^{-1}$. After peak (29.5\,d), clear differences persist for these viewing angles. In the \artisnlte simulation the high-velocity component remains strong and the low-velocity component has strengthened for all cases. In the \artiscl simulation the low-velocity feature strengthens further and a weak high-velocity absorption feature reappears. In both simulations, the velocities of the two components show little evolution with time. 

For the southern viewing angles ($135^\circ$ and $180^\circ$), the differences in the evolution of the NIR triplet between the simulations are less pronounced (see Figure~\ref{fig:spectra}). In both simulations, the NIR triplet appears as a single absorption feature, with the exception of the \artiscl spectra after peak (see below). The absorption velocities of the NIR triplet are also consistent: at $180^\circ$ the velocity shift is ${\sim}$13\,000 km\,$\mathrm{s}^{-1}$ for all epochs, indicating that Ca from the core detonation dominates the opacity. At $135^\circ$ the velocity decreases from ${\sim}$17,000 km\,$\mathrm{s}^{-1}$ before peak (9.5\,d) to ${\sim}$13\,000 km\,$\mathrm{s}^{-1}$ for the later epochs, implying a transition from Ca produced in the He-shell detonation to Ca from the core detonation dominating the opacity in both cases. The main difference emerges from peak onward (19.5 and 29.5\,d) when the \artisnlte simulation predicts an extended red wing to the NIR triplet associated with higher-velocity \ion{Ca}{ii} absorption, which is absent in the \artiscl simulation. The \artiscl simulation instead predicts no higher velocity absorption at peak with a weak separate higher velocity absorption component then appearing for both viewing angles after peak. 

The variations arise from the different evolution of the \ion{Ca}{ii} populations (see Figure~\ref{fig:CaII_ionisation_2D_maps}). In the \artisnlte simulation, the low-velocity population shows little change with time, while the high-velocity population declines gradually. The \artiscl simulation also shows little change in the low-velocity population from 9.5\,d to 19.5\,d, however, it then predicts a substantial increase at 29.5\,d. In contrast, the high-velocity population drops significantly between 9.5 and 19.5\,d, before also increasing strongly at 29.5\,d. This indicates increased recombination to \ion{Ca}{ii} after peak, consistent with the overall lower ionisation state and stronger recombination to singly ionised species predicted by the \artiscl simulation (see Section~\ref{subsec:2D_lightcurve_comparisons}).  

\subsubsubsection{\ion{Ti}{ii} absorption trough}
\noindent Both simulations show significant absorption and fluorescence from \ion{Ti}{ii} for all viewing angles. They also predict that the strength of the \ion{Ti}{ii} contribution varies substantially with viewing angle. Before and around peak (e.g. 9.5 and 19.5\,d in Figure~\ref{fig:spectra}) both simulations predict a very strong \ion{Ti}{ii} trough for the $0^\circ$ viewing angle which systematically weakens as we move towards $180^\circ$, where no clear trough is present in either simulation. This is a result of the greater abundance of Ti at high velocities towards the north pole of the ejecta (see Figure~\ref{fig:2D_ejecta_composition}), and the lower ionisation of these regions relative to the southern viewing angles in both simulations, leading to a greater \ion{Ti}{ii} population. After peak (29.5\,d) the \artiscl simulation predicts a significant reduction in the viewing angle variation of the feature with a strong \ion{Ti}{ii} trough now predicted for all viewing angles. In contrast, the \artisnlte simulation still predicts a strong viewing angle dependence in the feature. This is again related to the lower ionisation state of the \artiscl simulation: the significant recombination to singly ionised species for this epoch means the \ion{Ti}{ii} populations are large enough to produce strong troughs for all viewing angles, not just the directions where there is a greater amount of Ti ejected.  

Comparing equivalent viewing angles, the \artiscl simulation predicts a systematically stronger \ion{Ti}{ii} absorption trough at all epochs, with the largest differences appearing from peak onwards. This is a consequence of its lower ionisation state, which results in larger populations of singly ionised species. This leads to increased \ion{Ti}{ii} absorption, while stronger absorption from other singly ionised species, particularly \ion{Fe}{ii}, also contributes to the deeper troughs.

\subsubsubsection{\ion{O}{i} 7774\,\AA\ feature}
\noindent The \ion{O}{i} 7774\,\AA\ absorption feature\footnote{The line we refer to here as \ion{O}{i} 7774\,\AA, as is common in the literature, is actually a triplet with wavelengths 7772, 7774 and 7775\,\AA.} is clearly present for all viewing angles in the \artisnlte simulation at all epochs, while the \artiscl simulation predicts that this feature is only present before peak. Both simulations predict velocity shifts of ${\sim}$13,000--16,000 km\,$\mathrm{s}^{-1}$ for the feature, depending on the viewing angle. This is consistent with the feature forming in the most O-rich ejecta region (see Figure~\ref{fig:2D_ejecta_composition}).  

In the feature forming region \ion{O}{ii} is the dominant ionisation stage until peak in the \artiscl simulation (after which \ion{O}{i} becomes dominant), whereas \ion{O}{ii} remains dominant throughout the \artisnlte simulation. The lower level of the \ion{O}{i} 7774\,\AA\ transition is metastable and has a relatively high excitation energy of $\sim$9\,eV above the ground state,  suggesting that recombination can be an important mechanism for populating this level. Since \artisnlte determines the level populations using a full solution to the NLTE equations of statistical equilibrium (see Section~\ref{subsec:RT_method} for details), recombination from \ion{O}{ii} can contribute to populating the lower level of the \ion{O}{i} 7774\,\AA\ transition throughout the simulation. However, as excited level populations in \artiscl are calculated by applying the Boltzmann formula to each ion, the lower level of this transition is only sufficiently populated when temperatures are very high (see \citealt{kromer2009a} for details) and the feature is only predicted before peak.

\subsubsubsection{\ion{Si}{ii} 6355\,\AA\ feature}
\noindent The \artisnlte simulation predicts that the \ion{Si}{ii} 6355\,\AA\ absorption feature is strong and clearly visible for all viewing angles at all epochs. Comparing equivalent viewing angles, the \artiscl simulation predicts similar strengths and velocities for this feature at 9.5 and 19.5\,d. However, at 29.5\,d this feature has almost entirely disappeared in the \artiscl simulation for all viewing angles, despite a similar amount of absorption from \ion{Si}{ii} 6355\,\AA\ in both simulations. The different behaviour of the \ion{Si}{ii} 6355\,\AA\ feature is instead due to increased fluorescence from singly ionised IGEs (in particular \ion{Fe}{ii}) in the \artiscl simulation, which overwhelms the feature. This is again attributed to the lower ionisation state of the \artiscl simulation.

\subsubsection{Degree of viewing angle variation}
\label{subsec:viewing_angle_variation}

As discussed in Sections~\ref{subsec:2D_lightcurve_comparisons} and \ref{subsec:2D_spectra_comparisons}, there are significant variations in photometric and spectroscopic behaviour between the 2D \artisnlte and \artiscl simulations when comparing equivalent viewing angles. However, the viewing angle variation predicted in the synthetic observables of different explosion models is also key to evaluating their success relative to observations. Here we discuss to what extent using the NLTE treatment of \artisnlte versus the approximate NLTE treatment of \artiscl impacts the overall scale of viewing angle variations predicted by 2D radiative transfer simulations of our double detonation model at early times.

Both the 2D \artisnlte and \artiscl simulations predict the same photometric trends with viewing angle (see Figure~\ref{fig:band_lightcurves}). Moving from the north ($\theta=0^\circ$) to south ($\theta=180^\circ$) pole the models become brighter and more quickly evolving. Additionally, for both models, the U-band shows the greatest viewing angle variation, followed by the B-band, with the redder bands showing substantially less variation. 

The scale of the viewing angle dependence in the band light curves is generally similar between simulations, with the \artisnlte simulation predicting slightly more variation. This is highlighted by Figure~\ref{fig:bolUBV_viewing_angle_scatter}, which shows the viewing angle variation of derived bolometric and band light curve quantities (peak brightness, rise time to peak, and $\Delta \mathrm{{m}_{{15}}}$) for the 2D \artisnlte and \artiscl simulations.

Both simulations exhibit a similar amount of viewing angle variation in U-, B-, V- and R-band peak brightness. However, from ${\sim}$10\,days onward, the \artisnlte simulation shows more variation in I-band brightness (see Figure~\ref{fig:band_lightcurves}), resulting in greater viewing angle variation in the peak bolometric brightness of this simulation. The \artisnlte simulation also shows greater viewing angle variation in rise time, although the effect is modest in both cases: in U-band, where the variation is largest, the rise time only differs by ${\sim}$3 days from fastest to slowest rise for the \artisnlte simulation. 

While the variation in $\Delta \mathrm{{m}_{{15}_{U}}}$ is also greater for the \artisnlte simulation, it predicts less variation in $\Delta \mathrm{{m}_{{15}_{B}}}$, with both simulations showing very little scatter in $\Delta \mathrm{{m}_{{15}_{V}}}$. As a result, the magnitude of the viewing-angle variation in $\Delta \mathrm{{m}_{{15}_{bol}}}$ is similar between simulations. However, while the \artisnlte simulation predicts the slower-declining viewing angles show more variation from the equatorial, the \artiscl simulation predicts it is instead the faster declining viewing angles which show greater variation. 

As discussed above, the 2D \artisnlte simulation only shows a modest overall increase in the scale of the viewing angle variation of photometric quantities relative to the 2D \artiscl simulation. Before peak the 2D \artisnlte and \artiscl simulations also predict a comparable degree of spectroscopic variation with viewing angle. However, at later epochs, the \artisnlte simulation continues to exhibit substantial spectroscopic viewing angle variation, whereas the \artiscl spectra exhibit progressively less viewing angle dependence. This behaviour is particularly evident in the \ion{Ti}{ii} absorption trough and \ion{Ca}{ii} NIR triplet. This demonstrates the importance of using spectra in addition to photometric quantities when assessing viewing angle effects, as the spectroscopic viewing angle dependence may not be reflected in the light curves. 

We note that \cite{collins2022a} previously carried out an \artiscl simulation for the M10\_02 model investigated in this work. Their simulation used the atomic data set compiled by \cite{gall2012a} (see \citealt{collins2022a} for details). However, that atomic data set lacks the electron collision data and radiative transition rates for forbidden transitions required to fully exploit the NLTE capabilities of \artisnlte. All simulations in this work therefore use the atomic data set compiled by \textsc{cmfgen} (see Section \ref{subsec:RT_method}) which includes this additional information. This does, however, lead to some differences in the predicted viewing angle variation when comparing our \artiscl simulation to theirs: the brighter viewing angles show similar variation but the fainter viewing angles in the \cite{collins2022a} simulation predict more viewing angle variation, particularly in U- and B-band.  

\subsection{2D versus 1D \artisnlte comparisons for the equatorial direction}
\label{subsec:1D_2D_NLTE_comparisons}
The substantial viewing angle variation predicted by the 2D \artisnlte simulation (see e.g. Section~\ref{subsec:viewing_angle_variation}) cannot be reproduced by any single 1D model. However, our 2D \artisnlte simulation allows us to investigate how well a 1D model captures the effects of the corresponding line-of-sight in the multi-dimensional model. In the following, we compare the synthetic observables from the equatorial ($90^\circ$) viewing angle in the 2D \artisnlte simulation with those from the 1D \artisnlte simulation constructed based on this direction (see Section~\ref{subsec:ejecta_model}). Throughout this section, when discussing the 2D \artisnlte simulation, we are therefore referring to its equatorial properties. 

We select the equatorial direction for our comparisons, as this represents the most probable viewing angle in the 2D model for the observer. \cite{collins2025a} demonstrated for the M2a double detonation model of \cite{gronow2020a} that synthetic observables from a 1D model constructed based on the equatorial direction show smaller discrepancies relative to the corresponding multi-dimensional viewing angle than those constructed from polar directions. Therefore, the differences discussed are very likely a lower limit on those expected for a given direction in the model.

\subsubsection{Light curves}
\label{subsec:1D_2D_NLTE_lightcurve_comparisons}
From Figure~\ref{fig:band_lightcurves} we can see that the bolometric, U- and B-band light curves show clear differences between the equatorial viewing angle in the 2D \artisnlte simulation and the 1D \artisnlte simulation. The bolometric light curve of the 1D simulation rises more rapidly to peak, reaches a higher peak luminosity and declines more slowly after peak (see Table~\ref{tab:1D_2D_simulations_lightcurve_params}). It is therefore evident that the multi-dimensional structure directly impacts the observables: despite having a slightly smaller \nickel\ mass than the 2D model (0.51 compared to 0.55\,\msun), the 1D \artisnlte simulation predicts a greater bolometric luminosity around peak than the equatorial viewing angle of the 2D \artisnlte simulation. The different bolometric evolution is primarily driven by the U- and B-band light curves. In U-band the 1D simulation is brighter from ${\sim}$10\,d until the end of the simulations and ${\sim}$0.4\,mag brighter at peak. In B-band the 1D simulation rises faster to peak by over a day and is also ${\sim}$0.2\,mag brighter at peak. The V-, R- and I-band light curves are much more similar, with the most noticeable difference being the faster post-peak decline of the 1D simulation in these bands. Light curve differences are also reflected in clear variations in the derived photometric quantities when comparing the simulations (see Figure~\ref{fig:bolUBV_viewing_angle_scatter}). 

Both simulations display similar overall trends in colour evolution (see Figure~\ref{fig:colour_evolution}), with an initial evolution to bluer colours in U-B, B-V and R-I before evolving to redder colours and very little evolution in their V-R colours. Clear differences are however present, particularly in the blue bands. In U-B, the 1D simulation is redder by ${\sim}$0.5\,mag for the first 10\,days before showing a bluer U-B colour for the remainder of the simulation. In B-V, the two simulations predict similar colour evolution for the first ${\sim}$10\,days, but the 1D NLTE simulation predicts bluer colours than the 2D NLTE simulation in B-V after this time. As with the light curves, the differences in the colour evolution of the redder bands are minimal. 

Blueward of ${\sim}$4000\,\AA\ both simulations show very substantial line blanketing attributed to a combination of singly and doubly ionised IGEs and intermediate mass elements (IMEs) and singly ionised Ti and Cr. This absorption substantially suppresses the flux at these wavelengths, and the fluorescence of the absorbed radiation to redder wavelengths significantly impacts the optical spectral formation. It is differences in this fluorescence that primarily drive the differences in the light curves and colour evolution of the simulations. The majority of this fluorescence occurs for wavelengths covered by the U- and B-bands (${\sim}$3000--5500\,\AA) resulting in these bands showing the most prominent differences. No individual species shows very obvious differences between simulations; instead, subtle differences in the contributions of multiple species together lead to the variation between simulations. 

The colour differences between the \artisnlte and \artiscl simulations were primarily due to their different ionisation states. However, the 1D and 2D \artisnlte simulations predict a similar overall ionisation state. Instead the differences here are primarily driven by the greater bolometric luminosity of the 1D simulation, which leads to increased absorption and fluorescence, particularly for bluer wavelengths. 

\begin{table*}
\centering
\caption{Light curve rise times (in days), peak brightnesses and declines rates in bolometric, U-, B- and V-band for the equatorial viewing angle of the 2D \artisnlte simulation and the 1D \artisnlte simulation.}
\begin{tabular*}{\textwidth}{@{\extracolsep{\fill}}lcccccccccccc}
\toprule
Simulation 
& $\mathrm{t}_{\mathrm{rise}}^{\mathrm{bol}}$ 
& $\mathrm{M}_{\mathrm{peak}}^{\mathrm{bol}}$
& $\Delta \mathrm{m}_{15}^{\mathrm{bol}}$
& $\mathrm{t}_{\mathrm{rise}}^{\mathrm{U}}$  
& $\mathrm{M}_{\mathrm{peak}}^{\mathrm{U}}$ 
& $\Delta \mathrm{m}_{15}^
{\mathrm{U}}$
& $\mathrm{t}_{\mathrm{rise}}^{\mathrm{B}}$  
& $\mathrm{M}_{\mathrm{peak}}^{\mathrm{B}}$ 
& $\Delta \mathrm{m}_{15}^{\mathrm{B}}$
& $\mathrm{t}_{\mathrm{rise}}^{\mathrm{V}}$  
& $\mathrm{M}_{\mathrm{peak}}^{\mathrm{V}}$ 
& $\Delta \mathrm{m}_{15}^{\mathrm{V}}$ \\
\midrule
2D \artisnlte $90^\circ$  & $19.0$ & $-18.64$ & $0.87$ & $19.3$ & $-19.12$ & $1.40$ & $19.7$ & $-19.04$ & $0.99$ & $20.1$ & $-19.18$ & $0.80$ \\

1D \artisnlte & $18.4$ & $-18.82$ & $1.00$ & $19.3$ & $-19.52$ & $1.39$ & $18.5$ & $-19.26$ & $1.05$ & $19.5$ & $-19.14$ & $0.89$ \\
\bottomrule
\end{tabular*}
\label{tab:1D_2D_simulations_lightcurve_params}
\end{table*}

\subsubsection{Spectra}
\label{subsec:1D_2D_NLTE_spectra_comparisons}
Comparing the optical spectra predicted by the equatorial (90$^\circ$) viewing angle in the 2D \artisnlte simulation and the 1D \artisnlte simulation we see clear differences before, at and after peak (see Figure~\ref{fig:spectra}). Variations in the SED of the simulations have already been discussed in Section \ref{subsec:1D_2D_NLTE_lightcurve_comparisons} in the context of the band light curve colours. In the following we focus on differences in the evolution of the \ion{Ca}{ii} NIR triplet and the \ion{Ti}{ii} absorption trough.  

Before peak (e.g. 9.5\,d in Figure~\ref{fig:spectra}), both simulations predict \ion{Ca}{ii} NIR triplets dominated by very strong high-velocity absorption, with no clearly separated low-velocity component. In both cases, the \ion{Ca}{ii} population is greater at high velocities (see Figure~\ref{fig:CaII_ionisation_2D_maps}) resulting in very substantial optical depth from the \ion{Ca}{ii} NIR triplet (Figure~\ref{fig:Ca_II_Sobolev_depth_maps} demonstrates that high-velocity absorption is almost entirely saturated at this epoch in both simulations). Although there is also significant optical depth at low velocities, this absorption is masked by emission from the high-velocity \ion{Ca}{ii}. Before peak the NIR triplet has a velocity of ${\sim}$25\,000 km\,$\mathrm{s}^{-1}$ in the 2D \artisnlte simulation and ${\sim}$27\,000 km\,$\mathrm{s}^{-1}$ in the 1D \artisnlte simulation (our approach for determining velocity shifts is described in Section \ref{subsec:2D_spectra_comparisons}). As the \ion{Ca}{ii} populations are similar in both simulations at this epoch the difference is a result of variations in the velocity structure of Ca, which is artificially smoothed in the 1D model. 

Around peak (19.5\,d), the equatorial direction of the 2D \artisnlte simulation shows a strong high-velocity \ion{Ca}{ii} NIR triplet absorption component at the same velocity as before peak. No clearly separated low-velocity absorption component is present. Instead, the high-velocity feature exhibits an extended red wing associated with low-velocity absorption. In contrast, the 1D \artisnlte simulation predicts two distinct components: a high-velocity component (again at ${\sim}$27,000 km\,$\mathrm{s}^{-1}$) and low-velocity component (at ${\sim}$14,000 km\,$\mathrm{s}^{-1}$). These differences arise from variations in the \ion{Ca}{ii} population, and therefore the NIR triplet optical depth, in the high- and low-velocity Ca. Both simulations show a reduction in the \ion{Ca}{ii} population between 9.5\,d and 19.5\,d (Figure~\ref{fig:CaII_ionisation_2D_maps}). However, the decrease in high-velocity \ion{Ca}{ii} is much greater in the 1D simulation. As a result, the high- and low-velocity regions in the 1D model have similar \ion{Ca}{ii} populations and hence similar NIR triplet optical depths (see Figure~\ref{fig:Ca_II_Sobolev_depth_maps}). Despite these similar optical depths, the high-velocity absorption in the 1D model remains more prominent because emission from the high-velocity component again partially masks the low-velocity absorption. In the 2D simulation, the smaller reduction in high-velocity \ion{Ca}{ii} results in significantly larger optical depths at high velocities, producing both a stronger high-velocity absorption feature and stronger emission that obscures the low-velocity absorption more substantially.

After peak (29.5\,d), the equatorial direction of the 2D \artisnlte simulation again shows a strong high-velocity \ion{Ca}{ii} NIR triplet absorption component at the same velocity. However, a distinct low-velocity absorption feature with a velocity shift of ${\sim}$15\,000 km\,$\mathrm{s}^{-1}$ now also appears, although it remains weaker than the high-velocity component. In contrast, the 1D \artisnlte simulation predicts that the high-velocity component has weakened substantially with the low-velocity absorption instead dominating. We note that the high-velocity components in both simulations, and low-velocity component in the 1D \artisnlte simulation, show no significant velocity evolution across the three epochs. As at peak, the differences arise from the \ion{Ca}{ii} populations and the resulting NIR triplet optical depths. Relative to peak, the 1D \artisnlte simulation predicts a further decline in the high-velocity \ion{Ca}{ii} population and a significant increase at low velocities, leading to much greater optical depth in the NIR triplet at low velocities (Figure~\ref{fig:Ca_II_Sobolev_depth_maps}). The 2D \artisnlte simulation also shows an increase in the low-velocity \ion{Ca}{ii} population at this epoch. However, unlike the 1D model, it shows only a small reduction in high-velocity \ion{Ca}{ii}, so the NIR triplet optical depth at high velocities remains dominant. The variations in the high-velocity \ion{Ca}{ii} populations between the simulations are due to differences in the excited state \ion{Ca}{ii} photoionisation rate, with the 1D \artisnlte simulation predicting a higher \ion{Ca}{ii} photoionisation rate at high velocities than the equatorial direction of the 2D \artisnlte simulation. This is discussed in detail in Appendix \ref{appendix:Ca_II_photoionisation}. We do, however, note that 1D test simulations demonstrate that the exact strength and evolution of the high-velocity component of the \ion{Ca}{ii} NIR triplet is one of the more sensitive spectral features to Monte Carlo noise due to the small number of photon packets around the \ion{Ca}{ii} photoionisation edges in the simulations (see Appendix D). 

The 1D and 2D \artisnlte simulations also predict clear differences in the \ion{Ti}{ii} absorption trough. At 9.5\,d the 1D simulation predicts a deeper \ion{Ti}{ii} absorption trough. Both simulations predict a similar amount of absorption from \ion{Ti}{ii} at this epoch. It is instead variations in the spectral contributions of the \ion{Ca}{ii} H\&K lines, which appear as a P~Cygni feature to the blue of the \ion{Ti}{ii} trough, that drive this difference. At this epoch the 2D \artisnlte simulation predicts a stronger H\&K emission feature that overlaps with the \ion{Ti}{ii} absorption trough. This is not the case for the 1D simulation which therefore predicts a deeper absorption trough.

At 19.5\,d and 29.5\,d it is instead the 2D \artisnlte simulation that predicts the deeper \ion{Ti}{ii} absorption trough. Again, both simulations predict a similar amount of \ion{Ti}{ii} absorption around these wavelengths, with the difference instead driven by variations in IGE emission. In each case, there is substantial absorption from IGEs (primarily \ion{Co}{ii}, \ion{Co}{iii}, \ion{Fe}{ii} and \ion{Fe}{iii}) blueward of ${\sim}$3000\,\AA, with this radiation subsequently fluoresced to redder wavelengths. In the 1D simulation, there is substantially more emission from these iron group species around the deepest absorption of the \ion{Ti}{ii} trough, while more of this emission appears just redward of the \ion{Ti}{ii} trough in the 2D \artisnlte simulation. 

Both the 2D (equatorial direction) and 1D \artisnlte simulations predict a similar Ti ionisation state where \ion{Ti}{iv} is the dominant species in the inner regions (below ${\sim}$10\,000 km\,$\mathrm{s}^{-1}$) and \ion{Ti}{iii} is the dominant species in the rest of the ejecta, with the exception of the very outermost ejecta (above ${\sim}$27\,000 km\,$\mathrm{s}^{-1}$) at early times (until $\sim$15\,d) where \ion{Ti}{ii} is dominant. The strength of the \ion{Ti}{ii} absorption trough in the simulations is therefore not purely a diagnostic of the Ti ionisation state but can also be impacted by the contributions of IGEs in this spectral region. 

\subsection{Comparisons to observations}
\label{subsec:comparisons_with_observations}
In the following section, we compare our simulated spectra and light curves to the well-observed normal SN~2011fe. Our aim is not to present our 2D \artisnlte as an accurate description of SN~2011fe but rather to use this comparison to contextualise the impact of multi-dimensionality and NLTE effects on comparisons with observations i.e., where differences between simulations are small compared to offsets from observation, one might argue that the effects responsible for this differences are unlikely to be critical for the interpretation of data - on the other hand, where differences between calculations lead to substantial effects (on the scale of mismatches with data, or larger) it is clear evidence that the physics considered matters. To place our simulations in the context of the broader SNe~Ia population, we also compare to the B- and V-band width-luminosity distribution of normal SNe~Ia.

\begin{figure*}
	\includegraphics[width=0.9\linewidth,trim={0.0cm 0.0cm 0.0cm 0.0cm},clip]{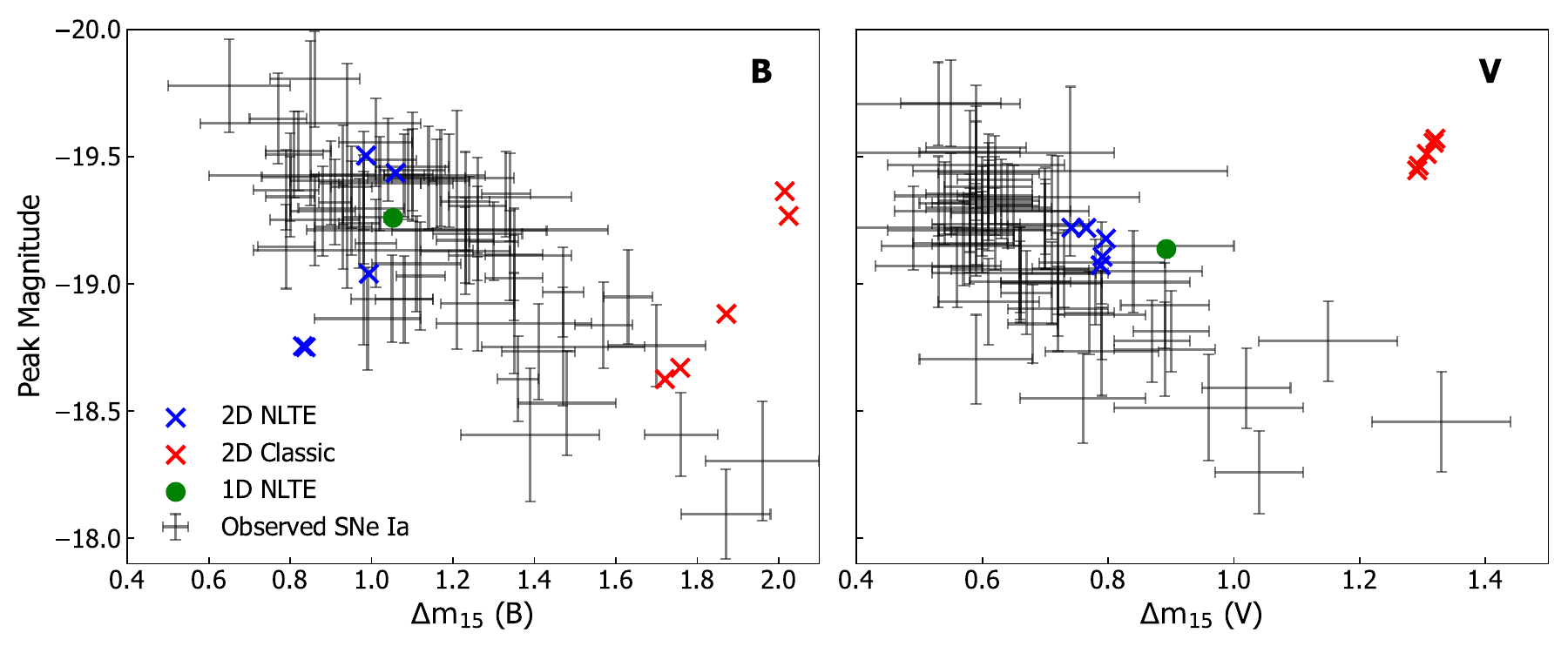}    
    \caption{Decline rate versus peak magnitude in B- and V-band for the five viewing angles of our 2D \artisnlte and \artiscl simulations considered in Figures~\ref{fig:band_lightcurves}, \ref{fig:spectra} and \ref{fig:bolUBV_viewing_angle_scatter}. The corresponding values for the 1D \artisnlte simulation and the sample of normal SNe~Ia from \citet{hicken2009a} are also plotted for reference.}
    \label{fig:B_V_deltam15_vs_peakmag_scatter}
\end{figure*}
\subsubsection{Light curves}
\label{subsubsec:Lightcurve_compared_observations}
The 2D \artisnlte simulation predicts much bluer light curves than the \artiscl simulation, particularly after peak, significantly improving agreement with SN~2011fe. This bluer post-peak evolution is consistent with the findings of previous 1D NLTE radiative transfer studies of both double detonation models \citep{collins2025a, boos2025a} and other SNe Ia explosion models (e.g.\,\citealt{shen2021a, blondin2022a}).

First focussing on the blue (U- and B-bands), the 2D \artisnlte and \artiscl simulations have similar rise times although the \artisnlte calculation rises slightly more rapidly (e.g. $\sim$2 d faster in both U- and B-bands). The rise of both models is similar to SN~2011fe and consistent with the normal SNe Ia sample of \cite{Liu2026a}. However, while the 2D \artisnlte light curves also provide a reasonable match to the post-peak decline of SN~2011fe, the \artiscl simulation predicts light curves that systematically decline much too rapidly for all viewing angles. Focussing on the the 90$^\circ$ viewing angle in our 2D \artisnlte simulation, which provides the best photometric agreement with SN~2011fe, we find a good match to the B-band light curve, although it declines too slowly after $\sim$35\,d. In the U-band however the simulation is slightly too faint and rises too quickly relative to SN~2011fe. Therefore, despite the systematic improvement over our 2D \artiscl simulation, no individual line-of-sight in our 2D \artisnlte simulation simultaneously matches both the U- and B-band light curve evolution of SN~2011fe. There are also clear differences between the equatorial direction in our 2D \artisnlte simulation (90$^\circ$ viewing angle) and our 1D \artisnlte simulation. The 1D \artisnlte simulation predicts U- and B-band light curves that are brighter at peak than the 2D \artisnlte simulation, improving agreement with SN~2011fe in U-band, but leading to poorer agreement in B-band. These differences are relevant on the scales of comparisons with SN~2011fe.

For all viewing angles in our 2D \artisnlte simulation, the V-band agreement with SN~2011fe is excellent, with the R-band light curves also showing reasonable agreement until $\sim$30\,d, after which they decline too quickly. In contrast, all viewing angles in our \artiscl simulation are much too bright in V-, R- and I-bands from $\sim$10\,d onwards.

One of the most striking differences between the simulations and SN~2011fe appears in the NIR bands. SN~2011fe shows a clear secondary maximum in I-band from $\sim$30\,d onwards (see Figure~\ref{fig:band_lightcurves}). This is not reproduced by any viewing angle in the 2D \artisnlte simulation, pointing to ejecta which are too highly ionised relative to normal SNe~Ia after peak. In contrast, the \artiscl simulation does predict a secondary I-band maximum, but it is both too bright and appears too early compared to SN~2011fe. The \artiscl simulation is therefore significantly under-ionised relative to normal SNe~Ia, particularly after peak, resulting in colours that are too red (see Section \ref{subsec:2D_lightcurve_comparisons} for details). This is further highlighted by the post-peak colour evolution of the \artiscl simulation (see Figure~\ref{fig:colour_evolution}). These results are consistent with the findings from the \artiscl simulation of the same model carried out by \cite{collins2022a} using a different atomic data set. 

While the higher ionisation state, and thus bluer colours of the 2D \artisnlte simulation improves the photometric agreement with normal SNe~Ia relative to the \artiscl simulation, the \artisnlte simulation is somewhat over-ionised. Nebular-phase radiative transfer simulations also predict over-ionisation relative to normal SNe~Ia for a variety of different explosion models in both 1D \citep{shingles2022a, blondin2023a} and 3D \citep{Pollin2026b, Pollin2026a}. Multi-dimensional NLTE radiative transfer simulations in the photospheric phase for different explosion models will be key to assessing if over ionisation is a generic prediction of realistic explosion models at all phases. If so, it may suggest that we are not accounting for a key process in our simulations that would act to reduce the ionisation (e.g. clumping, which enhances the local density of the ejecta leading to increased recombination rates; \citealt{wilk2018a, wilk2020a, mazzali2020a}).   

\subsubsection{Spectra}
\label{subsubsec:Spectra_compared_observations}
The southern viewing angles ($135^\circ$, $180^\circ$) of the 2D \artisnlte simulations provide a promising spectral match with SN~2011fe. For these viewing angles, the 2D \artisnlte  spectra show no strong line blanketing or prominent \ion{Ti}{ii} absorption (with the exception of the weak \ion{Ti}{ii} absorption trough predicted by the $135^\circ$ viewing angle after peak). The spectra also reproduce the strength of key IME and O features well at all epochs and provide an excellent match to the SED of SN~2011fe before peak (e.g. 9.5\,d in Figure~\ref{fig:spectra}). Some discrepancies, however, remain: there is a deficit of flux redward of $\sim$6000\,\AA\ after peak due to the absence of a secondary NIR maximum (see Sections \ref{subsec:2D_lightcurve_comparisons} and \ref{subsubsec:Lightcurve_compared_observations}) and the velocities of IME and O features remain too high for these viewing angles. In contrast to the promising agreement of the 2D \artisnlte simulation for these viewing angles, the 2D \artiscl simulation predicts relatively poor agreement with SN~2011fe, particularly due to excessive absorption blueward of $\sim$4000\,\AA\ after peak. The subsequent fluorescence, along with the overly prominent secondary NIR maximum predicted, produces an SED that is far too red compared to SN~2011fe. Strong absorption from \ion{Ti}{ii} and singly ionised IGEs and the disappearance of key IME and O features after peak further worsen the agreement with SN~2011fe. The choice of plasma conditions, therefore, directly impacts our conclusions when comparing to observations.
 
For the $0^\circ$--$90^\circ$ viewing angles both 2D simulations predict excessive line blanketing blueward of $\sim4000$\,\AA, strong \ion{Ti}{ii} absorption troughs, and spectral features from IMEs and O that are too high-velocity compared to SN~2011fe (e.g. \ion{Ca}{ii} NIR triplet, \ion{Si}{ii} 5972\,\AA\ and 6355\,\AA\ features, \ion{S}{ii} `W' at $\sim$5500\,\AA, and \ion{O}{i} 7774\,\AA\ feature). The \ion{Ca}{ii} NIR triplet, in particular, is much too strong and high-velocity before peak (9.5\,d) compared to SN~2011fe in both cases. However, relative to the \artiscl simulation, the \artisnlte simulation shows improved agreement with SN~2011fe for these viewing angles. In particular, the \artisnlte simulation predicts weaker line blanketing, a bluer SED and clear \ion{Si}{ii} 5972, \ion{Si}{ii} 6355\,\AA\ and \ion{O}{i} 7774\,\AA\  features at all epochs that disappear after peak in the \artiscl simulation (see Section \ref{subsec:2D_spectra_comparisons} for details).

Comparing the $90^\circ$ viewing angle in the 2D \artisnlte simulation and the 1D \artisnlte simulation to the spectral evolution of SN~2011fe, neither provides a good match. The spectra predicted by the simulations do, however, show clear differences in key spectral features that are relevant on the scale of comparisons with SN~2011fe. As discussed in Section \ref{subsec:1D_2D_NLTE_spectra_comparisons} the \ion{Ca}{ii} NIR triplet shows striking differences from peak onward. Most notable for comparisons with SN~2011fe is the post-peak behaviour: the 2D \artisnlte simulation predicts a NIR triplet dominated by a high-velocity absorption component with only a weak low-velocity contribution, whereas the 1D \artisnlte simulation predicts the opposite. As a result, the 1D model provides significantly better agreement with the \ion{Ca}{ii} NIR triplet observed in SN~2011fe after peak. The origin of these differences in the NIR triplet between simulations is discussed in detail in Appendix \ref{appendix:Ca_II_photoionisation}. The spectra also show clear differences in absorption at blue wavelengths. Before peak, the 2D \artisnlte simulation predicts weaker line blanketing and a less pronounced \ion{Ti}{ii} absorption trough compared to the 1D simulation, leading to improved agreement with SN~2011fe. However, from peak onward it is instead the 1D \artisnlte simulation that shows reduced line blanketing and a weaker \ion{Ti}{ii} feature, leading to improved agreement with SN~2011fe, particularly after peak at bluer wavelengths.

\subsubsection{Width-luminosity relation}
\label{subsubsec:viewing_relative_observations}
Figure~\ref{fig:B_V_deltam15_vs_peakmag_scatter} shows the B- and V-band width luminosity relation predicted by our 2D \artisnlte and \artiscl simulations, together with the values from the 1D \artisnlte simulation and a sample of normal SNe Ia for comparison. The slower decline rates predicted by the 2D \artisnlte simulation in B- and V-band, as well as the bluer light curve colours at peak, lead to significantly better agreement with the observed distribution relative to the 2D \artiscl simulation. The NLTE treatment therefore also has a significant impact on derived light curve quantities such as peak magnitude and decline rate. These differences are relevant when evaluating the consistency of explosion models relative to population wide correlations such as the width-luminosity relation. This is consistent with the findings of \cite{shen2021a} for 1D NLTE simulations of pure detonation explosion models.

Both the 2D \artisnlte and \artiscl simulations do, however, show a similar absolute viewing angle variation in their decline rate versus peak magnitude in both B- and V-band (see Figure~\ref{fig:B_V_deltam15_vs_peakmag_scatter}). The viewing angle variation predicted by both simulations is anti-correlated with the width-luminosity relation in B-band (i.e. brighter viewing angles tend to have more rapidly declining light curves). However, for both simulations the viewing angle variation is consistent with the width of the observed distribution. 

While derived photometric quantities provide a means to evaluate the success of explosion models relative to observations, we emphasise the importance of considering the spectroscopic evolution alongside such quantities. For example, the $90^\circ$ viewing angle in the 2D \artisnlte simulation falls within the width-luminosity relation in B-band while the two faintest viewing angles ($0^\circ$ and $45^\circ$) fall just below. However, as discussed in Section \ref{subsubsec:Spectra_compared_observations}, the peak spectra of these viewing angles show strong absorption from singly ionised IGEs and \ion{Ti}{ii} at blue wavelengths inconsistent with the spectra of normal SNe~Ia. 

\subsection{Implications for the explosion model}
\label{subsec:explosion_model_implications}
As discussed in Section~\ref{subsec:comparisons_with_observations}, the 2D \artisnlte simulation shows improved agreement with normal SNe Ia for all viewing angles relative to the 2D \artiscl simulation. Despite this, our results suggest that the 0.02\,\msun\ He-shell mass of the model investigated here is too large for a double detonation model to resemble normal SNe~Ia over the majority of viewing angles. This is a direct consequence of the ejecta structure predicted by the explosion model, which produces excessive He-shell burning products in the outer ejecta for the northern and equatorial viewing angles. This leads to extremely strong and high-velocity \ion{Ca}{ii} NIR triplets, excessive absorption from singly ionised IGEs and \ion{Ti}{ii} at blue wavelengths, and IME and O features that appear at velocities that are significantly too high (see Section \ref{subsubsec:Spectra_compared_observations} for details). We note that an increase in ionisation state could act to suppress the impact of these burning products. However, as the inner regions of the 2D \artisnlte simulation are already over-ionised relative to normal SNe~Ia (see Section~\ref{subsec:2D_lightcurve_comparisons}), this would require a mechanism that only increases the ionisation in the outer ejecta layers. 

The M10\_02 model is from the \cite{gronow2021a} sequence of accretion-induced double detonations that are assumed to ignite due to stable mass transfer from a He-rich companion. M10\_02 is the lowest He-shell mass model in the sequence, yet the northern and equatorial viewing angles do not resemble normal SNe~Ia. 
The specific 1\,\msun\ CO core and 0.02\,\msun\ He-shell configuration investigated here is therefore unlikely to represent a common SN~Ia progenitor channel. The improved agreement for the southern viewing angles, which are less impacted by He-shell burning products, instead indicates that lower He-shell masses are required to reproduce normal SNe~Ia for the majority of lines of sight. Dynamically driven double degenerate double detonation models \citep{guillochon2010a, pakmor2013a, pakmor2022a, boos2024a} and violent mergers \citep{pakmor2010a, pakmor2011a, pakmor2012a, sato2015a, sato2016a, pakmor2026a, Pollin2026b} represent a promising avenue to produce such lower He-shell mass explosions \citep{shen2024a}. We note that the prominent high velocity \ion{Ca}{ii} NIR triplet and strong line blanketing at blue wavelengths predicted for the northern and equatorial viewing angles are more characteristic of SNe Ia previously proposed to arise from double-detonations involving thick ($\gtrsim$0.1\,\msun) He-shells (e.g. SN~2018byg \citealt{de2019a} and SN~2020jgb \citealt{liu2023b}), a scenario thought to be intrinsically rare \citep{liu2023b}. 

\section{Summary and Conclusions}
\label{sec:conclusions}
We have presented synthetic observables from a 2D NLTE photospheric-phase radiative transfer simulation of the M10\_02 double detonation explosion model of \cite{gronow2021a}, consisting of a 1\,\msun\ CO core and 0.02\,\msun\ He-shell. In agreement with previous work, our results show that both multi-dimensionality (e.g. \citealt{kromer2010a, shen2021b, collins2022a}) and NLTE effects (e.g. \citealt{dessart2014b, shen2021a, collins2025a, boos2025a}) are already important in the photospheric phase: both have effects on scales relevant to comparisons with observations before peak, with their impact increasing over time. This first multi-dimensional \artisnlte simulation of the photospheric phase for any explosion model demonstrates both the feasibility and importance of such simulations. 

We compared to an \artiscl simulation of the same 2D ejecta model that adopted an approximate treatment for NLTE effects. The full NLTE treatment led to a bluer SED, slower optical light curve declines, changes to the NIR light curve evolution, weaker absorption from singly ionised IGEs, systematically weaker \ion{Ti}{ii} absorption troughs and substantial differences in a number of other key spectral features. These differences result in the 2D \artisnlte simulation showing better agreement with normal SNe~Ia for all viewing angles, primarily due to its higher ionisation. This difference in ionisation is most apparent in the behaviour of the NIR light curves. The \artiscl simulation predicts a secondary NIR maximum that is significantly stronger and earlier than observed, indicating the ionisation state decreases too rapidly. In contrast, no viewing angle in the 2D \artisnlte simulation produces a secondary NIR maximum, indicating the ionisation state does not decrease sufficiently after $\sim$30\,d for a secondary maximum to form. This finding is consistent with previous NLTE radiative transfer simulations of the nebular phase which predict the ejecta are over-ionised at late times \citep{shingles2022a, blondin2023a, Pollin2026b, Pollin2026a}. 

Our 2D \artisnlte and \artiscl simulations have confirmed that differences in the NIR maximum predicted can provide useful constraints on the underlying ionisation state (see e.g. \citealt{kasen2006a}). Beyond the secondary NIR maximum, which is now routinely observed by photometric surveys, recent observational advances are also providing a wealth of high quality infrared spectroscopic data that can be compared with our radiative transfer simulations. The SOXS instrument on the ESO NTT \citep{schipani2018a} is enabling rapid NIR spectroscopic follow-up, while photospheric-phase NIR-to-MIR observations from the \textit{James Webb Space Telescope} \citep{gardner2006a} are now becoming available (e.g. \citealt{kwok2025a}). The infrared spectroscopic signatures predicted by our 2D \artisnlte simulation, including the NIR He signatures predicted by previous 1D \artisnlte double detonation simulations \citep{collins2023a,callan2025a}, will be investigated in future work. 

The 2D \artisnlte simulation predicted substantial viewing angle variation in both light curves and spectra, particularly at blue wavelengths, with peak brightness variations of 1.5 and 0.7 mag in the U- and B-bands, respectively. The magnitude of viewing angle variations is not substantially impacted by NLTE effects up to peak. However, after peak, the \artisnlte simulation retains much more significant spectroscopic variation with viewing angle than the 2D \artiscl simulation. NLTE effects, therefore, also have an impact on the scale of viewing angle variations, particularly after peak. 

We compared the equatorial viewing angle of our 2D \artisnlte simulation to a 1D \artisnlte simulation constructed based on that direction. While the 1D simulation is generally representative of the 2D viewing angle, there are clear differences: the 1D model predicts brighter U- and B-band light curves, changes in the line blanketing at blue wavelengths, a bluer SED post peak and substantial variations in the evolution of the \ion{Ca}{ii} NIR triplet and \ion{Ti}{ii} absorption trough. These differences are mainly associated with the bluer radiation field of the 1D model, which has a higher bolometric luminosity than the 2D equatorial viewing angle, despite having a slightly smaller \nickel\ mass than the 2D model (0.51 compared to 0.55\,\msun). Although this modest increase in the luminosity of the 1D model has little impact on the overall ionisation state compared to the 2D model, it has a significant impact on the observables predicted. As a result, the 1D model cannot fully reproduce the behaviour of the corresponding line-of-sight in the multi-dimensional simulation.

The evolution of the \ion{Ca}{ii} NIR triplet differs between the 1D and 2D \artisnlte simulations due to variations in the excited state \ion{Ca}{ii} photoionisation rate at high velocities. This highlights the importance of a detailed treatment of not only ground state but also excited state photoionisation, as employed by \artisnlte. For the \ion{Ti}{ii} absorption trough, the differences are primarily due to varying contributions of other species around the absorption trough (\ion{Ca}{ii} H\&K lines before peak; singly and doubly ionised IGEs from peak onwards). This demonstrates that the strength of the trough may not simply be a diagnostic of \ion{Ti}{ii} ionisation.

For the 2D \artiscl simulation presented here and previous multi-dimensional \artiscl double detonation simulations (see \citealt{kromer2010a, collins2022a}), strong line blanketing at blue wavelengths from \ion{Ti}{ii} and singly ionised IGEs drives much of the differences with normal SNe~Ia. The reduced line blanketing predicted by the 2D \artisnlte simulation for all viewing angles, therefore, results in improved agreement with normal SNe Ia. However, for the northern and equatorial directions of the 2D \artisnlte simulation the line blanketing is still excessive. An increase in ionisation would decrease the fraction of IGEs and Ti in the singly ionised state in the outer ejecta, potentially reducing the impact of these He-shell burning products. However, the inner ejecta regions are already over-ionised relative to normal SNe~Ia, so this would require a process that only increases the ionisation of the outer ejecta. The strong line blanketing is therefore most likely a direct consequence of the ejecta structure of the explosion model, suggesting that He-shell masses of $\sim$0.02\,\msun\ are too large for double detonation models to resemble normal SNe~Ia over most viewing angles. The southern viewing angles of our 2D \artisnlte simulation, which are less impacted by He-shell burning products, do, however, show promising spectroscopic agreement with normal SNe Ia. This demonstrates that while some lines of sight in such models can resemble normal SNe~Ia, lower He-shell masses are required for this agreement to extend across the majority of viewing angles. 

The simulations presented here demonstrate the importance of simultaneously treating multi-dimensionality and NLTE effects in photospheric-phase radiative transfer. While we have focused on a 2D double detonation model for this initial study, multi-dimensionality is inherent to a variety of hydrodynamic explosion simulations of SNe Ia \citep{pakmor2024a}. Our findings, therefore, motivate further multi-dimensional NLTE radiative transfer simulations of the photospheric phase to determine the impact of this simultaneous treatment across a range of explosion models.

\section*{Acknowledgements}
FPC and SAS acknowledge funding from STFC grant ST/X00094X/1. This work used the DiRAC Memory Intensive service (Cosma8) at Durham University, managed by the Institute for Computational Cosmology on behalf of the STFC DiRAC HPC Facility (www.dirac.ac.uk). The DiRAC service at Durham was funded by BEIS, UKRI and STFC capital funding, Durham University and STFC operations grants. DiRAC is part of the UKRI Digital Research Infrastructure. The authors gratefully acknowledge the Gauss Centre for Supercomputing e.V. (www.gauss-centre.eu) for funding this project by providing computing time on the GCS Supercomputer JUWELS at J\"ulich Supercomputing Centre (JSC). LJS acknowledges funding by the European Union (ERC, HEAVYMETAL, 101071865).
The work of FKR and AH is supported by the Klaus Tschira Foundation and 
by the Deutsche Forschungsgemeinschaft (DFG, German Research
Foundation) -- RO 3676/7-1, project number 537700965.
FKR acknowledges funding by the European Union (ERC, ExCEED, project number 101096243). Views and opinions expressed are however those of the authors only and do not necessarily reflect those of the European Union or the European Research Council Executive Agency. Neither the European Union nor the granting authority can be held responsible for them.
AH is a fellow of the International Max Planck Research School for Astronomy and Cosmic Physics at the University of Heidelberg (IMPRS-HD) and acknowledges financial support from IMPRS-HD.
CEC is funded by the European Union’s Horizon Europe
research and innovation programme under the Marie Skłodowska-Curie grant
agreement No.~101152610.
NumPy \citep{harris2020a}, SciPy \citep{virtanen2020a}, Matplotlib \citep{hunter2007a}  and \href{https://zenodo.org/records/8302355} {\textsc{artistools}}\footnote{\href{https://github.com/artis-mcrt/artistools/}{https://github.com/artis-mcrt/artistools/}} \citep{artistools2025a} were used for data processing and plotting.

\section*{Data Availability} 
The spectra and light curves presented here will be made available on the Heidelberg supernova model archive HESMA\footnote{\href{https://hesma.h-its.org}{https://hesma.h-its.org}}~\citep{kromer2017a}.

\bibliographystyle{mnras}
\bibliography{references} 


\appendix
\section{Adaptive NLTE solver}
\label{appendix:NLTE_solution_developments}

The NLTE solver implemented in \artis\ by \cite{shingles2020a} required that for each element all ionisation states of that element present in the simulation were included in the solution. In certain cases, this approach can result in the NLTE solver calculating unphysical level population values due to numerical issues. In extreme cases, the handling of this resulted in partition function values that overflowed the float32 range, causing the simulation to fail.

These numerical issues are most commonly associated with ions that have very low populations and therefore tend to occur in cells where the ionisation state is dominated by either the highest or lowest ionisation stages included in the solution (i.e., when highly ionised states dominate the elemental population, the lowest ionisation state often has a very low population and vice versa). Neither our previous 3D nebular-phase \artisnlte simulations nor our early-phase 1D \artisnlte simulations were commonly impacted by these numerical issues. However, our early-time multi-dimensional \artisnlte simulations occasionally encountered such numerical issues with the NLTE solution ($\sim 1.5$\% of the time). This is a result of the larger dynamic range of conditions in the ejecta of multi-dimensional NLTE simulations at early phases. Relative to the nebular phase, the ejecta conditions at early phases evolve more rapidly, a wider velocity range of ejecta material is relevant to the spectral formation and a significant number of ionisation stages of key species are relevant (we include five ionisation stages for the IGEs for our 2D \artisnlte simulation presented here). For 1D simulations the more smoothly varying ejecta structure relative to multi-dimensional simulations means that even at early times it is much less likely for cells to be in the more extreme ionisation conditions that lead to these numerical problems.

\begin{figure}
	\includegraphics[width=1.0\linewidth,trim={0.0cm 0.0cm 0.0cm 0.0cm},clip]{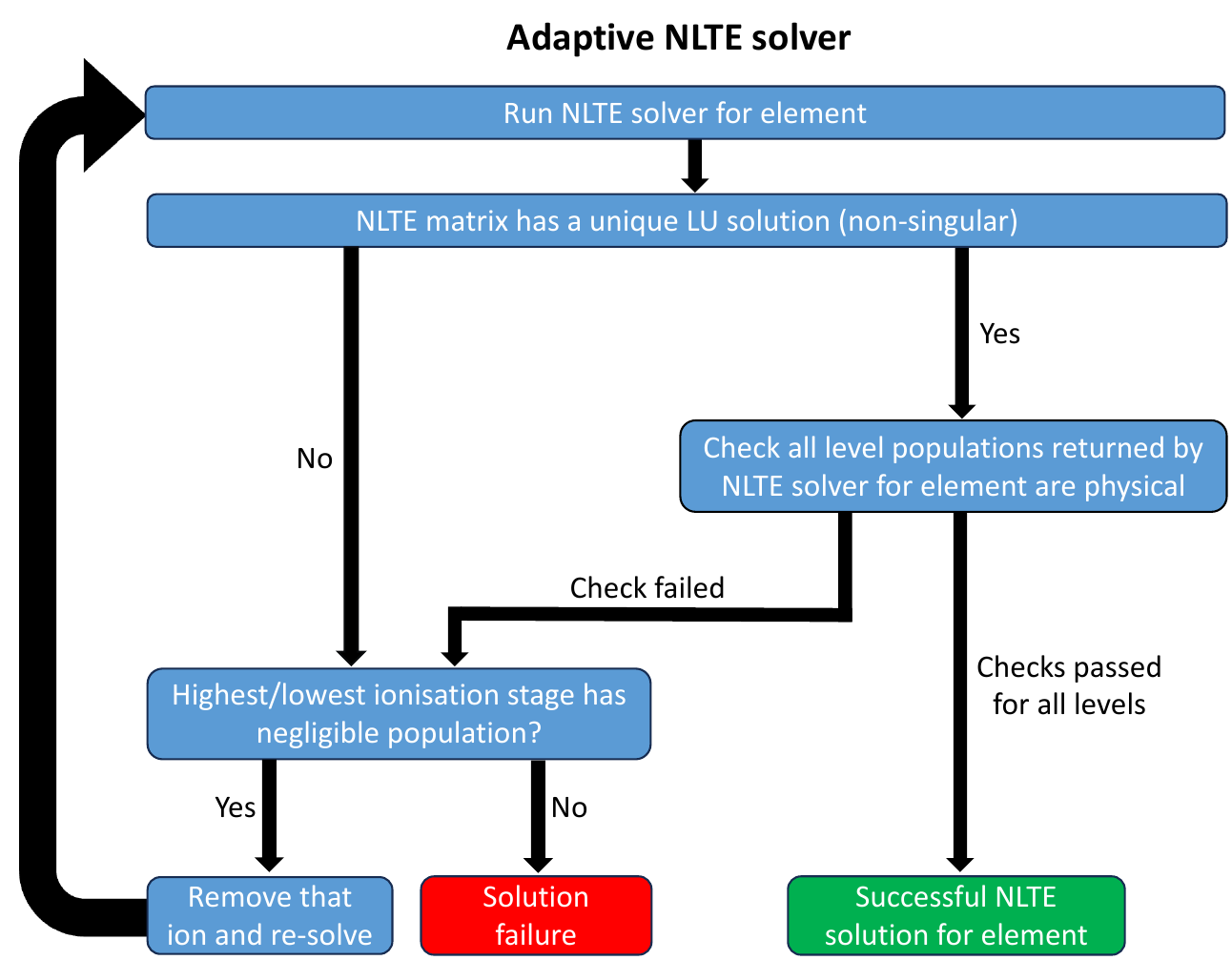} 
    \caption{Flowchart summarising the operation of the updated NLTE solver in \textsc{artis}.}
    \label{fig:adaptive_NLTE_solver_flow_chart}
\end{figure}

We have modified the NLTE solver to overcome these numerical issues, enabling the first early-phase multi-dimensional \artisnlte simulation presented here. Specifically, we implemented an adaptive solution (visualised by the flowchart in Figure~\ref{fig:adaptive_NLTE_solver_flow_chart}) that allows the range of ions included in the NLTE solution to be dynamically reduced by removing ions with negligible populations that are responsible for unphysical level populations in the NLTE solution. These updates were first introduced in the v2025.08.01 release of \artis \citep{artis_code_version_v2025.08.01}.

In our new implementation, if the NLTE matrix has no unique solution or the solver gives unphysical level populations, we first check the population of the top ion. If it has a negligible population we then re-solve the NLTE matrix with this ion removed to attempt to obtain a successful solution for the element. If the top ion has a population that is too high to be safely removed from the solution, we instead check the population of the bottom ion and follow the same procedure. Ion stages are iteratively stripped following this procedure until a successful NLTE solution is calculated for the element i.e. one that has populations that are physical for all levels. In the case that both the highest and lowest ionisation stages have non-negligible populations, this method cannot be used to recover a successful solution. 

These updates ensure the numerical stability of multi-dimensional NLTE simulations at early phases. We have verified that the changes described here do not impact the synthetic spectra predicted for 1D early-time and 1D nebular-phase \artisnlte simulations. The adaptive NLTE solver has also been successfully used to ensure the numerical stability of 3D nebular-phase \artisnlte simulations \citep{Pollin2026b,Pollin2026c}.

\section{Computational resource requirements}
\label{appendix:computational_resources}
The run time of our early-phase simulations is dominated by the propagation of the Monte Carlo packets due to the large number of interactions that each packet undergoes. Our 2D \artisnlte simulation required 1.7 million CPU hours, our 2D \artiscl simulation required 570\,k CPU hours and our 1D \artisnlte simulation (which used fewer packets than the 2D simulations) required 300\,k CPU hours to complete. The 1D and 2D \artisnlte simulations were performed on 3072 CPU cores while the 2D \artiscl simulations used 1536 cores. All simulations used 2 × 64-core AMD EPYC 7H12 processors (128 cores per node; 1 TB RAM per node).

\section{\ion{Ca}{ii} Optical Depth Comparisons}
\label{appendix:Ca_II_Sobolev_depth_plots}
\begin{figure*}
    \centering
    
    \includegraphics[width=0.49\textwidth, trim={0cm 0cm 0cm 0cm}, clip]{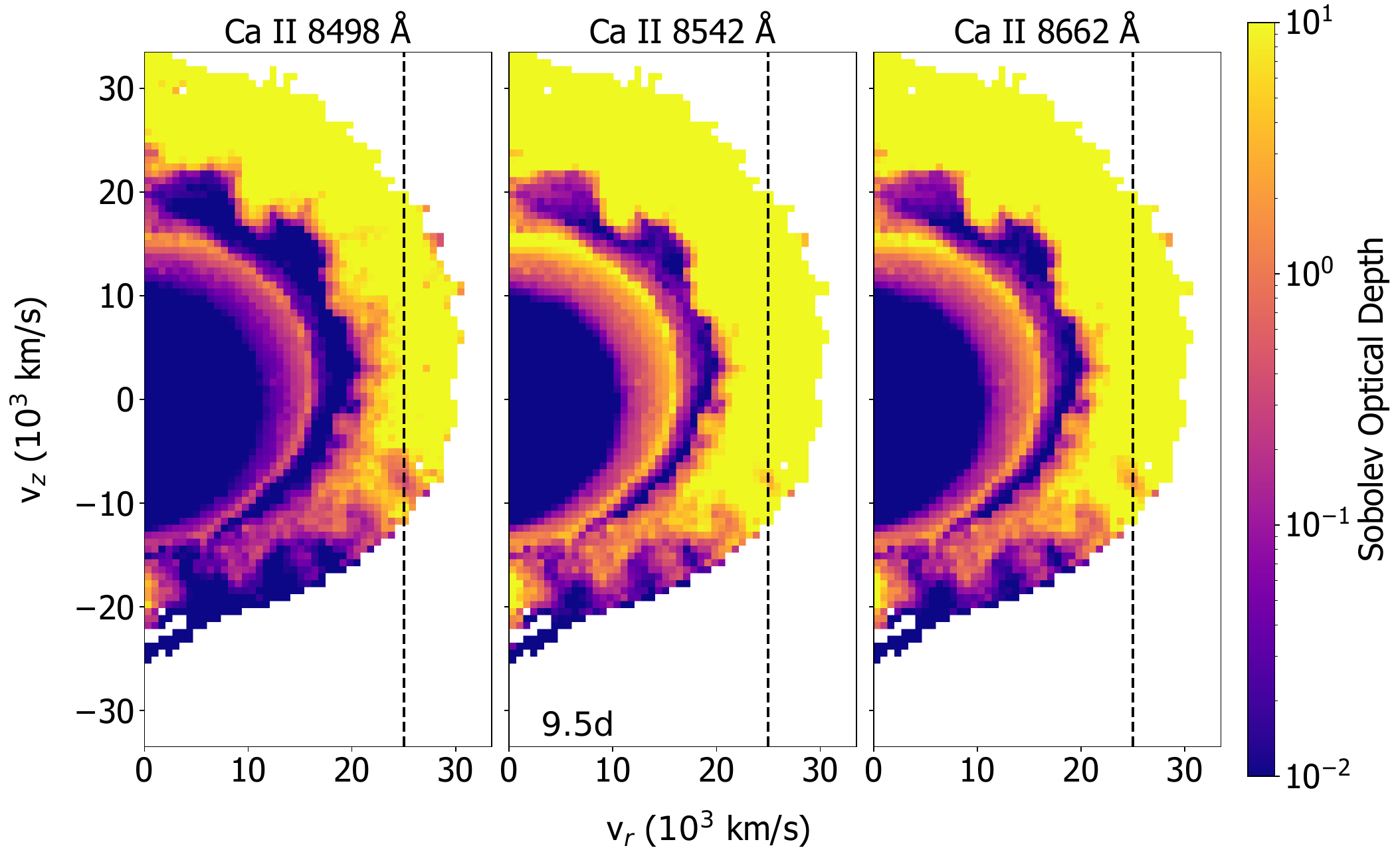}
    \includegraphics[width=0.49\textwidth, trim={0cm 0cm 0cm 0cm}, clip]{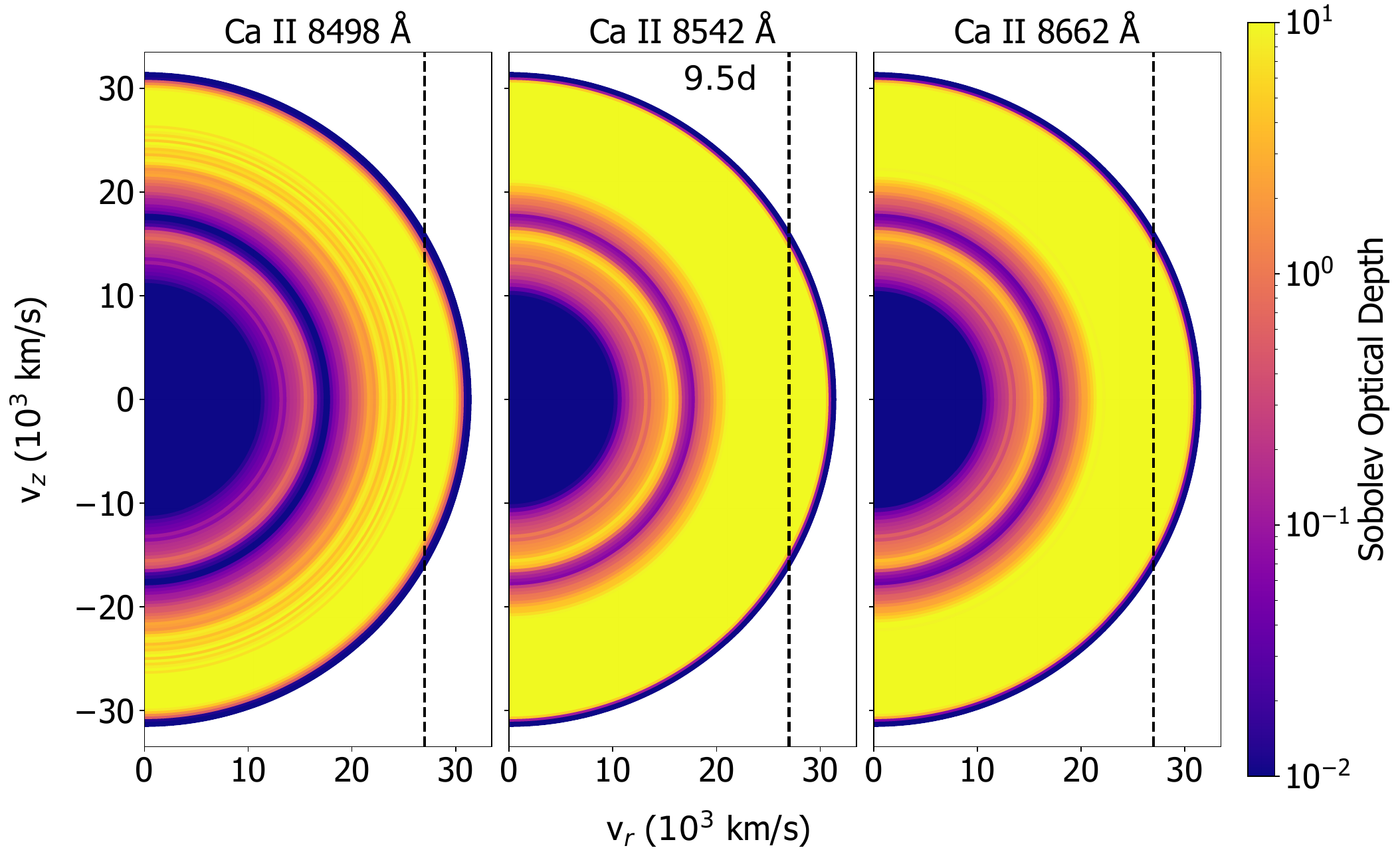}
    
    \vspace{0.0em}
    
    \includegraphics[width=0.49\textwidth, trim={0cm 0cm 0cm 0cm}, clip]{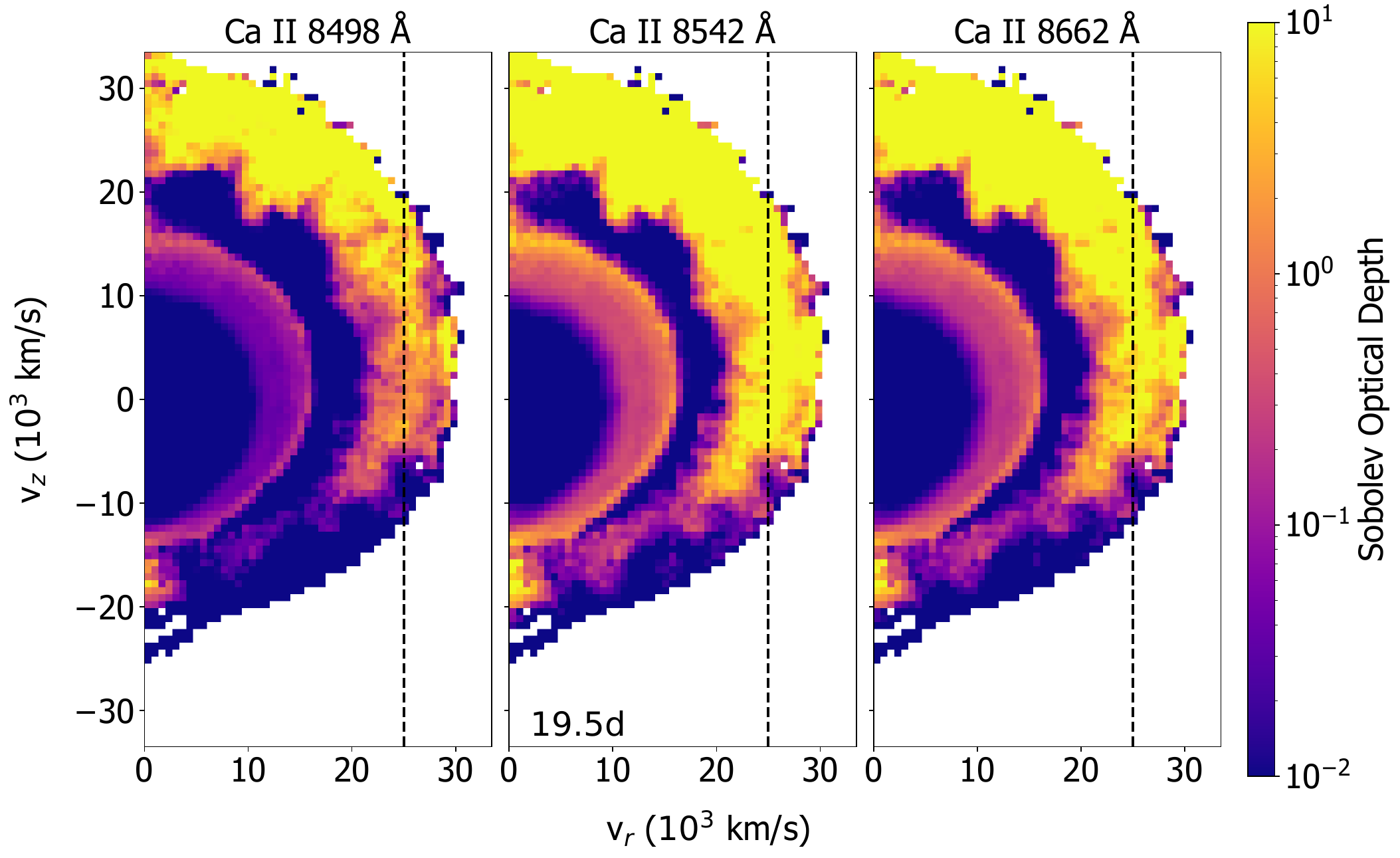}
    \includegraphics[width=0.49\textwidth, trim={0cm 0cm 0cm 0cm}, clip]{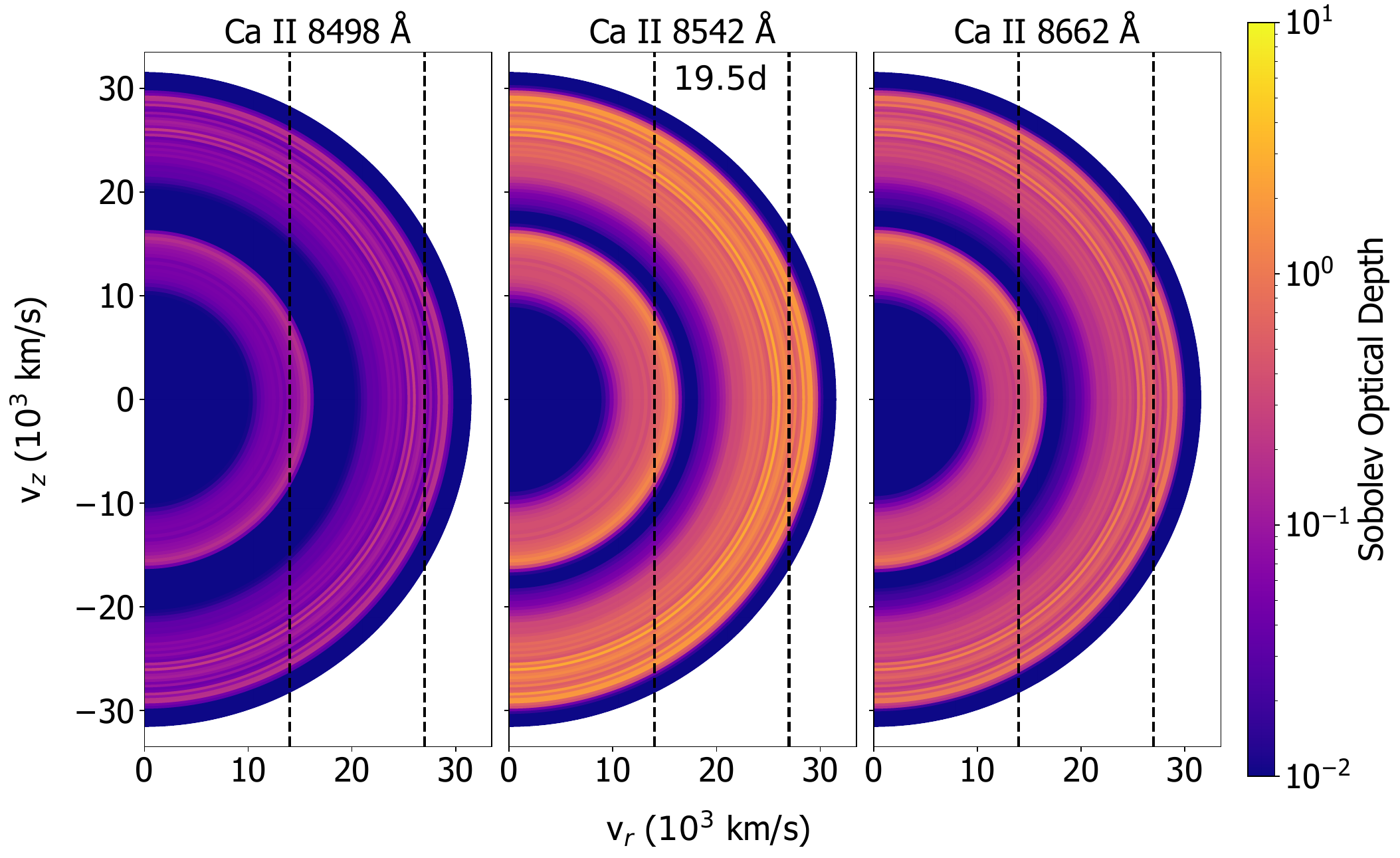}
    
    \vspace{0.0em}
    
    \includegraphics[width=0.49\textwidth, trim={0cm 0cm 0cm 0cm}, clip]{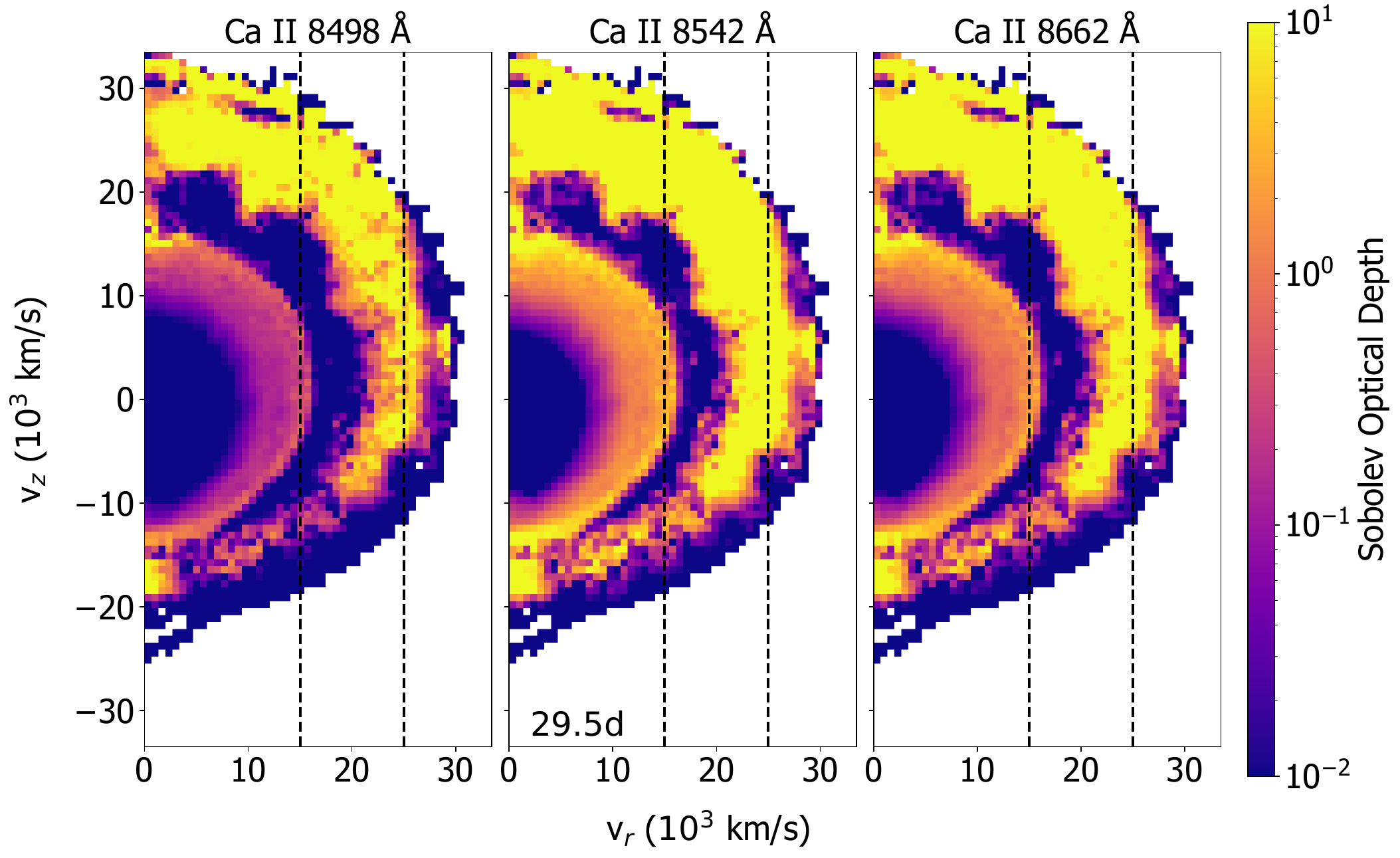}
    \includegraphics[width=0.49\textwidth, trim={0cm 0cm 0cm 0cm}, clip]{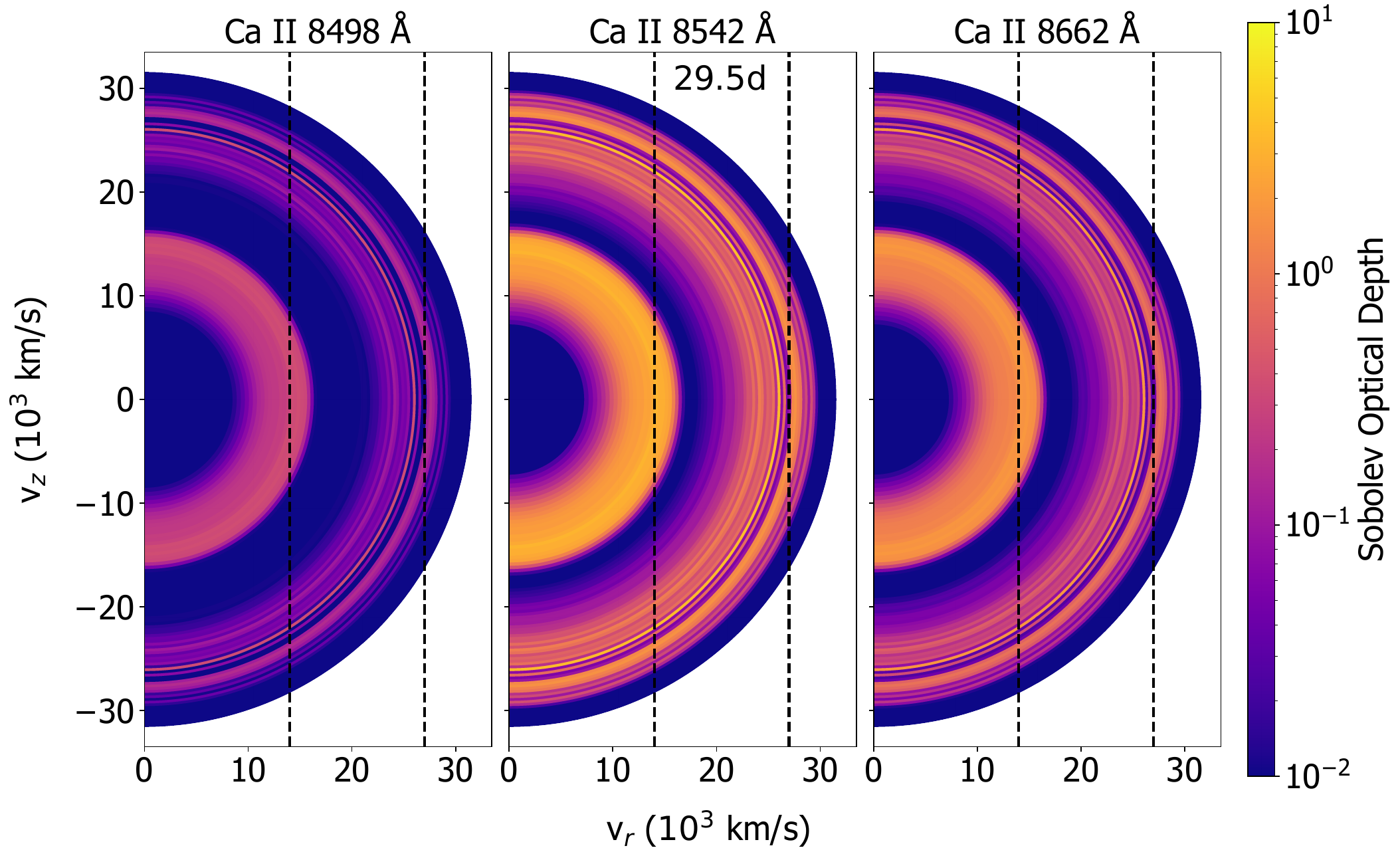}
    
    \caption{Sobolev optical depths for the spectral lines in the \ion{Ca}{ii} NIR triplet at 9.5, 19.5 and 29.5\,d after explosion (top to bottom) for our 2D (left three columns) and 1D (right three columns) \artisnlte simulations. The velocities of the high- and low-velocity components of the NIR triplet in the 1D simulation and the equatorial direction of the 2D simulation are indicated by dashed lines; where no low-velocity absorption feature is present, its velocity is not marked.}
    \label{fig:Ca_II_Sobolev_depth_maps}
\end{figure*}

In Figure~\ref{fig:Ca_II_Sobolev_depth_maps} we show the Sobolev optical depths for the lines in the \ion{Ca}{ii} NIR triplet before, around and after peak for our 2D and 1D \artisnlte simulations. 

\section{Importance of excited-state photoionisation for \ion{Ca}{ii} population and NIR triplet feature in NLTE simulations}
\label{appendix:Ca_II_photoionisation}
Both the 2D and 1D \artisnlte simulations predict that \ion{Ca}{iii} is generally the dominant Ca ion, with the exception of velocities below ${\sim}$10\,000 km\,$\mathrm{s}^{-1}$ before peak, where \ion{Ca}{iv} or \ion{Ca}{v} instead dominate. In both simulations \ion{Ca}{ii} remains a substantially sub-dominant species of Ca throughout. Despite this, the evolution of the \ion{Ca}{ii} population, particularly at high velocities, differs markedly and drives the variations in the \ion{Ca}{ii} NIR triplet features discussed in Section \ref{subsec:1D_2D_NLTE_spectra_comparisons}. 

Both simulations agree that photoionisation is the dominant ionisation process for \ion{Ca}{ii}, with photoionisation from excited levels dominating over the ground state photoionisation rate for both low and high-velocity \ion{Ca}{ii}. In both cases the majority of the \ion{Ca}{ii} population is contained within the ground state and the first two excited levels (the metastable D-states). The populations of the third and fourth excited levels (P-states) are typically ${\sim}$10-100 times lower with the remaining excited states a further ${\sim}$100 times (or more) less populated. The photoionisation edges of the ground and first four excited levels occur at 1044, 1218, 1219, 1417 and 1422\,\AA\ respectively. The simulations predict that the radiation field is very suppressed at these short wavelengths due to significant absorption from both Fe-group (primarily \ion{Co}{iii}, \ion{Co}{iv} and \ion{Fe}{iii}) and intermediate mass (primarily \ion{S}{iii}, \ion{S}{ii}, \ion{Si}{iii}) elements, resulting in a low ground state photoionisation rate. Moving to redder wavelengths the strength of the radiation field increases sharply in both simulations, dramatically increasing the number of photons available to photoionise these more highly excited levels. For many cases, this effect outweighs the impact of the lower populations of the excited levels: excited states with populations more than a factor of 10$^{4}$ lower than the ground state can have higher photoionisation rates than the ground state. 

In the high-velocity Ca ejecta regions of both the 2D (equatorial direction) and 1D \artisnlte simulations, the radiation field is strongly suppressed blueward of 2000\,\AA\ at early times, resulting in a \ion{Ca}{ii} photoionisation rate dominated by highly excited levels (those above the ground, D- and P-states). However, from ${\sim}$15\,d onwards, while photoionisation from these highly excited states continues to dominate in the equatorial direction of the 2D \artisnlte simulation, photoionisation from the D-states becomes dominant in the 1D \artisnlte simulation, resulting in a significantly higher \ion{Ca}{ii} photoionisation rate. This behaviour is driven by the bluer radiation field predicted by the 1D \artisnlte simulation at high velocities at these times, which increases the number of photons available to photoionise the D-states. This is related to the greater bolometric luminosity of the 1D simulation from $\sim$10\,d onwards, with this extra radiation primarily emitted at blue wavelengths: the U- and B-band light curves predicted by the 1D \artisnlte simulation are significantly brighter than those of the equatorial direction of the 2D \artisnlte simulation. We note that both simulations predict a similar \ion{Ca}{ii} photoionisation rate at low velocities throughout, with the D-states dominating the photoionisation rate in both cases.

The greater \ion{Ca}{ii} photoionisation rate at high velocities for the 1D simulation leads to a much more substantial reduction in the \ion{Ca}{ii} population over time compared to the 2D simulation (see Figure~\ref{fig:CaII_ionisation_2D_maps}). As a result, there are substantial differences in the evolution of the \ion{Ca}{ii} NIR triplet predicted by the simulations (see Section~\ref{subsec:1D_2D_NLTE_spectra_comparisons}). Excited state photoionisation is therefore an important process that can lead to clear spectroscopic differences. However, we note that for the high-velocity ejecta regions, the strong suppression of the radiation field at shorter wavelengths results in relatively few Monte Carlo packets around the bluest \ion{Ca}{ii} photoionisation edges, meaning these rates could be somewhat impacted by Monte Carlo noise. Increasing the packet sampling in such short wavelength regions, for example through the packet frequency biasing techniques used in the Monte-Carlo radiative transfer code \textsc{jekyll} \citep{ergon2018a}, is therefore an area of interest for future work. 


\bsp	
\label{lastpage}
\end{document}